\documentclass{article}

\usepackage{arxiv}
\usepackage{amsmath, amssymb, amsthm}
\usepackage[utf8]{inputenc} % allow utf-8 input
\usepackage[T1]{fontenc}    % use 8-bit T1 fonts

\usepackage{hyperref}       % hyperlinks
\hypersetup{
    colorlinks=true,
    citecolor = blue,
    linkcolor=blue,
    filecolor=blue,      
    urlcolor=black,
}

\usepackage[capitalise]{cleveref}

\usepackage{url}            % simple URL typesetting
\usepackage{booktabs}       % professional-quality tables
\usepackage{amsfonts}       % blackboard math symbols
\usepackage{nicefrac}       % compact symbols for 1/2, etc.
\usepackage{microtype}      % microtypography
\usepackage{lipsum}		% Can be removed after putting your text content
\usepackage{graphicx}
\usepackage[compress]{cite}
\usepackage{doi}
\usepackage{float}
\usepackage{multicol}
\usepackage{multirow}
\usepackage{caption}
\usepackage{subcaption}

\usepackage{soul}
\usepackage{todonotes}

\usepackage{siunitx}

\newcommand{\vecJ}{\mathbf{J}}
\newcommand{\vecK}{\mathbf{K}}

\newcommand{\vecM}{\mathbf{M}}

\newcommand{\vecb}{\mathbf{b}}

\newcommand{\vecn}{\mathbf{n}}

\newcommand{\vecu}{\mathbf{u}}

\newcommand{\Oh}{\mathcal{O}}

\newcommand{\Ro}{\mathcal{R}_0}

\newcommand{\de}{\mbox{d}}

\usepackage[aboveskip=1pt,skip=3pt,labelfont=bf,labelsep=period,justification=raggedright,singlelinecheck=off]{caption} %skip to control spacing between figure and the caption for single panel figures
\emergencystretch=\maxdimen
\usepackage[title]{appendix}
\newtheorem{proposition}{Proposition}[section]

\title{A Behaviour and Disease Model of Testing and Isolation}

\author{
\href{https://orcid.org/0000-0003-2373-4384}{\includegraphics[scale=0.06]{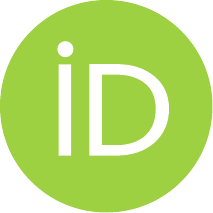}\hspace{1mm}Matthew Ryan} \\
	Commonwealth Scientific and Industrial Research Organisation (CSIRO), \\
	Adelaide, Australia \\
	\texttt{matt.ryan@csiro.au} \\
	\AND
	\href{https://orcid.org/0000-0001-6453-7745}{\includegraphics[scale=0.06]{orcid.pdf}\hspace{1mm}Roslyn I. Hickson} \\
	James Cook University and \\
    Commonwealth Scientific and Industrial Research Organisation (CSIRO), \\
	Townsville, Australia \\
	\texttt{Roslyn.Hickson@csiro.au} \\
	\AND
    \href{https://orcid.org/0000-0002-2992-2004}{\includegraphics[scale=0.06]{orcid.pdf}\hspace{1mm}Edward M. Hill}\\
    Civic Health Innovation Labs and 
    Institute of Population Health and\\
    NIHR Health Protection Research Unit in Emerging and Zoonotic Infections, \\
    University of Liverpool \\
    Liverpool, United Kingdom \\
	\texttt{Edward.Hill@liverpool.ac.uk} \\
    \AND
    \href{https://orcid.org/0000-0001-5835-8062}{\includegraphics[scale=0.06]{orcid.pdf}\hspace{1mm}Thomas House}\thanks{Corresponding author: thomas.house@manchester.ac.uk }\\
	Department of Mathematics, University of Manchester \\
    Manchester, United Kingdom \\
	\texttt{thomas.house@manchester.ac.uk} \\
    \AND
    \href{https://orcid.org/0000-0001-5189-2844}{\includegraphics[scale=0.06]{orcid.pdf}\hspace{1mm}Valerie Isham} \\
	Department of Statistical Science, University College London\\
    London, United Kingdom\\
	\texttt{v.isham@ucl.ac.uk} \\
	\AND
    \href{https://orcid.org/0000-0001-6365-5491}{\includegraphics[scale=0.06]{orcid.pdf}\hspace{1mm}Dongni Zhang} \\
	Department of Health, Medicine and Caring Sciences, Linköping University\\
      Linköping, Sweden\\
	\texttt{dongni.zhang@liu.se} \\
	\AND
	\href{https://orcid.org/0000-0003-2693-5093}{\includegraphics[scale=0.06]{orcid.pdf}\hspace{1mm}Mick G. Roberts} \\
    New Zealand Institute for Advanced Study, \\
	Massey University, Auckland, New Zealand \\
	\texttt{M.G.Roberts@massey.ac.nz}
}

\date{}

\begin{document}

\vspace*{0.25cm}

\maketitle

% \newpage
\vspace{3cm}

\begin{abstract}
There has been interest in the interactions between infectious disease dynamics and behaviour for most of the history of mathematical epidemiology. This has included consideration of which mathematical models best capture each phenomenon, as well as their interaction, but typically in a manner that is agnostic to the exact behaviour in question. 
Here, we investigate interacting behaviour and disease dynamics specifically related to decisions around testing and isolation.
To carry out our investigation we extend an existing ``behaviour and disease'' (BaD) model by incorporating the dynamics of symptomatic testing and isolation, including the influence of positive tests on perception of infection risk. We provide a dynamical systems analysis of the ordinary differential equations that define this model, providing theoretical results on its behaviour early in a new outbreak (particularly its basic reproduction number) and endemicity of the system (its steady states and associated stability criteria). We then supplement these findings with a numerical analysis to inform how temporal and cumulative outbreak metrics depend on the model parameter values for epidemic and endemic regimes. We observe novel model outputs such as epidemics that have more observed cases detected through increased testing, but are less objectively severe in terms of total number of infections.
\end{abstract}


\section*{Plain language summary}

Throughout the history of infectious disease outbreaks, people have modified their behaviour in response to risk. Such behaviours can include vaccination, seeking treatment, and non-pharmaceutical interventions like hand-washing and social distancing. There have been various attempts to include behavioural response quantitatively in models of epidemics, but these have typically not specified the exact behaviour. In this article, we develop and analyse a joint mathematical model of infection and specific decisions around testing and isolation.
This epidemiological-behavioural interaction is of particular interest as, prospectively, it is well-placed to be informed by real-world data temporally monitoring test results and compliance with testing policy. As the presented interdisciplinary modelling approach can accommodate further extensions (including, but not limited to, adding testing capacity, decay in behavioural effects and multiple pathogen variants), we hope that our work will encourage further modelling studies integrating specific measured behaviours and disease dynamics that may reduce the health and economic impacts of future epidemics.

\clearpage

\section{Introduction}\label{Intro}

\subsection{Background}

From early in the history of infectious disease modelling, it was clear that the broad
mathematical approaches used to model pathogen transmission could also be used to
model social transmission such as the spread of rumours or behaviours, and that there were
important specific differences between social and viral spreading, theoretically and empirically
\cite{Daley:1965,Centola:2010}. As data availability and quality from outbreaks improved,
empirical evidence for the impact of behavioural response on infection spread increased,
particularly during the 2004 SARS epidemic \cite{Cori:2009}.

A prominent 2009 paper by Funk \textit{et al.} \cite{Funk:2009} influenced much subsequent research by providing a framework to consider the interaction of viral and behavioural spreading.
In the last 15 years there has been significant progress on modelling the interaction of behaviour and epidemics
including theoretical analyses \cite{Poletti:2009, Kiss:2010}, consideration of social network structure
and demographics \cite{Kamp:2010}, and attempts to elucidate prototypical models combining both mechanisms \cite{Perra:2011}. An early review of the field was published in 2010 \cite{Funk:2010}, a dedicated monograph on behavioural dynamics was published in 2013 \cite{Manfredi:2013}, and a further review paper in 2015 \cite{Wang:2015}. 
Spreading of behaviour continues to inspire mathematical developments, for example development of differential equation systems with cubic and higher-order interaction terms \cite{House:2011}, use of multiplex network theory \cite{daSilva:2019},
and investigation of paradoxical behaviour \cite{Kolok:2025}. The COVID-19 pandemic, seeing significant impacts from behaviour, also saw attempts to advance this field \cite{Gozzi:2024}.

Despite this progress, important
challenges remain in developing a generalised model of epidemiological and behavioural dynamics, including balancing model complexity, interpretability and capability to validate them with empirical data \cite{Hill:2024}. And at the same time, few models have attempted to capture \textit{specific} behaviours, preferring to adopt an intervention-agnostic approach.

\subsection{Outline of this study}

Since testing and isolation are known to have both a strong behavioural component and a significant impact on disease transmission \cite{Marshall:2022,Shearer:2024}, we define and analyse an extension to the Behaviour and Disease (BaD) model of Ryan \textit{et al.} \cite{Ryan:2024, Ryan:2025} reflecting behavioural dynamics specific to (symptomatic) testing and isolation. 

Specifically, we assume that the perception of illness threat reflects the rate of positive testing for disease, and thus we take the rate of behavioural uptake to be a function of that of positive testing, leading to different dynamics from previous models. In terms of data availability, it is also possible to monitor both test results and compliance with testing policy over time using community surveys \cite{Eales:2024}, opening up possibilities for direct fitting of our model to data.

We structure the remainder of the paper as follows. We initially present our model that incorporates dynamic testing behaviour and infectious disease transmission processes (\cref{sec:methods_model}). We then conduct theoretical analyses of the model (\cref{sec:theoretical_analysis}) to give insight on the early dynamics and endemicity of the system; specifically, we derive its basic reproduction number $\mathcal{R}_{0}$ (\cref{sec:methods_R0}), analyse stability criteria (\cref{sec:methods_stability}) and compute its steady states (\cref{sec:methods_steady_states}). Note that derivation of quantities such as final size in the absence of waning immunity are not believed to be easily available for this class of models using existing methods (R\"{o}st, personal communication relating to the model published in Sayyar \textit{et al.} \cite{Sayyar:2023}). Accordingly, to investigate how temporal and cumulative outbreak metrics depend on model parameter values we describe the results of numerical simulations (\cref{sec:numerics}). We conclude by discussing the findings and their implications (\cref{sec:discussion}).

While our analysis is intended to demonstrate overall features of our model, we note that most novelty arises from the modelled dependence of behavioural uptake due to perception of illness threat on positive testing rates. We will see that it is possible to simulate scenarios where increasing linear dependence of behavioural uptake on positive testing (captured by a parameter we call $\omega_2$) can lead to epidemics that have higher reported numbers of cases, but are objectively better mitigated in terms of having lower numbers of infections. This phenomenon is shown in Figure \ref{fig:phase_epidemic_r0D} (b,e,h).

\section{Behaviour and Disease (BaD) model for symptomatic testing}\label{sec:methods_model}

\begin{figure}[htb!]
	\centering
	\includegraphics[width=0.6\textwidth]{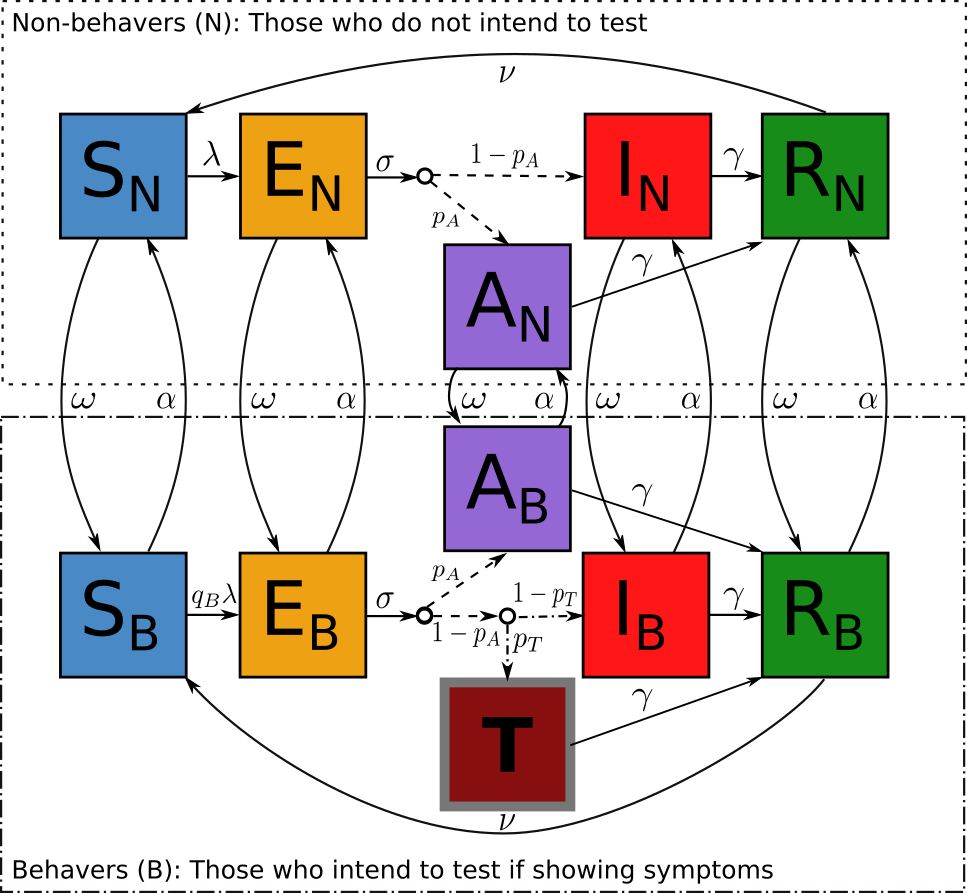}
	\caption{\textbf{A compartmental diagram illustrating the BaD model for symptomatic testing.} We distinguish between individuals who do not seek testing when symptomatic, referred to as \textit{non-behavers} (labelled $N$) and those who intend to seek testing when symptomatic, referred to as \textit{behavers} (labelled $B$). Changes in testing behaviour are governed by the rates of behavioural uptake ($\omega$) and abandonment ($\alpha$). Uptake due to perception of illness threat is with respect to those in the highlighted compartment $T$. The epidemiological states are susceptible ($S$), exposed ($E$), asymptomatic infectious (or pauci-symptomatics) labelled $A$, symptomatic infectious who are labelled $I$ if they do not seek \textit{or} do not receive a positive test and $T$ if they seek \textit{and} receive a positive test, and those who are recovered ($R$). The transitions between epidemiological states follow a standard SEIRS framework, with waning immunity where recovered individuals can become susceptible again. We define the model parameters in Table~\ref{tab:model_parameter}.}
	\label{fig:compartmental_diagram}
\end{figure}

In this section we introduce our mathematical model, which extends the BaD model of Ryan \textit{et al.} \cite{Ryan:2024} to incorporate dynamics of symptomatic testing. Ryan \textit{et al.}'s BaD model considers a susceptible-infectious-recovered-susceptible infection process paired with a non-specific visual protective health behaviour.  We extend this work in two ways. First, we consider more complex pathogen dynamics, including asymptomatic infection, as described below. This is primarily a step towards our main aim and second extension of previous work, which is to investigate the specific protective behaviour of symptomatic testing. Both generalisations are shown in the compartmental diagram for our model, Figure~\ref{fig:compartmental_diagram}.

A key concept behind our model is the distinction between individuals who, at any particular time, do not intend to seek testing when symptomatic, referred to as \textit{non-behavers} (labelled $N$) and those who may seek testing when symptomatic, referred to as \textit{behavers} (labelled $B$). In general, these intentions may change over time. We emphasise, however, that while we use the nomenclature of
behavers and non-behavers for consistency with the other literature, we are considering $B$ individuals as those who hold an \emph{intention} to test if symptomatic, and so may fail to do this if, for example, there is a lack of testing capacity. This means we parameterise the model in terms of the \emph{effectiveness} of testing as an intervention, including the real-world factors that can obstruct access to testing, rather than its \emph{efficacy} under ideal conditions that is primarily determined by test sensitivity.

Our pathogen dynamical framework is built upon the standard SEIRS compartmental framework: susceptible individuals ($S$) become exposed ($E$) at a rate $\lambda$ (defined below), progress to an infectious state ($I$, $A$ or $T$ as explained below) at a rate $\sigma$ and then recover ($R$) at a rate $\gamma$, with a non-zero rate of waning immunity $\nu$ meaning that recovered individuals can become susceptible again. The rate $\lambda$ captures the force of infection, whereas $\sigma^{-1}$, $\gamma^{-1}$ and $\nu^{-1}$ capture the average latent, infectious, and immune periods, respectively.
To account for symptomatic testing behaviour, we split the infectious class into asymptomatics (or pauci-symptomatics) labelled $A$ with probability $p_A$, and symptomatics. Symptomatics are further subdivided and are labelled $T$ with probability $p_T$ if they seek \textit{and} receive a positive test, and $I$ otherwise. The value $p_A$ captures the probability of being asymptomatic, whereas $p_T$ captures the effectiveness of the test, as defined previously.
The transitions between epidemiological states follow the standard SEIRS framework, while the changes in testing behaviour are governed by the rates of behavioural uptake ($\omega$) and abandonment ($\alpha$). 

Let $B(t)$ denote the proportion of the population at time $t$ that intend to test if showing symptoms, and $N(t)$ denote the proportion that do not. We will henceforth mostly suppress explicit dependence on $t$ for notational compactness.
For each state class $X \in \{S, E, I, A, R\}$, $X_{N}$ and $X_{B}$ denote the proportions of non-behavers and behavers, respectively. Additionally, let the proportion of symptomatic individuals who seek testing and test positive be $T$. 

We adopt similar functional forms for the rates of behavioural uptake and behaviour loss as in the original BaD model \cite{Ryan:2024}, with one key difference in behavioural uptake. The rate of behavioural uptake is governed by
\begin{equation}\label{eqn:omBT}
\omega(B,T)=\phi_1(B)+\phi_2(T)+\omega_3=\omega_1 B+\omega_2 T+\omega_3+\Oh(B,T)^2
\end{equation}
where $\phi_1$ ($\omega_1$) represents the influence of social contagion on testing behaviour, $\phi_2$ ($\omega_2$) accounts for behavioural uptake driven by the perception of illness threat, and $\omega_3$ captures spontaneous adoption of the behaviour.  Here, perception of illness threat is influenced by the prevalence of positive tests in the population ($T$, highlighted in Figure \ref{fig:compartmental_diagram}) as opposed to the prevalence of infection ($I_N + I_B + T$) in the BaD model \cite{Ryan:2024}, which also did not capture asymptomatics.
The rate of behaviour loss is 
\begin{equation}\label{eqn:alN}
\alpha(N)=\psi (N)+\alpha_2=\alpha_1 N+\alpha_2+\Oh(N)^2
\end{equation}
where $\psi$ ($\alpha_1$) represents the influence of social contagion on not seeking testing and $\alpha_2$ captures spontaneous abandonment of the behaviour. Note that while, for a
complex contagion model, the quadratic and higher-order terms on the right-hand sides of Equations \eqref{eqn:omBT} and \eqref{eqn:alN} might be large and even provide the dominant
contribution to $\omega$ and $\alpha$, we will only include linear terms in our analysis when an explicit form is required.

The BaD model for symptomatic testing is described by the following differential equations with the parameters detailed in Table~\ref{tab:model_parameter}. 
\begin{align}
%\label{dN}
\label{eqn:Sn}
  \frac{\de S_N}{\de t}  & =-\lambda S_N +\nu R_N-\omega(B,T)S_N+\alpha(N)S_B \\
\label{eqn:En}
   \frac{\de E_N}{\de t}    &  = \lambda S_N-\sigma E_N -\omega(B,T)E_N+\alpha(N)E_B  \\% \nonumber \\
\label{eqn:An}
     \frac{\de A_N}{\de t}    &  = p_A \sigma E_N -\gamma A_N-\omega(B,T)A_N+\alpha(N)A_B \\%   \nonumber \\
\label{eqn:In}
        \frac{\de I_N}{\de t}    &  = \left(1-p_A\right) \sigma E_N -\gamma I_N-\omega(B,T)I_N+\alpha(N)I_B \\%  \nonumber \\
\label{eqn:Rn}
         \frac{\de R_N}{\de t}    &  = \gamma (A_N+ I_N)-\nu R_N-\omega(B,T)R_N+\alpha(N)R_B   \\
\label{eqn:Sb}
  \frac{\de S_B}{\de t}  & =-q_B \lambda S_B+\nu R_B+\omega(B,T)S_N-\alpha(N)S_B  \\
\label{eqn:Eb}
   \frac{\de E_B}{\de t}    &  = q_B \lambda S_B-\sigma E_B +\omega(B,T)E_N-\alpha(N)E_B   \\%\nonumber \\
\label{eqn:Ab}
     \frac{\de A_B}{\de t}    &  = p_A \sigma E_B -\gamma A_B+\omega(B,T)A_N-\alpha(N)A_B \\ % \nonumber \\
\label{eqn:Ib}
        \frac{\de I_B}{\de t}    &  = \left(1-p_A\right) \left(1-p_T\right)\sigma E_B -\gamma I_B+\omega(B,T)I_N-\alpha(N)I_B \\%   \nonumber \\
\label{eqn:T}
         \frac{\de T}{\de t}    &  = \left(1-p_A\right) p_T \sigma E_B -\gamma T  \\%\nonumber \\
\label{eqn:Rb}
         \frac{\de R_B}{\de t}    &  = \gamma (A_B+ I_B+ T)-\nu R_B+\omega(B,T)R_N-\alpha(N)R_B. %\nonumber
\end{align}

The force of infection on non-behavers is given by
$$\lambda=\beta\left(I_N+I_B+q_A\left(A_N+A_B\right)+q_T T\right)$$
whereas for behavers it is $q_{B}\lambda$.  Here, $\beta$ is the transmission rate for the pathogen while the values $q_B, q_A$ and $q_T$ capture the relative reduction in infectiousness for behavers, asymptomatics, and test-and-isolators respectively.

\begin{table}[htb!]
    \centering
        \caption{\textbf{Behavioural and epidemiological parameters for the model.}  The parameters and functions defined by Greek letters are in units of $\text{time}^{-1}$, which have been scaled to such that the infectious period is one.  All other parameters are dimensionless.  Where possible, the last column contains references for the parameter choices. We chose all other parameters to take \textit{a priori} reasonable values with a non-trivial impact on the dynamics.
        }
         \label{tab:model_parameter}
         \renewcommand\arraystretch{1.05}
    \begin{tabular}{l l S[table-format=1.4] r}
    \\
    \hline
       {\bf Parameter}  &  {\bf Description} & {\bf Value} & {\bf Ref.}\\
       \hline
       $\lambda(t)$ & the force of infection &&\\
      
       \hspace{4mm}$\beta$ & the transmission rate & 4.151& \cite{Karimizadeh:2023}\\
       
        \hspace{4mm}$q_{A}$ & the reduction in transmission due to being asymptomatic & 0.58 & \cite{Byambasuren:2020}\\
       
       \hspace{4mm}$q_{T}$ & the reduction in transmission due to (imperfect) isolation & 0.25 &\\

       \hspace{4mm}$q_{B}$ & the reduction in transmission due to other changes in behaviour from being willing to test & 0.5 & \cite{Ryan:2024}\\
       \hline
       $\sigma$ & the transition rate from $E$ to $I$ (average latency period is $1/\sigma$) & 2.5 &\cite{CDNA:2024}\\
          \hline
       $\gamma$ & the recovery rate (average infectious period is $1/\gamma$) & 1.0 &\\
          \hline
         $\nu$ & the rate of waning immunity (average immune period is $1/\nu$) & 0.05 & \cite{Stein:2023}\\
          \hline 
        $p_{A}$ & probability of not developing symptoms & 0.18  &\cite{Byambasuren:2020}\\
          \hline
       $p_{T}$ & test effectiveness for symptomatic individuals & 0.9 &\\
          \hline
        $\omega(t)$ & the rate of 
        testing behavioural uptake &\\ 

            \hspace{4mm}$\omega_1$ & the social transmission rate & 0.25 &\\

            \hspace{4mm}$\omega_2$ & rate of response to perceived illness threat & 7.0 &\\

            \hspace{4mm}$\omega_3$ & spontaneous uptake rate & 0.0125 &\\
                
          \hline
        $\alpha(t)$ & the rate of 
        testing behaviour abandonment & \\ 

            \hspace{4mm}$\alpha_1$ & the social abandonment rate & 0.2 &\\

            \hspace{4mm}$\alpha_2$ & spontaneous abandonment rate & 0.1 &\\
            
          \hline
    \end{tabular}
\end{table}

For non-behavers, we can express the total proportion as 
$N=S_N+E_N+A_N+I_N+R_N.$
Similarly, for behavers, the total proportion is $B =S_B+E_B+A_B+I_B+T+R_B.$
In the absence of infection ($S_{N}=N$ and $S_{B}=B$) and ignoring higher order terms, the dynamics become
\begin{align*}
\frac{\de N}{\de t}  & =-\omega(B,0)N+\alpha(N)B  \\
    &   =-\left(\omega_1 B+\omega_3\right) N+\left(\alpha_1 N+\alpha_2 \right)B
\end{align*}
with $B= 1-N$.
Linearising about small initial behaviour ($B(0)\ll 1$) we have $N\approx 1$ and, if $\omega_3 = 0$, behaviour spreads through social pressure if $\omega_1 > \alpha_1 + \alpha_2$.  Following Ryan \textit{et al.} \cite{Ryan:2024}, we define the social reproduction number to be 
\[
\Ro^{B} = \frac{\omega_1}{\alpha_1 + \alpha_2}\, .
\]
The steady state solves
\begin{equation}\label{eqn:F_N}
    F(N)=\left(\alpha_1-\omega_1\right) N^2-\left(\alpha_1-\alpha_2-\omega_1-\omega_3\right) N-\alpha_2=0\, .
\end{equation}
As $F(0)=-\alpha_2<0$ and $F(1)=\omega_3>0$, there exists a unique solution $N_0 \in (0,1)$ with $F(N_0)=0$. Notice that if $\omega_3 >0$ then $B_0 > 0$ regardless of choice of $\omega_1$, that is, if $\omega_3 >0$ then $(N, B) = (1, 0)$ is not a steady state of the system.

\section{Theoretical analysis} \label{sec:theoretical_analysis}
In this section, we first derive the basic reproduction number $\mathcal{R}_{0}$ (\cref{sec:methods_R0}), 
then establish stability conditions for the infection-free equilibrium (\cref{sec:methods_stability}), 
and finally characterise the endemic steady states (\cref{sec:methods_steady_states}).

\subsection{The basic reproduction number}\label{sec:methods_R0}

\subsubsection{In the absence of behaviour}

Consider when $\omega_3=0$ and $\Ro^{B} < 1$.  Then behaviour will be absent from the model and the infection-free steady state has $S_N = 1$ and $S_B = 0$. In this situation, the model simplifies to the standard SEIRS model with asymptomatics, and so the epidemic \textit{takes off} if
$$\Ro^{D}=\frac{\left(p_Aq_A+1-p_A\right)\beta}{\gamma}>1\, .$$
We call $\Ro^{D}$ the behaviour-free reproduction number.  Note the threshold $\Ro^{D}$ also holds if $q_B = q_T = 1$, that is, there is no effect on the transmission dynamics from cautious behaviour or isolation.

\subsubsection{In the presence of behaviour}

We now consider the full model where both behaviour and infection are present (Equations \eqref{eqn:Sn}-\eqref{eqn:Rb}). Here we calculate the next-generation matrix, which will turn out to be equal to
\begin{equation}
\vecK =\left(\begin{array}{cc} \Lambda_N N_0& \Lambda_B N_0 \\ q_B\Lambda_N B_0 & q_B\Lambda_B B_0\end{array}\right), \label{nextGenWithB}
\end{equation}
where $B_0$ and $N_0$ are the proportions of the population behaving and not respectively at the start of the epidemic, $q_b$ is the reduction in transmission due to other changes in behaviour from being willing to test, and $\Lambda_B$ and $\Lambda_N$ are the forces of infection produced by behavers and non-behavers respectively. The dominant Eigenvalue of this matrix will give the basic reproduction number that we present and explain in Equation~\eqref{eqn:R0_to_R0d} below.

We now derive the result above by considering the different transmission and transition events possible. The model has two \textit{states-at-infection}, $E_N$ and $E_B$. Let $t_N(X)$ denote the expected time an individual spends in compartment $X$ following \textit{state-at-infection} $E_N$. Similarly, let $t_B(X)$ represent the corresponding time spent in compartment $X$ following \textit{state-at-infection} $E_B$.

Defining $\alpha_0 = \alpha(N_0)$ and $\omega_0 = \omega(B_0, 0)$, we derive expressions for the expected time spent in each compartment.  Consider an individual with \textit{state-at-infection} $E_N$ spending time in the asymptomatic state $A_N$.  This individual will exit the exposed state after $\sigma^{-1}$ time units, on average, enter the asymptomatic state at a rate $p_A \sigma$, and will spend on average $\gamma^{-1}$ time units in the asymptomatic state.  However, there are several possible paths this individual may take to move from $E_N$ to $A_N$.  They may move directly from $E_N$ to $A_N$, or they may pass through $E_B$ (possibly alternating between $E_B$ and $E_N$  some number of times) to reach $A_N$.  For an individual starting in $E_N$, the probability they exit $E_N$ into $A_N$ is
\[
    \frac{\sigma}{\sigma + \omega_0} p_A\sum_{i=0}^{\infty} \left\{\frac{\alpha_0\omega_0}{(\sigma + \alpha_0)(\sigma + \omega_0)} \right\}^i= \frac{\sigma + \alpha_0}{\sigma + \omega_0 + \alpha_0} p_A
\]
where the term $p_A\sigma/\left(\sigma+\omega_0\right)$ captures the probability of exiting epidemiologically out of $E_N$ into $A_N$, and the infinite sum accounts for any behavioural loops to $E_B$. Alternatively, the probability that they exit the exposed state via $E_B$ into $A_B$ is
\[  
    \left\{1-  \frac{\sigma + \alpha_0}{\sigma + \omega_0 + \alpha_0}\right\} p_A = \frac{\omega_0}{\sigma + \omega_0 + \alpha_0} p_A\, .
 \]
This individual will then transition from $A_B$ to $A_N$ with probability 
\[
    \frac{\alpha_0}{\alpha_0 + \gamma}.
\]
Finally the individual transitions epidemiologically from $A_N$ to $R_N$ (again allowing for behavioural loops to $A_B$) with probability
\[
     \frac{\gamma + \alpha_0}{\gamma + \omega_0 + \alpha_0}\,.
\]
Thus, combining the behavioural and epidemiological aspects and simplifying, we find
\begin{align}
\label{eqn:tEnAn}
    t_N(A_N) 
    & = \frac{p_A\left[\alpha_0\left(\sigma+\alpha_0+\omega_0\right)+\gamma\left(\sigma+\alpha_0\right)\right]}{\gamma\left(\sigma+\alpha_0+\omega_0\right)\left(\gamma+\alpha_0+\omega_0\right)}\, .
\end{align}
Applying the same process to all other infectious states, for individuals with \textit{state-at-infection} $E_N$ we have
\begin{align}
\label{ENN}
  t_N(I_N)   &  =\frac{\left(1-p_A\right)\left[\alpha_0\left(\sigma+\alpha_0+\omega_0\right)+\gamma\left(\sigma+\alpha_0\right)-p_T\alpha_0\omega_0\right]}{\gamma\left(\sigma+\alpha_0+\omega_0\right)\left(\gamma+\alpha_0+\omega_0\right)} 
\end{align}
and
\begin{align}
\label{ENB}
 t_N(A_B)   &  =\frac{p_A\omega_0\left(\gamma+\sigma+\alpha_0+\omega_0\right)}{\gamma\left(\sigma+\alpha_0+\omega_0\right)\left(\gamma+\alpha_0+\omega_0\right)} \\
  t_N(I_B)   &  =\frac{\left(1-p_A\right)\omega_0\left[\sigma+\alpha_0+\left(1-p_T\right)\left(\gamma+\omega_0\right)\right]}{\gamma\left(\sigma+\alpha_0+\omega_0\right)\left(\gamma+\alpha_0+\omega_0\right)} \nonumber \\
  t_N(T)   &  =\frac{\left(1-p_A\right)p_T\omega_0}{\gamma\left(\sigma+\alpha_0+\omega_0\right)}\, . \nonumber
\end{align}
Similarly, for individuals with \textit{state-at-infection} $E_B$, we have
\begin{align}
\label{EBN}
 t_B(A_N)   &  =\frac{p_A\alpha_0\left(\gamma+\sigma+\alpha_0+\omega_0\right)}{\gamma\left(\sigma+\alpha_0+\omega_0\right)\left(\gamma+\alpha_0+\omega_0\right)} \\
  t_B(I_N)   &  =\frac{\left(1-p_A\right)\alpha_0\left[\gamma+\alpha_0+\left(1-p_T\right)\left(\sigma+\omega_0\right)\right]}{\gamma\left(\sigma+\alpha_0+\omega_0\right)\left(\gamma+\alpha_0+\omega_0\right)} \nonumber
\end{align}
and
\begin{align}
\label{EBB}
 t_B(A_B)   &  =\frac{p_A\left[ \omega_0\left(\sigma+\alpha_0+\omega_0\right)+\gamma(\sigma + \omega_0)\right]}{\gamma\left(\sigma+\alpha_0+\omega_0\right)\left(\gamma+\alpha_0+\omega_0\right)} \\
  t_B(I_B)   &  =\frac{\left(1-p_A\right)\left[ \omega_0(\sigma + \alpha_0 + \omega_0) + \gamma(\sigma + \omega_0) - p_T(\sigma + \omega_0)(\gamma + \omega_0) \right]}{\gamma\left(\sigma+\alpha_0+\omega_0\right)\left(\gamma+\alpha_0+\omega_0\right)} \nonumber \\
  t_B(T)   &  =\frac{\left(1-p_A\right)p_T\left(\sigma+\omega_0\right)}{\gamma\left(\sigma+\alpha_0+\omega_0\right)}\, . \nonumber
\end{align}
If an individual with \textit{state-at-infection} $E_N$ or $E_B$ results in a total (over time) force of infection $\Lambda_N$ or $\Lambda_B$ respectively on non-behavers $N$, then
\begin{align*}
 \Lambda_N   & = \beta\left\{t_N(I_N)+t_N(I_B)+q_A\left(t_N(A_N)+t_N(A_B)\right)+q_T t_N(T)\right\}
 \end{align*}
 and
 \begin{align*}
 \Lambda_B     &   = \beta\left\{t_B(I_N)+t_B(I_B)+q_A\left(t_B(A_N)+t_B(A_B)\right)+q_T t_B(T)\right\}.
\end{align*}
By substituting Equations \eqref{eqn:tEnAn}--\eqref{EBB}, the terms $\Lambda_{N}$ and $\Lambda_{B}$ can be further expressed as
\begin{align}
	\label{eqn:lambda_n_simple}
	\Lambda_{N} & = \frac{\beta}{\gamma}\left\{ p_N(1-p_A) + q_A p_A + (1-p_N)(1-p_A)\left[ (1-p_T) + q_T p_T \right] \right\} \\
	\label{eqn:lambda_b_simple}
	\Lambda_{B} & = \frac{\beta}{\gamma}\left\{ (1-p_B)(1-p_A) + q_A p_A + p_B(1-p_A)\left[ (1-p_T) + q_T p_T \right] \right\}
\end{align}
where
$$p_N=\frac{\sigma+\alpha_0}{\sigma+\alpha_0+\omega_0} \qquad \text{and} \qquad
p_B=\frac{\sigma + \omega_0}{\sigma+\alpha_0+\omega_0}$$
are the proportions of time that individual with \textit{state-at-infection} $E_N$ will spend in $N$ before becoming infectious and an individual with \textit{state-at-infection} $E_B$ will spend in $B$ before becoming infectious, respectively.  

From Equation \eqref{eqn:lambda_n_simple}, we can observe how different factors contribute to the force of infection from individuals with \textit{state-at-infection}  $E_N$. The first term $p_N(1-p_A)$ represents the proportion of those individuals who become infectious and symptomatic while in state $N$. The second term $q_A p_A$ represents the proportion of those who become asymptomatic (note that this term combines contributions from both behaviour states). The expression $(1-p_N)(1-p_A)$ in the third and fourth terms describes the proportion of individuals who transition from state $N$ to state $B$ before becoming symptomatic while in $B$;  the third term component $(1-p_T)$ represents those who do not receive a positive test, which could be due to imperfect test sensitivity or to not testing despite having formed the intention to do so, while the fourth term component $q_T p_T$ represents those who receive a positive test.  
A similar interpretation applies to Equation \eqref{eqn:lambda_b_simple}, describing infections due to those with \textit{state-at-infection}  $E_B$.

Finally, the next generation matrix can be written as
$$\vecK =\left(\begin{array}{cc} \Lambda_N N_0& \Lambda_B N_0 \\ q_B\Lambda_N B_0 & q_B\Lambda_B B_0\end{array}\right).$$
This is the result advertised in Equation~\eqref{nextGenWithB} above.

As $\det(\vecK)=0$, the basic reproduction number of the BaD SEIRS model for symptomatic testing is given by
\begin{equation*}
    \Ro= \Lambda_N N_0+q_B\Lambda_B B_0
\end{equation*} 
where $\Lambda_N, \Lambda_B$ are given by Equations \eqref{eqn:lambda_n_simple} and \eqref{eqn:lambda_b_simple}, $N_0$ solves Equation \eqref{eqn:F_N} and $B_0 = 1-N_0.$
Note that $\Lambda_N$ and $\Lambda_B$ can be rewritten as
\begin{align*}
    \Lambda_N &= \Ro^{D} - \frac{\beta(1-p_N)(1-p_A)(1-q_T) p_T}{\gamma} 
    \end{align*}
 and
 \begin{align*}
    \Lambda_B & = \Ro^{D} - \frac{\beta p_B(1-p_A)(1-q_T) p_T}{\gamma}
\end{align*}
where $\Ro^{D}$ is the behaviour-free reproduction number. Defining
\[
    \Delta := \frac{\beta (1-p_A)(1-q_T) p_T}{\gamma}
\]
enables us to express $\Ro$ in a form that makes explicit the reductions in transmission due to the different behavioural aspects of the model, such that
\begin{equation}
\label{eqn:R0_to_R0d}
    \Ro= \underbrace{N_0 \Ro^{D}}_{\text{(i)}} 
    - \underbrace{(1-p_N)  N_0 \Delta}_{\text{(ii)}} 
    + \underbrace{q_B B_0 \Ro^{D}}_{\text{(iii)}} 
    - \underbrace{q_B p_B  B_0 \Delta}_{\text{(iv)}}\, .
\end{equation}
In Equation \eqref{eqn:R0_to_R0d}: (i) captures the spread of infection from non-isolators (those in $A$ and $I$) to non-behavers (those in $N$); (ii) captures the reduction in infections in non-behavers caused by isolation (those in $T$); (iii) captures the spread of infection from non-isolators to behavers (those in $B$), and; (iv) captures the reduction in infections in behavers caused by isolation.  Note that in the limiting case where $B_0$ approaches $0$, $p_N$ approaches $1$ and Equation \eqref{eqn:R0_to_R0d} shows that $\Ro$ approaches $\Ro^{D}$.

\subsubsection{The effective reproduction number}

The effective reproduction number can be defined using Equation \eqref{eqn:R0_to_R0d} such that
\begin{equation}
    \label{eqn:Reff}
    \mathcal{R}(t) = S_N(t) \Ro^{D}  - (1-p_N(t)) S_N(t) \Delta + q_B S_B(t) \Ro^{D} - q_B p_B(t) S_B(t) \Delta
\end{equation}
where
\begin{eqnarray*}
p_N(t)  =\frac{\sigma+\alpha(N(t))}{\sigma+\alpha(N(t))+\omega(B(t), T(t))} & \text{and} &
p_B(t) =\frac{\sigma + \omega(B(t), T(t))}{\sigma+\alpha(N(t))+\omega(B(t), T(t))}\, .
\end{eqnarray*}
The effective reproduction number $\mathcal{R}(t)$ is a time-dependent quantity representing the instantaneous number of new infections that each current infective is producing, accounting for depletion of the susceptible population.  When the system starts with initial conditions such that for early times $t$, $S_N(t) \approx N_0$ and $S_B(t)\approx B_0$, we note from comparison of 
\eqref{eqn:Reff} and \eqref{eqn:R0_to_R0d} that $\mathcal{R}(t) \approx \Ro$.  

Alternatively, consider where $S_N(0) \approx 1$ and $S_B(0)$ is negligible.  This captures situations where, for example, test availability may be extremely limited in early stage of an epidemic or the importance of testing is not widely known for a new pathogen.  For these conditions, we observe that
\[
    \mathcal{R}(0) \approx \Ro^{D} - (1-p_N(0)) \Delta
\]
where 
\[
    1-p_N(0) \approx \frac{\omega_3}{\sigma + \alpha_1 + \alpha_2 + \omega_3}
\]
captures the reduction in the behaviour-free reproduction number due to a small number of people spontaneously testing in the early stage of the epidemic.

\subsection{Stability analysis}\label{sec:methods_stability}

Here we investigate the stability properties of the system. Our focus is the infection-free equilibrium, for which we derive the conditions ensuring stability. The main result is stated in below, with full details provided in the Appendices.\\

\begin{proposition}\label{thm:stability}
Let $E_{B0}$ denote the infection-free equilibrium of the BaD model described in Section \ref{sec:methods_model}, with behaviour at its disease-free level
$(S_N,S_B)=(N_0,B_0)$ where $B_0=1-N_0$ and $N_0$ solves Equation \eqref{eqn:F_N}. $E_{B0}$ is stable if and only if $\Ro <1$ where $\Ro$ is defined in Section \ref{sec:methods_R0} (Equation \eqref{eqn:R0_to_R0d}).
\end{proposition}
\begin{proof}
See Appendix~\ref{append:stability}.
\end{proof}
This result shows that, in the absence of infection, the BaD model with testing follows the standard epidemic threshold principle: 
if $\Ro<1$, small outbreaks die out and the system remains at the infection-free equilibrium, while $\Ro>1$ makes this equilibrium unstable. In the latter case, the system moves towards the endemic steady states which we describe in the following subsection.

\subsection{The endemic steady states}
\label{sec:methods_steady_states}

We now characterize the endemic steady states of the model. The key result is summarized in the following proposition, with detailed derivations provided in the Appendices.\\

\begin{proposition}\label{thm:steady_states}
For each compartment $X$, let $X^*$ denote its steady state and write $X^* = X_N^* + X_B^*$. Let the endemic prevalence of symptomatic cases be $O^* = I^* + T^*$. For a given endemic infection prevalence $I^*$ and testing prevalence $T^*$, the remaining steady states are determined by 
\begin{align} 
S_N^* &= 1 - (E^* + A^* + O^* + R^*) - S_B^* & 
S_B^* &= B^* - (E_B^* + A_B^* + I_B^* + T^* + R_B^*) \label{eqn:SnSb}\\ 
E_N^* &= \frac{\gamma}{(1-p_A)\sigma} O^* - E_B^* & 
E_B^* &= \frac{q_B \lambda^* S^* + \big(\omega^* - q_B(\omega^*+\sigma)\big)E^*}{(1-q_B)(\alpha^*+\omega^*+\sigma)} \label{eqn:EnEb}\\ 
A_N^* &= \frac{p_A}{1-p_A} O^* - A_B^* & 
A_B^* &= \frac{p_A \sigma E_B^* + \omega^* A^*}{\gamma + \alpha^* + \omega^*} \label{eqn:AnAb}\\ 
I_N^* &= I^* - I_B^* & 
I_B^* &= \frac{(1-p_A)(1-p_T)\sigma E_B^* + \omega^* I^*}{\gamma + \alpha^* + \omega^*} \label{eqn:InIb}\\ 
R_N^* &= \frac{\gamma}{\nu(1-p_A)} O^* - R_B^* & 
R_B^* &= \frac{\gamma(A_B^* + I_B^* + T^*) + \omega^* R^*}{\alpha^* + \omega^* + \nu} \label{eqn:RnRb}
\end{align} 
where the endemic behaviour prevalence $B^* = 1-N^*$ with $N^*$ being the unique solution in $(0,1)$ of the quadratic equation 
\begin{equation}\label{eqn:G_N}
    G(N) = (\omega_1 - \alpha_1)N^2 - (\omega_2T^* + \omega_1 + \omega_3 + \alpha_2 - \alpha_1(1-T^*))N + \alpha_2(1-T^*) = 0
\end{equation}
and the equilibrium rates are given by 
\begin{align} \label{eqn:lambdastar} 
\lambda^* & = \beta \left(I^* + q_A A^* + q_T T^* \right)\\ \label{eqn:omegastar} 
\omega^* & = \omega_1 B^* + \omega_2 T^* + \omega_3 \\ \label{eqn:alphastar} 
\alpha^* & = \alpha_1 N^* + \alpha_2. \end{align}

Finally, $I^*$ and $T^*$ are determined by the simultaneous steady-state constraints
\begin{equation}\label{eqn:constraint_1}
    T^* = \frac{(1-p_A)p_T \sigma}{\gamma} E_B^*, 
\end{equation}
\begin{equation}\label{eqn:constraint_2}
    (\lambda^* + \omega^*) S_N^* = \alpha^* S_B^* + \nu R_N^*. 
\end{equation}
\end{proposition}
\begin{proof}
See Appendix~\ref{append:steady_states}.
\end{proof}
This result shows that the endemic equilibrium is defined through a system of coupled equations reflecting the feedback loop between infection dynamics and testing behaviour. Due to this complexity, the endemic states cannot be solved analytically in a straightforward way. We therefore explore their properties numerically in Section~\ref{sec:results_endemic_regime}.

\section{Numerical investigations}\label{sec:numerics}

Having obtained analytical insights on the early outbreak dynamics and stability criterion of the system, we next apply numerical methods to gather epidemiological and behavioural insights in epidemic and endemic infection regimes.

For the epidemic regime ($\nu=0$), we explore numerically the impact of key model parameters on the dynamics without waning immunity (SEIR-like infection structure), through exploration of the final size, peaks in testing behaviour, and phase diagrams (\cref{sec:results_epidemic_regime}). For the endemic regime ($\nu>0$), we use a model with waning immunity (SEIRS-like infection structure) to explore the impacts on the steady states and their stability (\cref{sec:results_endemic_regime}).

Given our focus on the uptake of willingness to test and isolate, the key model parameters we vary are the testing effectiveness ($p_T$), isolation effectiveness ($1-q_T$), initial proportion of the population willing to test and isolate ($B(0)$) by varying spontaneous uptake ($\omega_3$), the behaviour-free reproduction number ($\Ro^D$) by varying the transmission rate ($\beta$), and the social reproduction number ($\Ro^{B}$) by varying the social transmission rate ($\omega_1$). We further explore the impacts of changing each aspect of the uptake rate on the transient dynamics through phase diagrams without waning immunity.

We list the baseline parameters for our numerical investigation in \cref{tab:model_parameter}. We chose the transmission rate, asymptomatic rate, latent period, and immune period to represent a COVID-like illness.  Specifically, we set $\beta$ such that the basic reproduction number is $\Ro=3.28$\cite{Karimizadeh:2023}, we fixed $p_A$ and $q_A$ at $0.18$ and $0.58$ respectively \cite{Byambasuren:2020}, and fixed $\sigma$ and $\nu$ at $2.5$ \cite{CDNA:2024} and $0.05$ \cite{Stein:2023} respectively to correspond to a scaled infectious period where $\gamma =1$ representing an average infection period of $10$ days \cite{CDNA:2024}.  We selected the behavioural parameters making up $\alpha$ and $\omega$ to replicate waves in testing behaviour observed over multiple SARS-CoV-2 variants in real data such as those seen in Figure 1 of Eales \textit{et al.} \cite{Eales:2025}.  The transient dynamics for the baseline parameter values of Table \ref{tab:model_parameter} are found in Supplementary Figure \ref{fig:transient}. We fixed the reduction in susceptibility due to risk aversion at $q_B=0.5$, similarly to previous behaviour and disease models \cite{Ryan:2024}. Lastly, we fixed the parameters $p_T$ and $q_T$ at $0.9$ and $0.25$ respectively at baseline, and varied these throughout the numerical explorations.
 
We ran all numerical simulations with initial conditions of $S_B(0)=B_0$, $I_N(0)=10^{-6}$, $S_N(0) = 1 - S_B(0) - I_N(0)$ and all other compartments empty, where $B_0$ is the disease-free equilibrium. The exception to this was when exploring the transitory dynamics without waning immunity, when we instead set $S_B(0) = 10^{-6}$.  This choice was made to ensure the effective reproduction number at time zero was approximately comparable between simulations when investigating the phase diagrams.
We developed and ran code using Python 3.10.12, which we have made open-source and publicly available via GitHub  at \url{https://github.com/Matthew-Ryan1995/BaD_testing_and_isolation}.

\subsection{Epidemic regime with no waning immunity}
\label{sec:results_epidemic_regime}

Here we report the impact of key model parameters on the dynamics without waning immunity (i.e.\ a single-epidemic regime, with $\nu=0$), through exploration of the symptomatic final size, peaks in testing behaviour and the effect of behavioural uptake rates on the epidemiological-behavioural dynamics.

\subsubsection{Symptomatic final size}

\begin{figure}[htbp]
    \centering
    \begin{subfigure}[b]{0.3\textwidth}
    \caption{}
    \includegraphics[width=\textwidth]{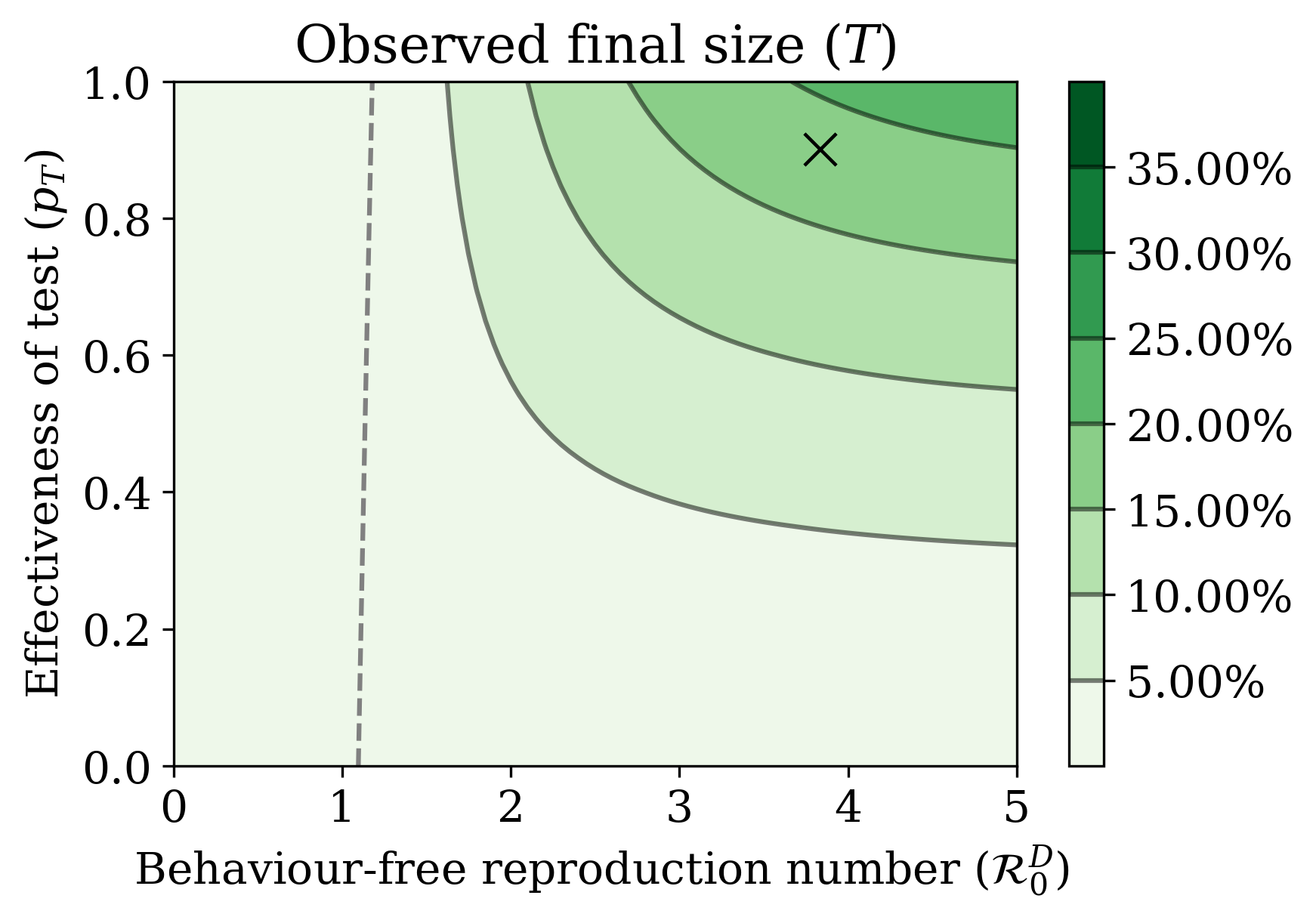}
    \end{subfigure}
    \hfill
    \begin{subfigure}[b]{0.3\textwidth}
    \caption{}
    \includegraphics[width=\textwidth]{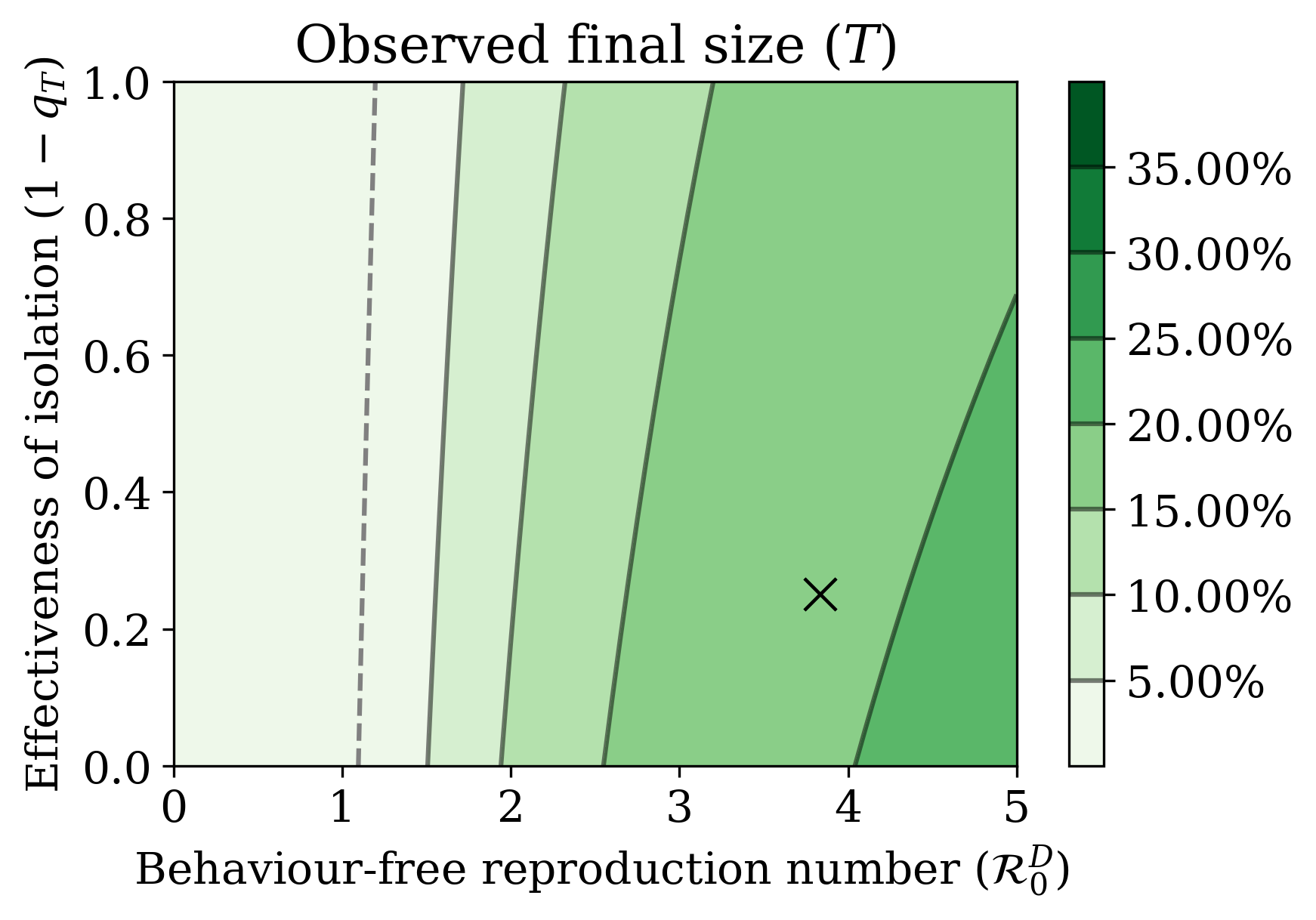}
    \end{subfigure}
    \hfill
    \begin{subfigure}[b]{0.3\textwidth}
    \caption{}
    \includegraphics[width=\textwidth]{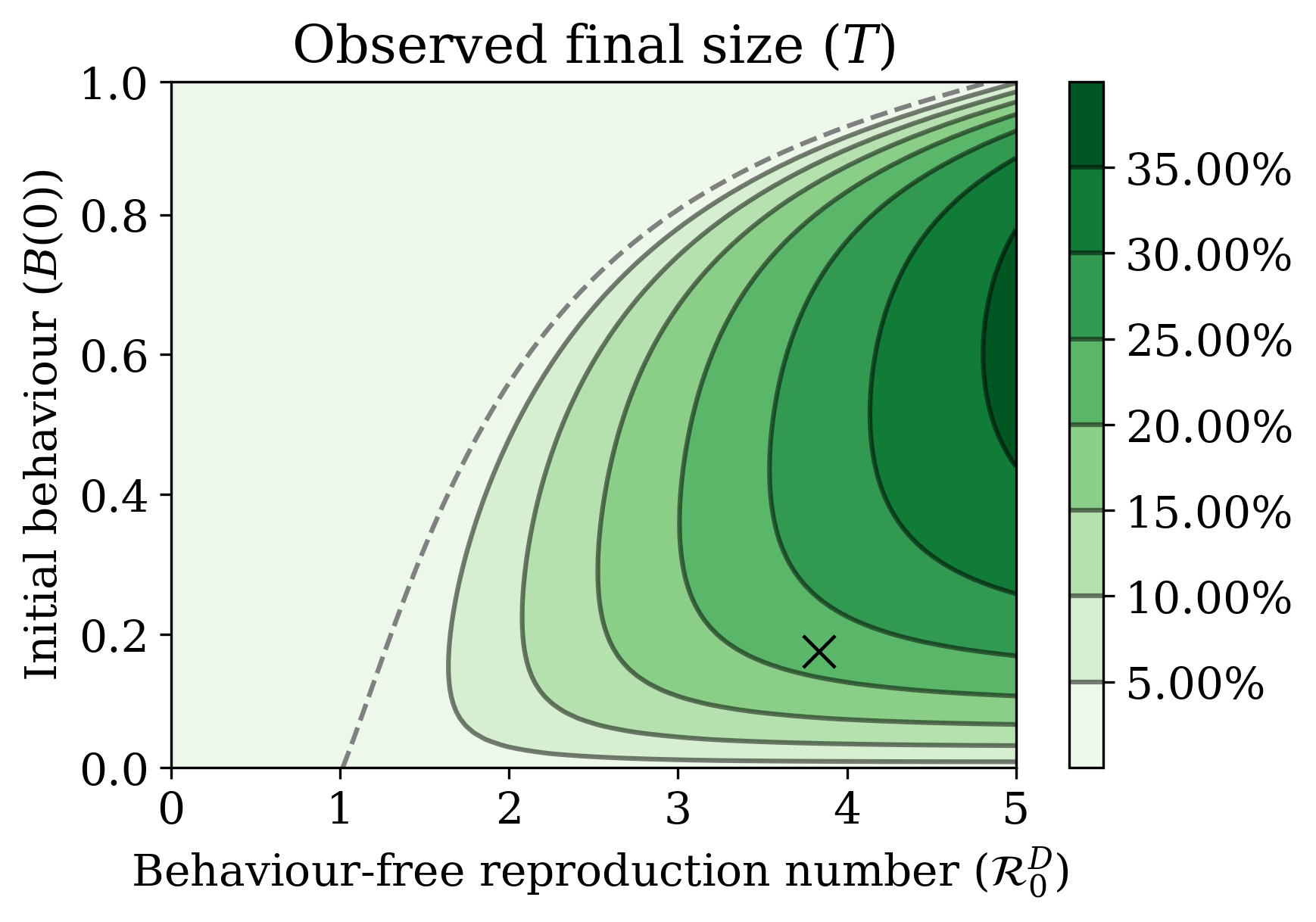}
    \end{subfigure}
    \hfill

    \begin{subfigure}[b]{0.3\textwidth}
    \caption{}
    \includegraphics[width=\textwidth]{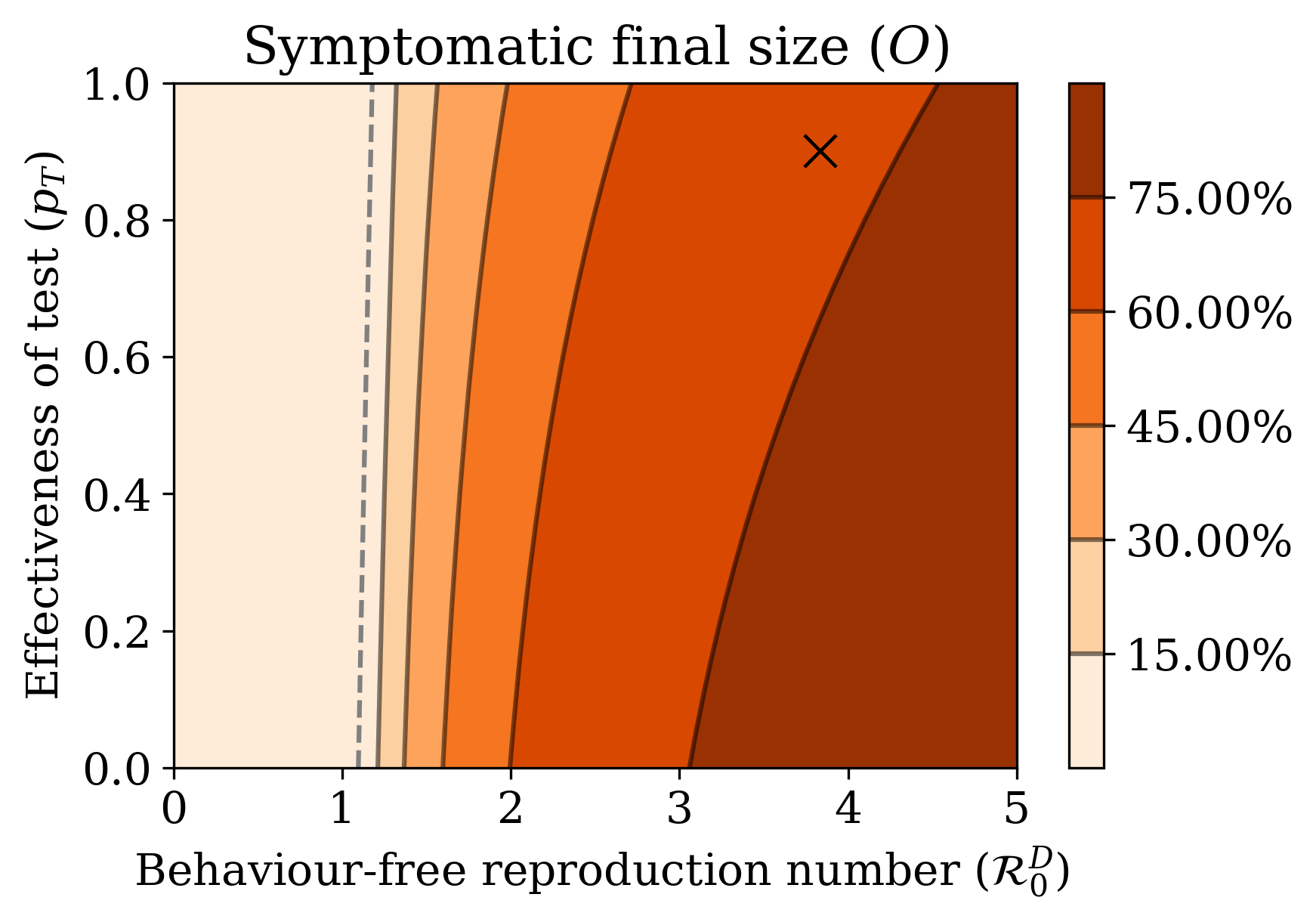}
    \end{subfigure}
    \hfill
    \begin{subfigure}[b]{0.3\textwidth}
    \caption{}
    \includegraphics[width=\textwidth]{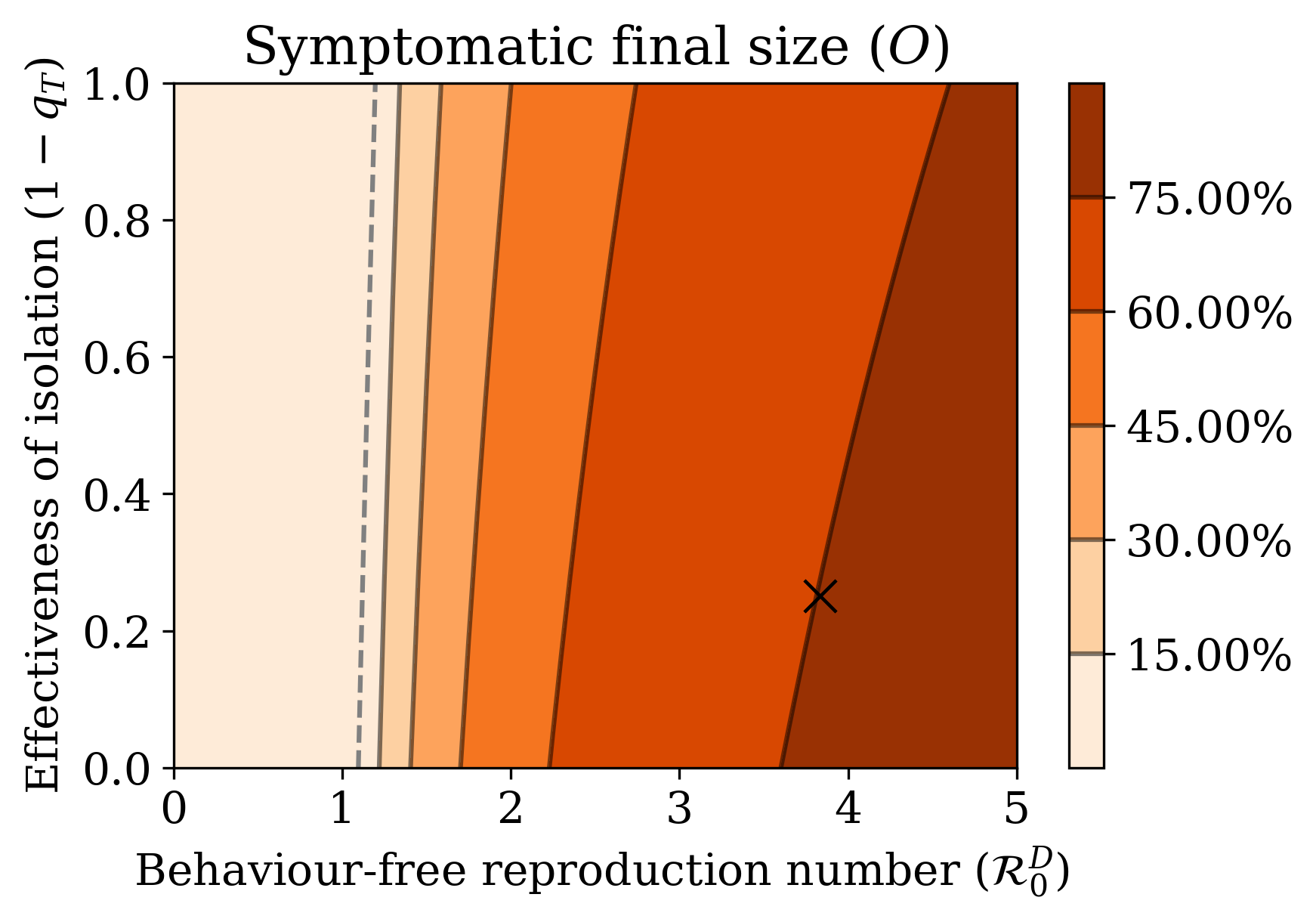}
    \end{subfigure}
    \hfill
    \begin{subfigure}[b]{0.3\textwidth}
    \caption{}
    \includegraphics[width=\textwidth]{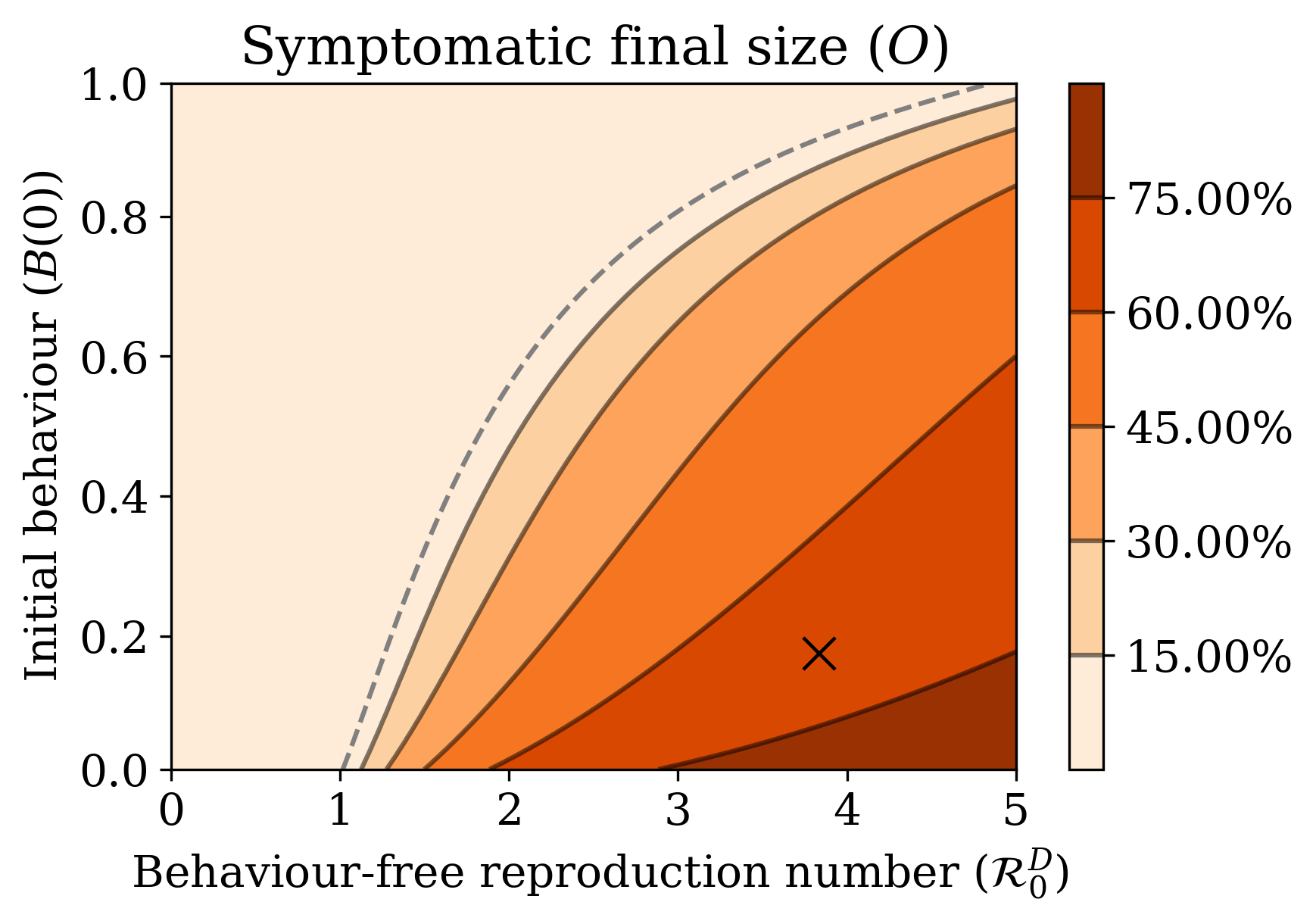}
    \end{subfigure}
    \hfill

    \begin{subfigure}[b]{0.3\textwidth}
    \caption{}
    \includegraphics[width=\textwidth]{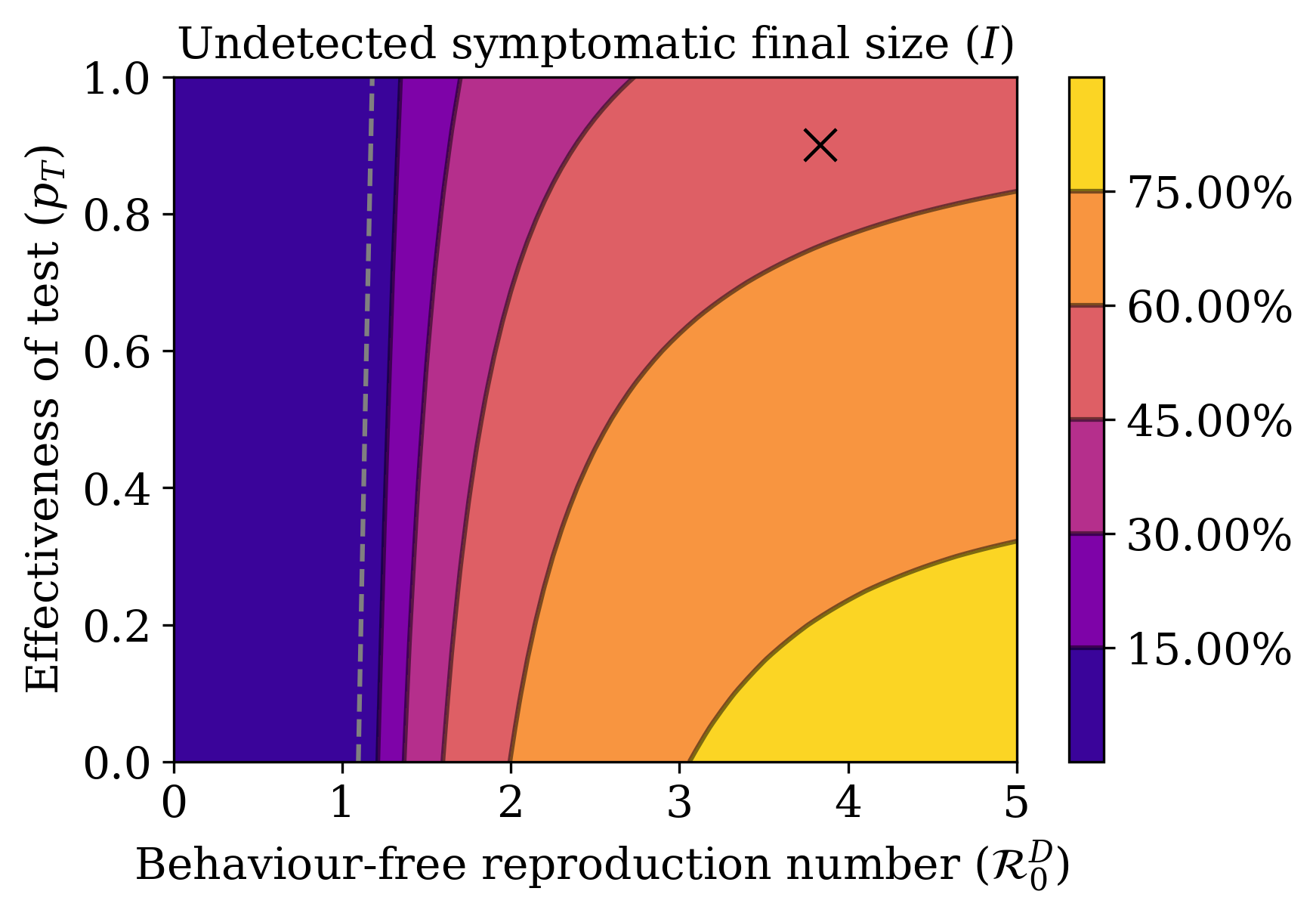}
    \end{subfigure}
    \hfill
    \begin{subfigure}[b]{0.3\textwidth}
    \caption{}
    \includegraphics[width=\textwidth]{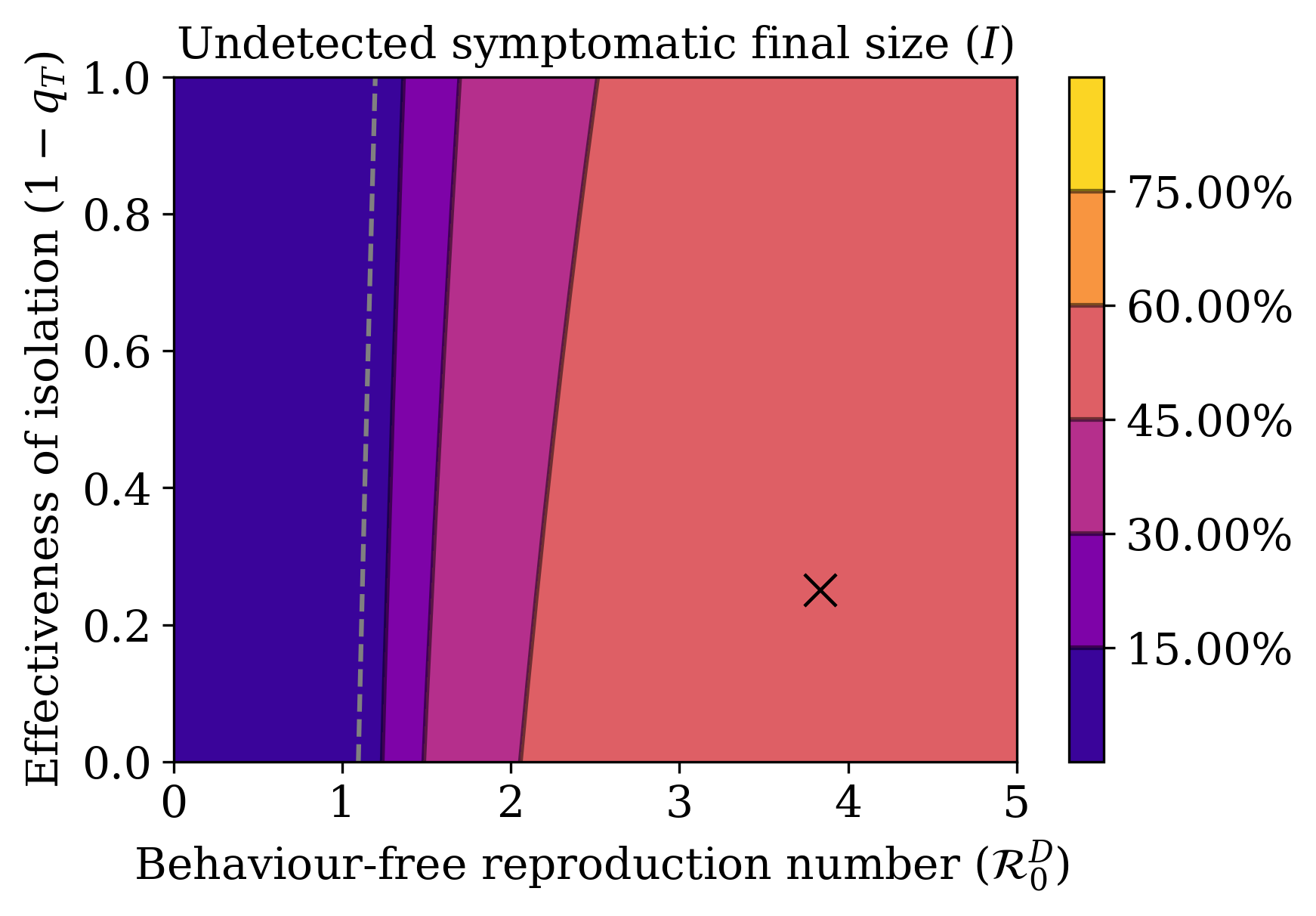}
    \end{subfigure}
    \hfill
    \begin{subfigure}[b]{0.3\textwidth}
    \caption{}
    \includegraphics[width=\textwidth]{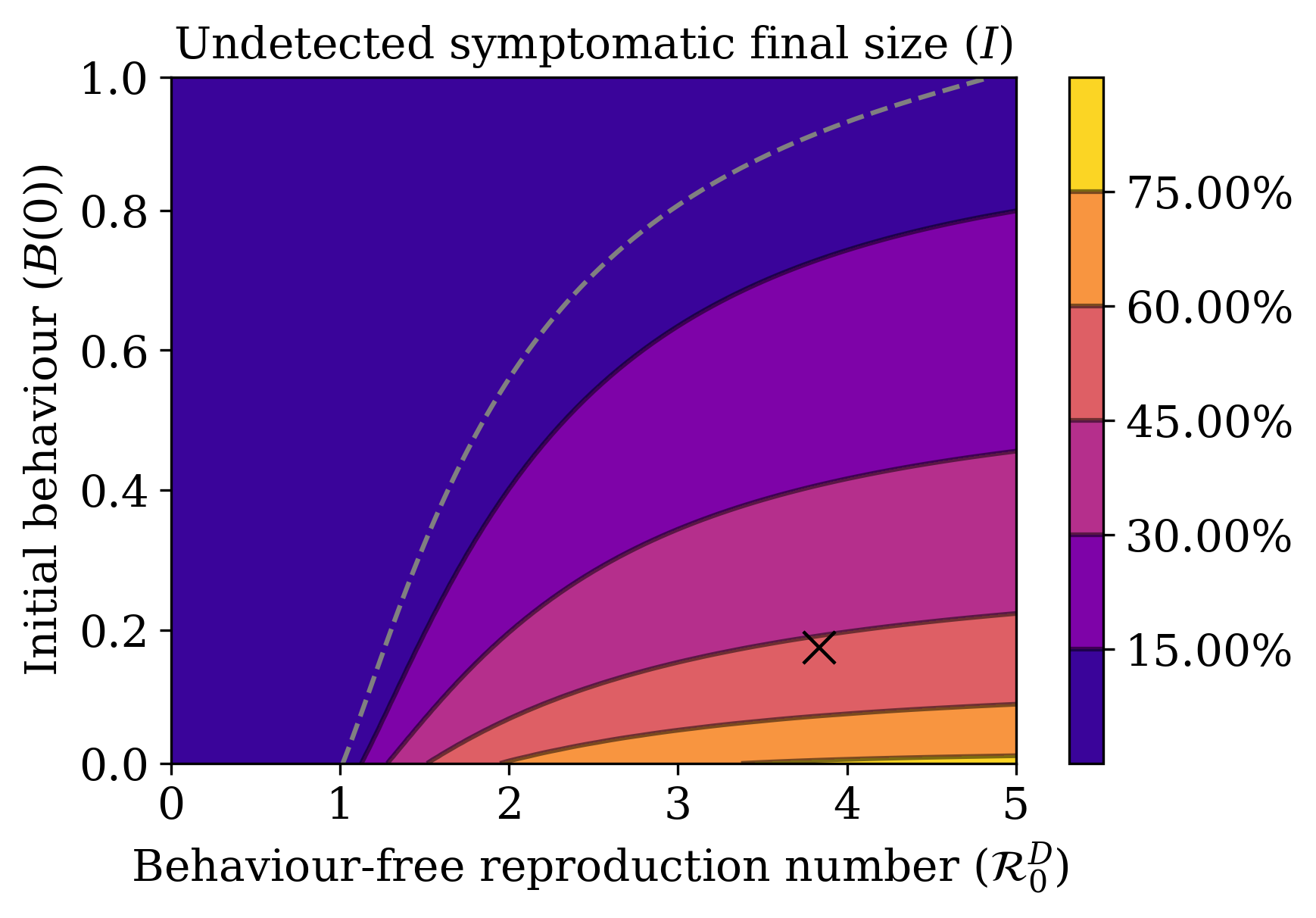}
    \end{subfigure}
    \hfill
    
    \caption{\textbf{Impact of test effectiveness (${p_T}$), isolation effectiveness ($1-{q_T}$), initial condition of behaviour (${B(0)}$) and the behaviour-free reproduction number (${\Ro^{D}}$) on observed symptomatic (${T}$), undetected symptomatic (${I}$), and total symptomatic (${O=I+T}$) final sizes. } The colour bars represent the final size as a percentage of the population and are consistent across each row.  Note, we vary the behaviour-free reproduction number instead of $\Ro$ due to the dependence of $\Ro$ on the vertical axes.  The dashed grey lines show $\Ro=1$ and the black cross indicates the baseline parameter values from Table \ref{tab:model_parameter}.  
    }
    \label{fig:sweep_finalSize}
\end{figure}

The testing parameters (effectiveness of testing, $p_T$, and  effectiveness of isolation, $1-q_T$) exhibit distinct influences over the observed ($T$) and true symptomatic ($O=I+T$) final sizes of the epidemic (Figure \ref{fig:sweep_finalSize}, rows one and two).  Increasing the effectiveness of the test increases the proportion of the observed final size and interacts non-linearly with the innate infectiousness of the pathogen, $\Ro^D$.  Specifically, even with a completely effective test ($p_T=1$), if $\Ro^D$ is small enough we will \emph{observe} a negligible final size, whereas the true impact of the disease can be significant (Figure \ref{fig:sweep_finalSize}, (a), (d), (g)).  Unsurprisingly, as the effectiveness of the test increases (i.e.\ higher values of $p_{T}$) and we observe a larger final size, the unobserved final size (Figure \ref{fig:sweep_finalSize} (g)) decreases.  The effect of isolation in reducing infection ($q_T$: column two) has an opposite impact on the final size.  Although the effect is not strong for these parameter values, increasing the effect of isolation (by reducing $q_T$) can reduce the observed final size (Figure \ref{fig:sweep_finalSize} (b)) and the true symptomatic final size (Figure \ref{fig:sweep_finalSize} (e)).  This relationship is driven by less contact with the susceptible community as isolation improves.

Initial behaviour in the population ($B(0)$; column three of Figure \ref{fig:sweep_finalSize}) has a strong and non-linear effect on the symptomatic final size of the epidemic, both observed and undetected.  Increasing the initial proportion of the population willing to test and isolate at the start of the epidemic can reduce the final size to a negligible value by driving the reproduction number $\Ro$ down.  For example, an initial proportion of the population willing to test and isolate exceeding $60\%$ can suppress an outbreak when the behaviour-free reproduction number
$\Ro^D$ is up to $2.1$.
 
These results suggest that having a population primed to be willing to test and isolate may have a substantial influence in controlling a new epidemic.  For example, if a new pathogen has been observed in a different population, efforts to increase test availability and testing willingness in the population may provide a strong preparedness plan.  A strong influence of the feedback between perception of illness threat and final size is also observed (Figure \ref{fig:sweep_finalSize}, (c), (f)).  Specifically, for a more infectious disease ($\Ro^D > 4$), increasing the initial populace willing to test and isolate increases the amount of infection seen throughout the epidemic as $T$ increases.  However, the actual symptomatic final size ($O$) is reduced due to the behavioural feedback mechanism.

Infection peaks show qualitatively similar patterns (Supplementary Figure \ref{fig:sweep_peak}).

\subsubsection{Peaks in testing behaviour}

\begin{figure}[htbp]
    \centering
    \begin{subfigure}[b]{0.45\textwidth}
    \caption{}
    \includegraphics[width=\textwidth]{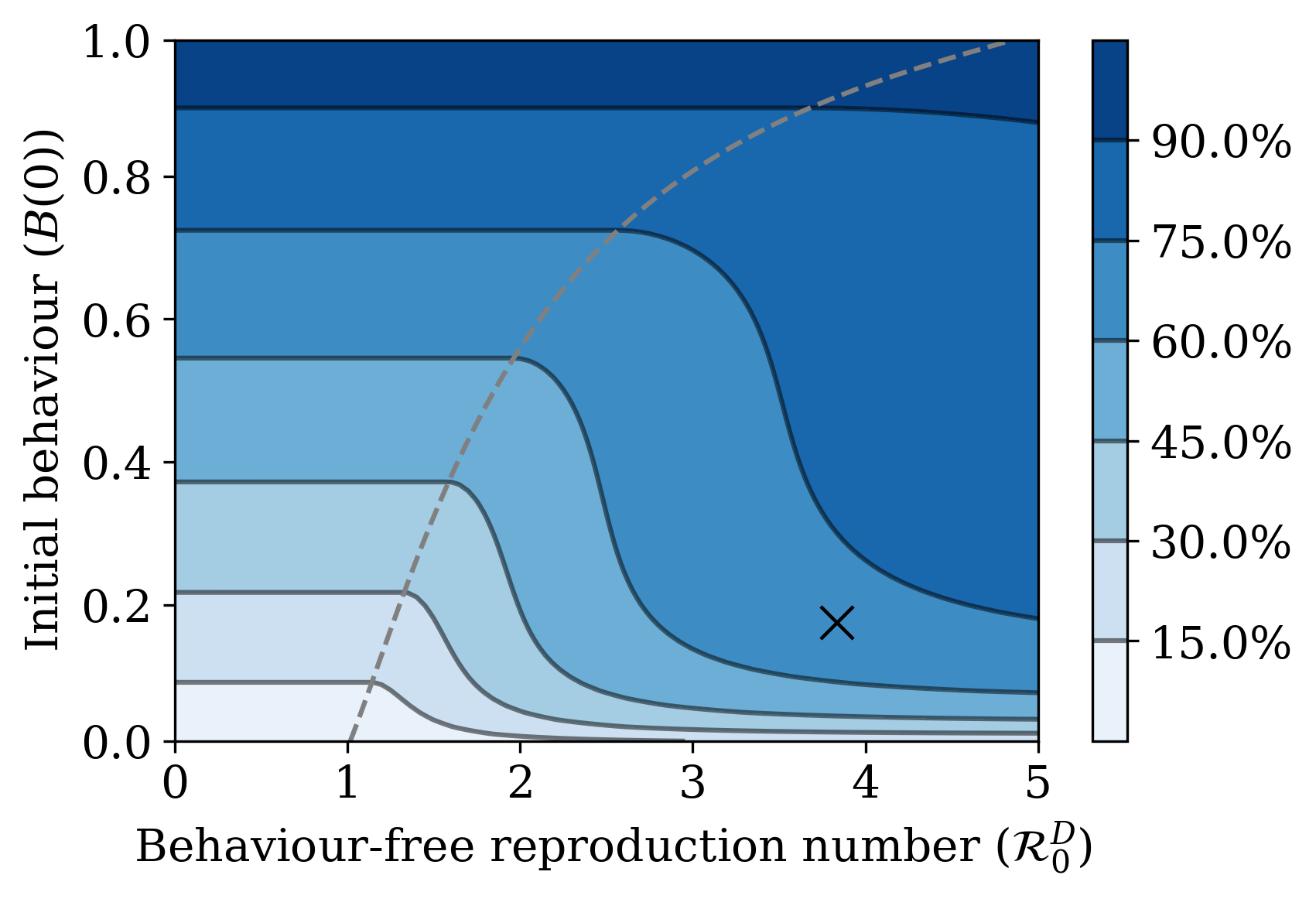}
    \end{subfigure}
        \hfill
    \begin{subfigure}[b]{0.45\textwidth}
    \caption{}
    \includegraphics[width=\textwidth]{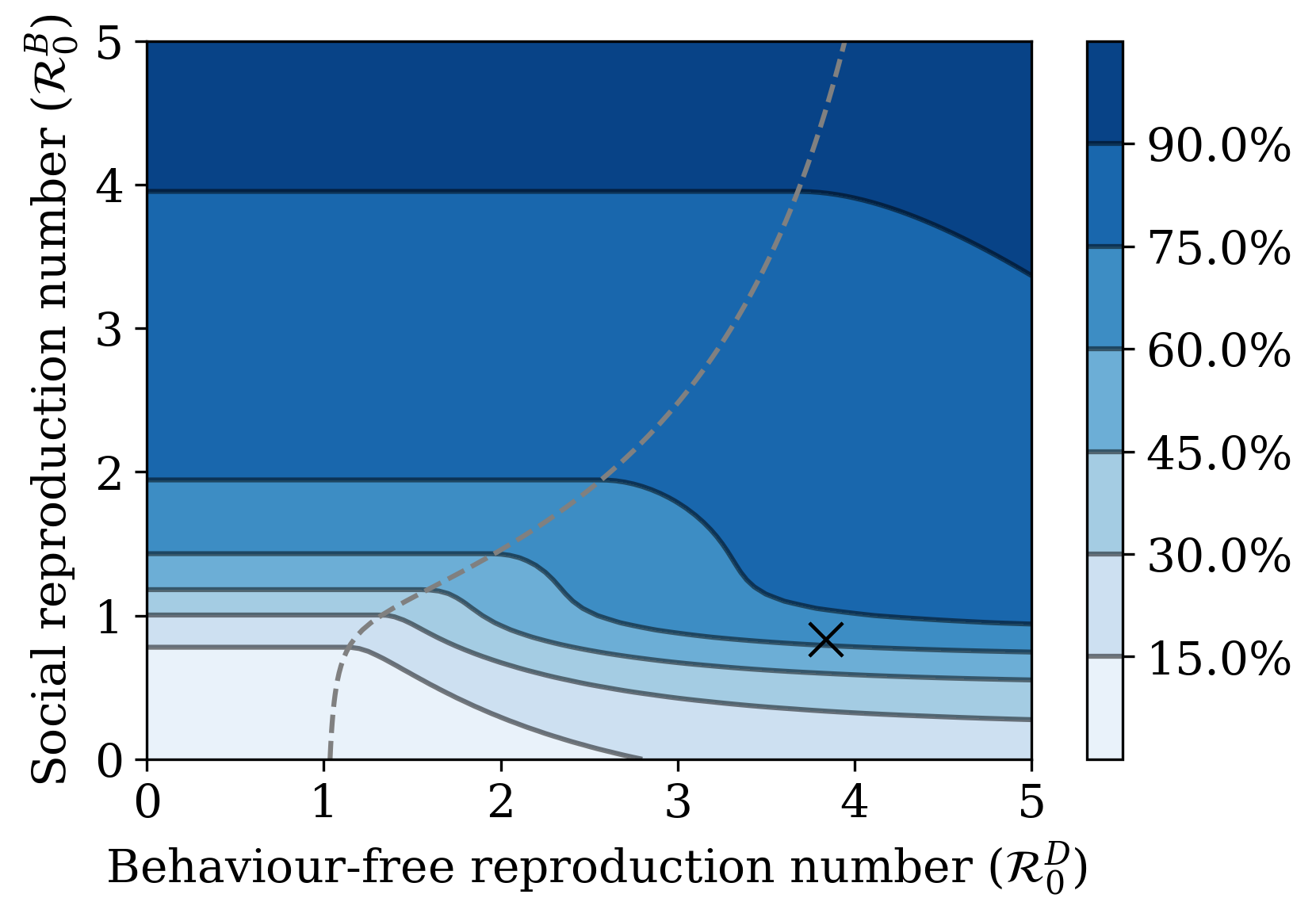}
    \end{subfigure}
    \hfill
    \begin{subfigure}[b]{0.45\textwidth}
    \caption{}
    \includegraphics[width=\textwidth]{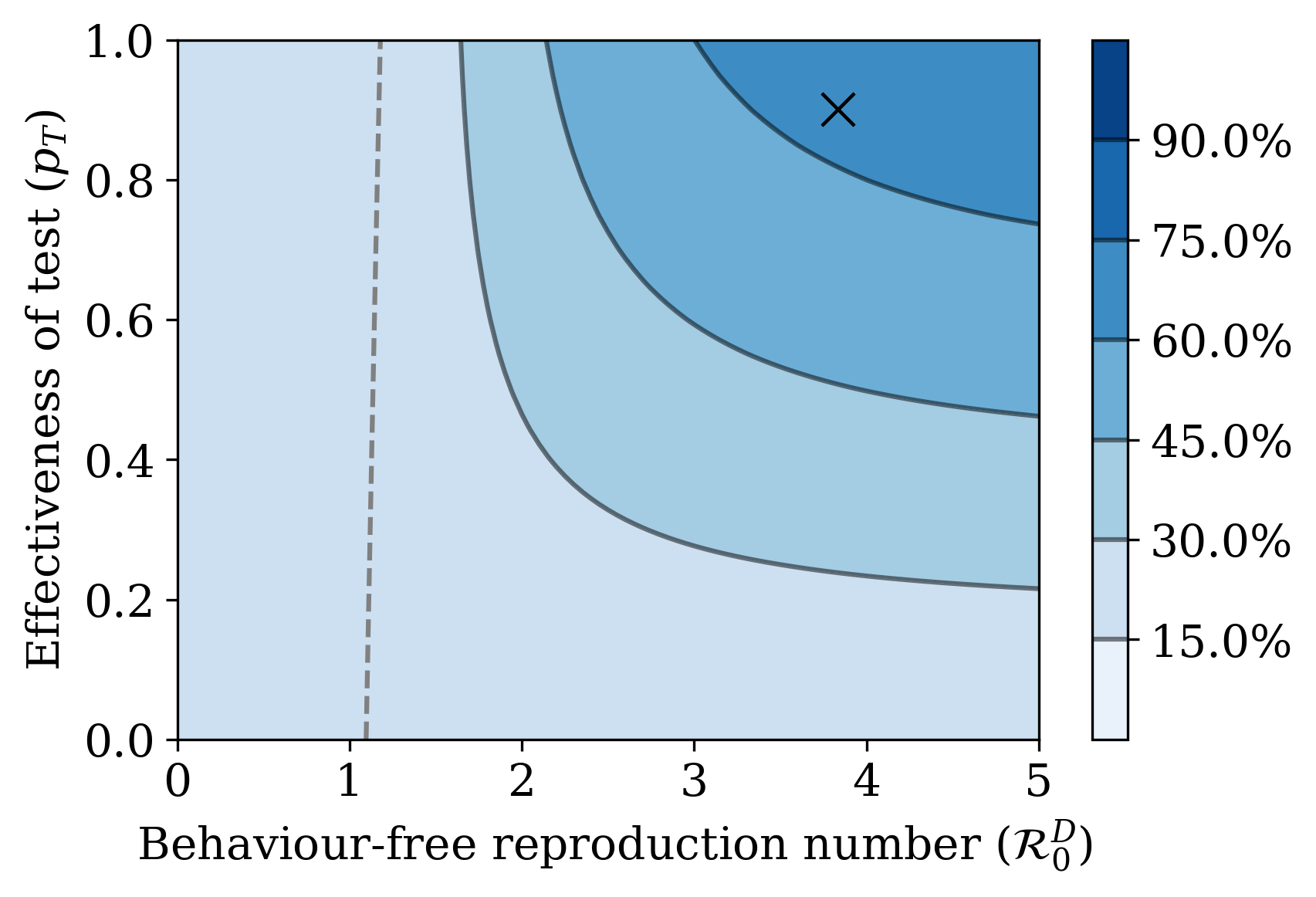}
    \end{subfigure}
    \hfill
    \begin{subfigure}[b]{0.45\textwidth}
    \caption{}
    \includegraphics[width=\textwidth]{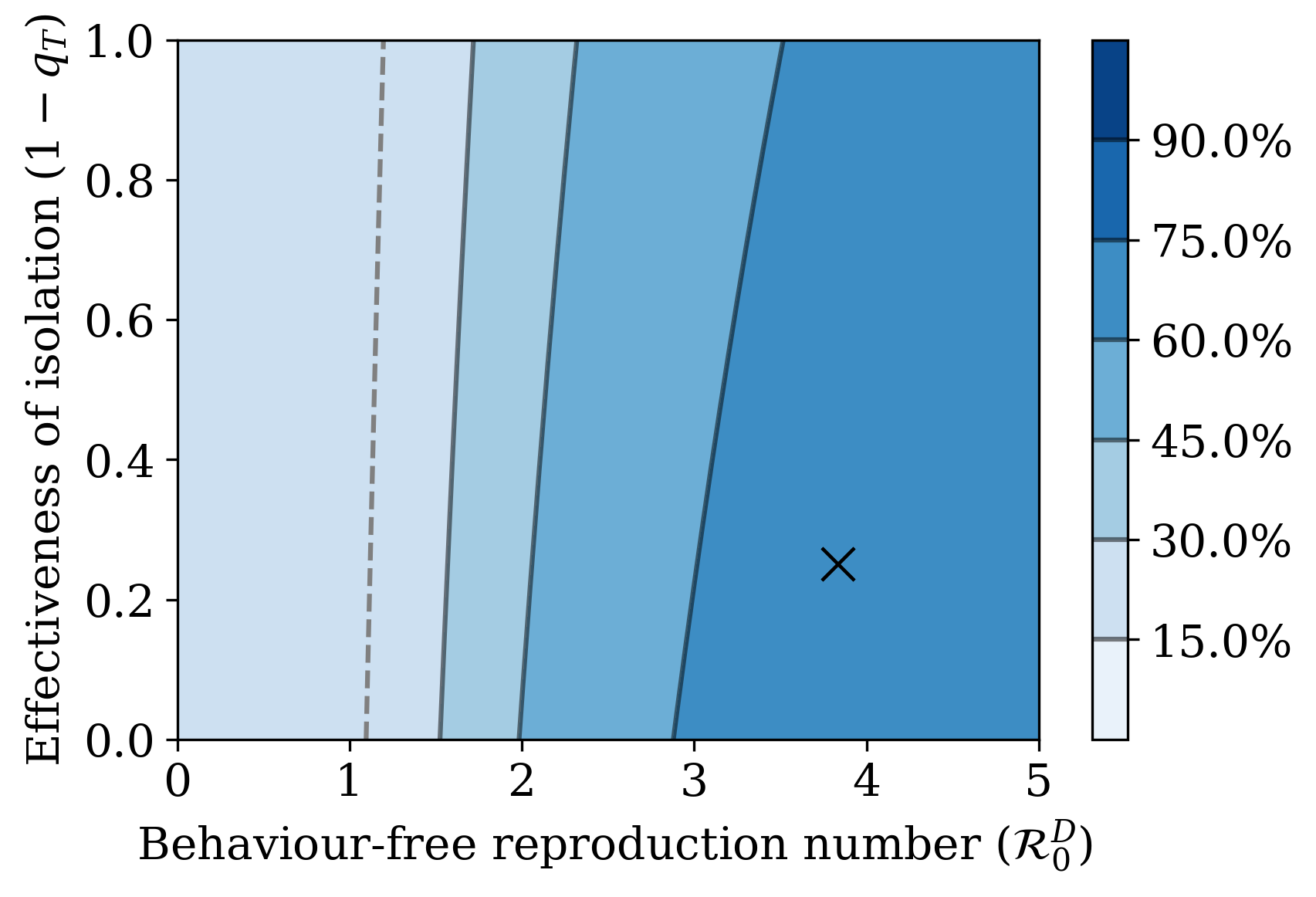}
    \end{subfigure}

    \caption{\textbf{The impact of key model parameters on the peak population proportion willing to test (${B}$) for different behaviour-free reproduction numbers (${\Ro^D}$).}  \textbf{(a)} The initial proportion willing to test and isolate ($B(0)$) varies through changing spontaneous uptake ($\omega_3$); \textbf{(b)} test effectiveness ($p_T$) varies; \textbf{(c)} isolation effectiveness ($1-q_T$) varies, and; \textbf{(d)} the social reproduction number ($R_0^B$) varies through changing the social influence of behaviour ($\omega_1$).  Note we vary the behaviour-free reproduction number instead of $\Ro$ because of the dependence of $\Ro$ on the vertical axes. The dashed grey line shows $\Ro=1$ and the black cross shows the baseline parameter values from Table \ref{tab:model_parameter}. 
    }
    \label{fig:B_peaks}
\end{figure}

The peak of behaviour over the epidemic describes the maximum proportion of the population at a single time point willing to test and isolate if showing symptoms (Figure \ref{fig:B_peaks}).   Given the feedback mechanism between perception of illness threat and testing behaviour, unsurprisingly increasing the infectiousness of the pathogen of interest ($\Ro^{D}$) increases the peak behaviour observed over the epidemic. Changing the initial population willing to test and isolate ($B(0)$) and the social reproduction number ($\Ro^{B}$) have similar qualitative effects on peak behaviour (Figure \ref{fig:B_peaks}, (a), (b)).  When the basic reproduction number $\Ro<1$ (grey dashed line), we observe the peak of behaviour is kept constant at the disease-free equilibrium $B_0$.  When $\Ro>1$, we observe a `waterfall' like contour where peak behaviour rapidly increases with the behaviour-free reproduction number $\Ro^D$.  This behaviour is driven by the perception of illness threat $\omega_2$ and is more pronounced when altering $\omega_3$ through $B(0)$ than when altering $\omega_1$ through $\Ro^{B}$.   A notable and interesting relationship occurs with the effectiveness of the test $p_T$ (Figure \ref{fig:B_peaks} (c)), which shows a non-linear relationship between pathogen infectiousness and increased behaviour.  Since the feedback mechanism in the model depends on the observed number of cases $T$, we only observe an increase in the peak behaviour from this mechanism when the testing effectiveness is sufficiently large. Finally, when altering the testing parameters $p_T$ and $q_T$, observe the visual correlation between peak behaviour (Figure \ref{fig:B_peaks}, (c), (d)) and detected cases (Figure \ref{fig:sweep_finalSize}, (a), (b)).  This correlation underpins the relationship between behaviour and testing through the perception of illness threat feedback mechanism.

\subsubsection{Effect of behavioural uptake rates on the dynamics}

\begin{figure}[htbp]
    \centering
    \begin{subfigure}[b]{0.32\textwidth}
    \caption{}
    \includegraphics[width=\textwidth]{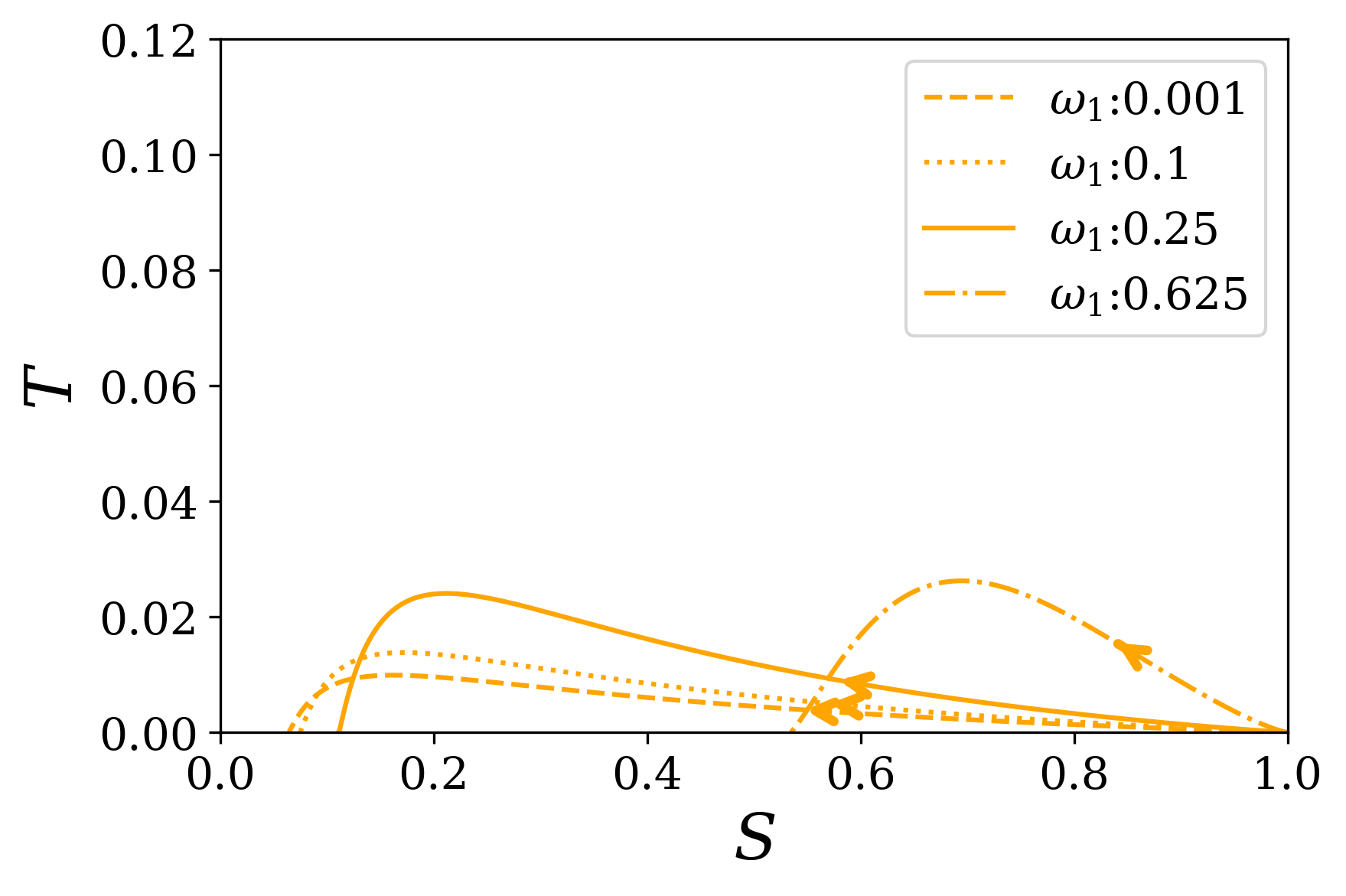}
    \end{subfigure}
    \hfill
    \begin{subfigure}[b]{0.32\textwidth}
    \caption{}
    \includegraphics[width=\textwidth]{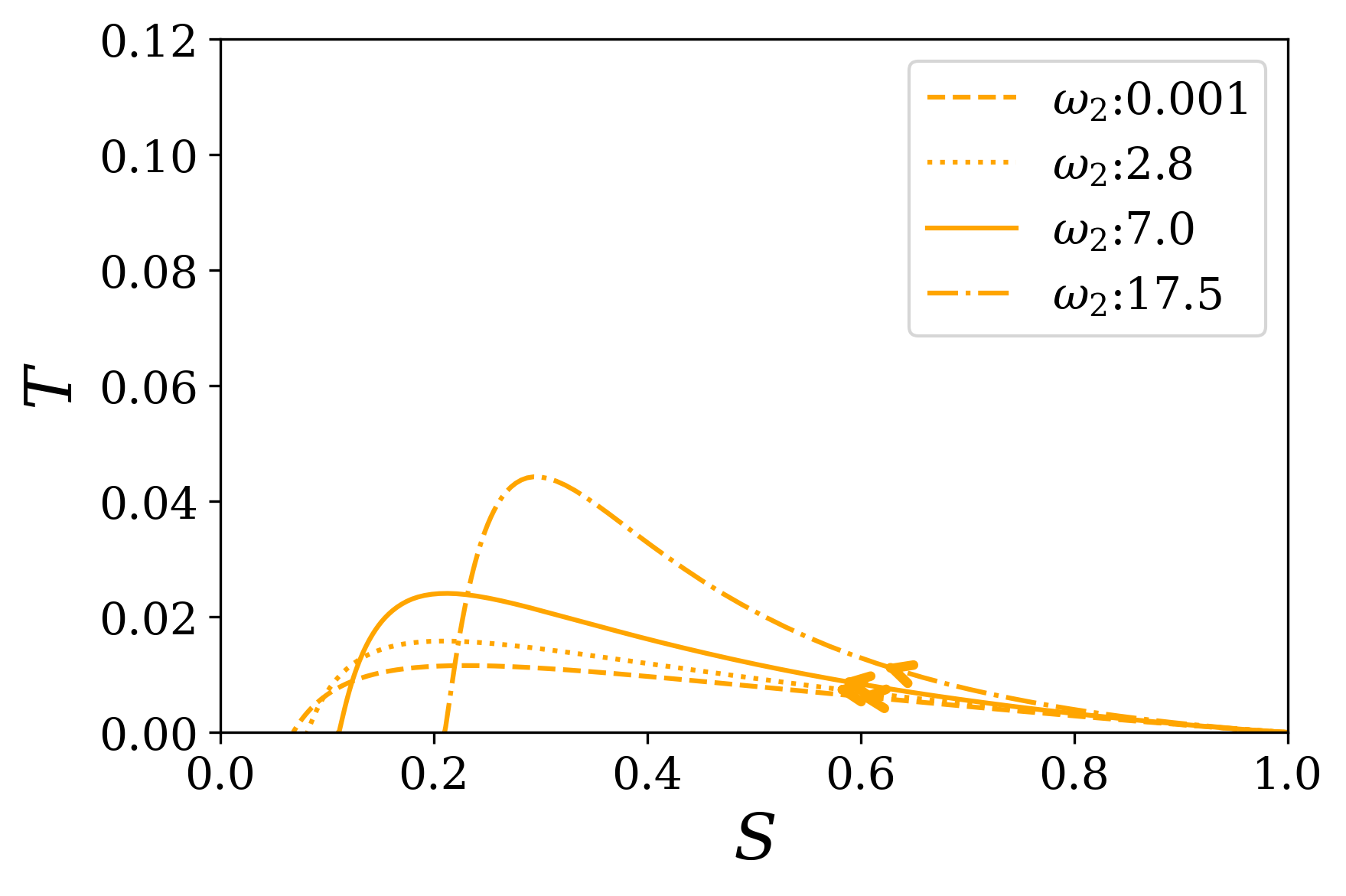}
    \end{subfigure}
    \hfill
    \begin{subfigure}[b]{0.32\textwidth}
    \caption{}
    \includegraphics[width=\textwidth]{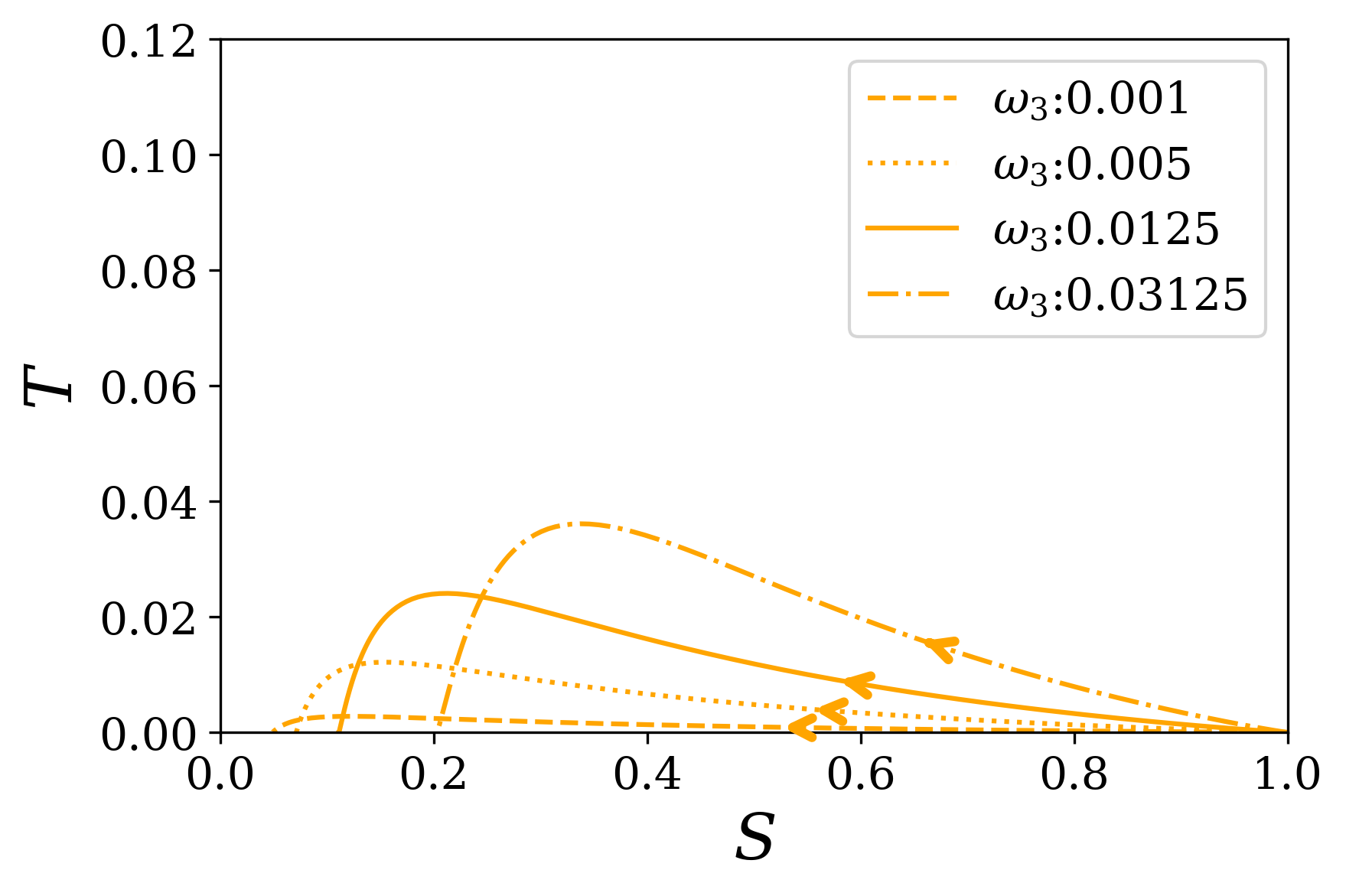}
    \end{subfigure}
    \hfill
    
    \begin{subfigure}[b]{0.32\textwidth}
    \caption{}
    \includegraphics[width=\textwidth]{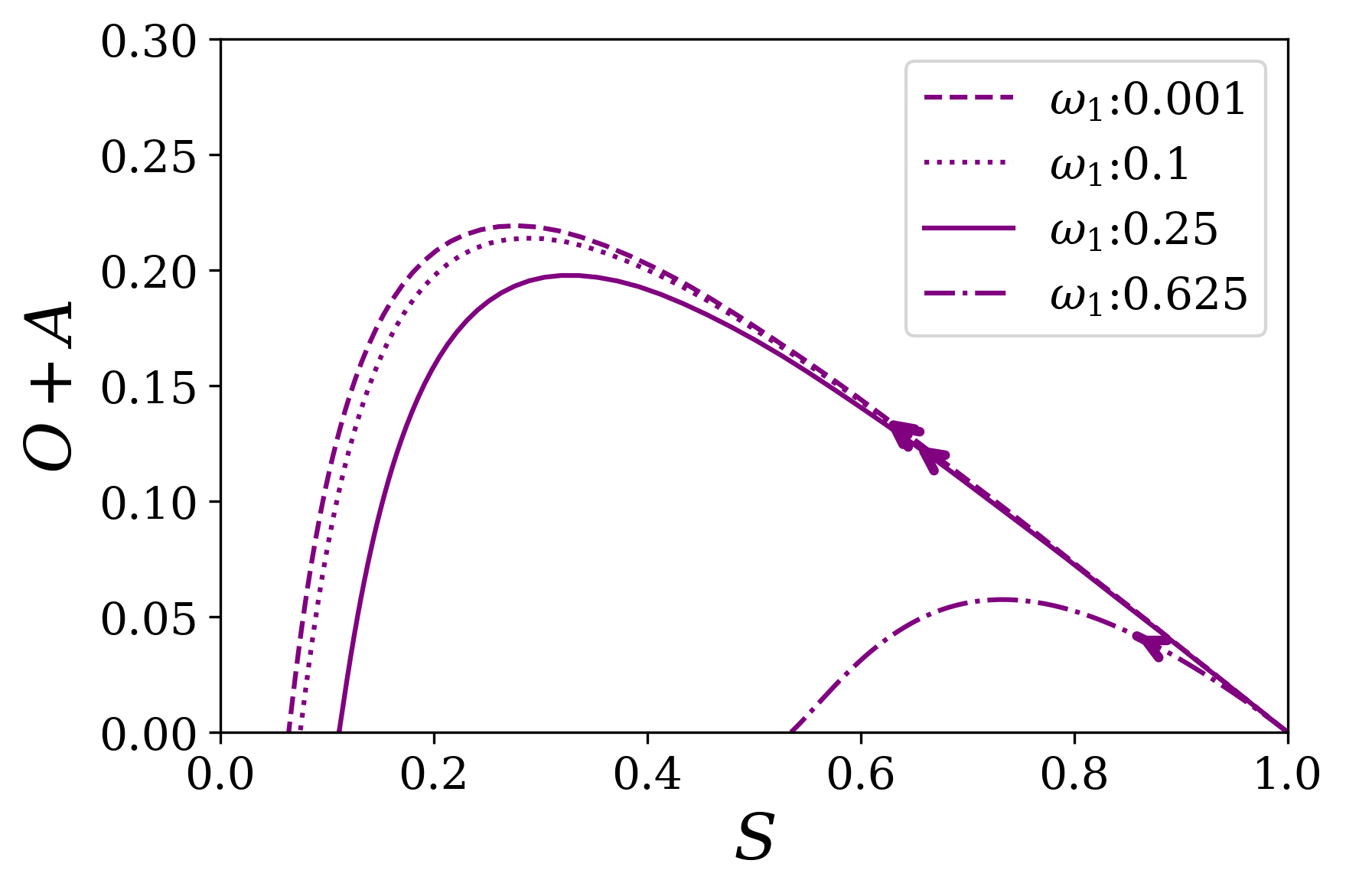}
    \end{subfigure}
    \hfill
    \begin{subfigure}[b]{0.32\textwidth}
    \caption{}
    \includegraphics[width=\textwidth]{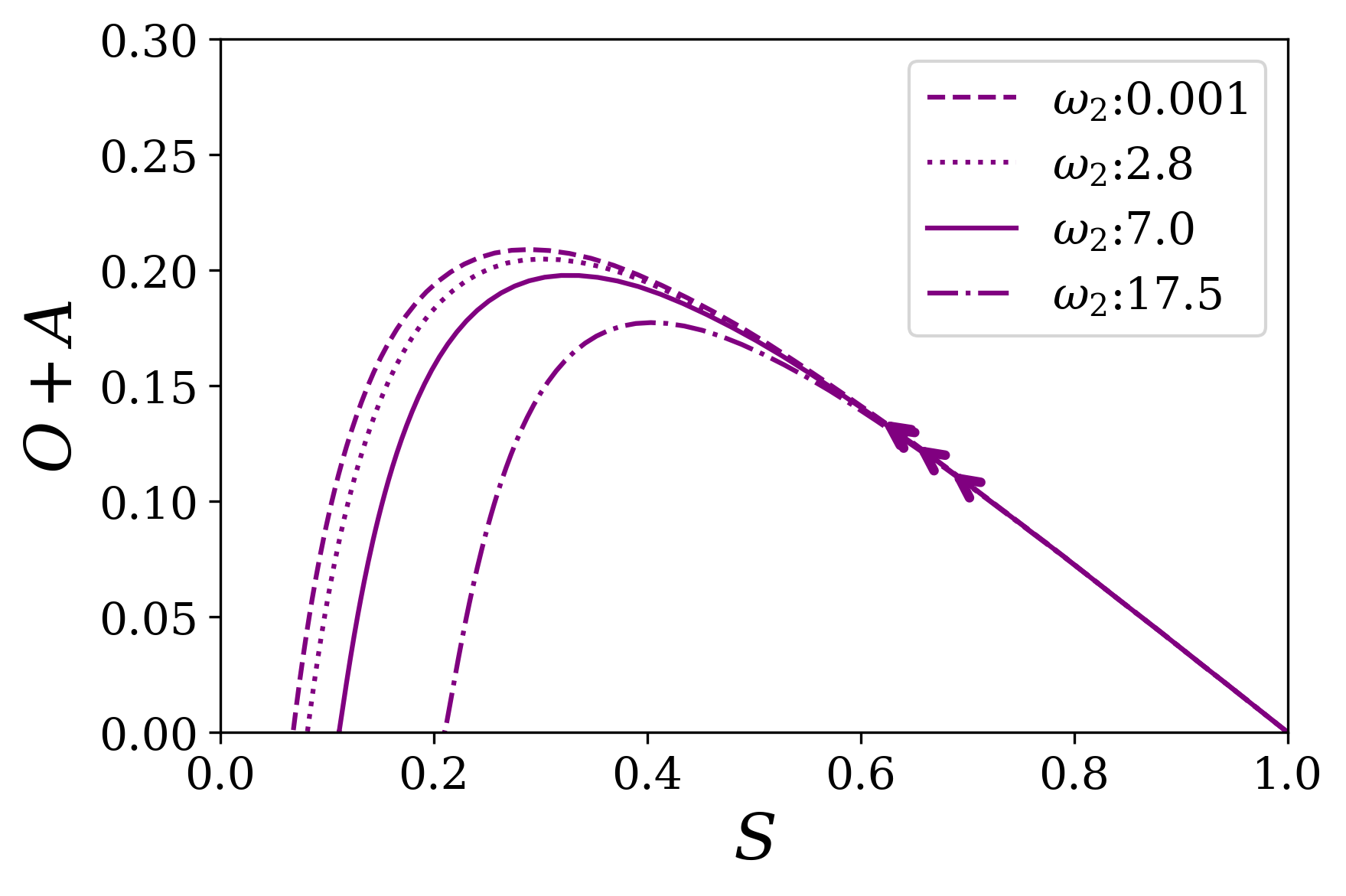}
    \end{subfigure}
    \hfill
    \begin{subfigure}[b]{0.32\textwidth}
    \caption{}
    \includegraphics[width=\textwidth]{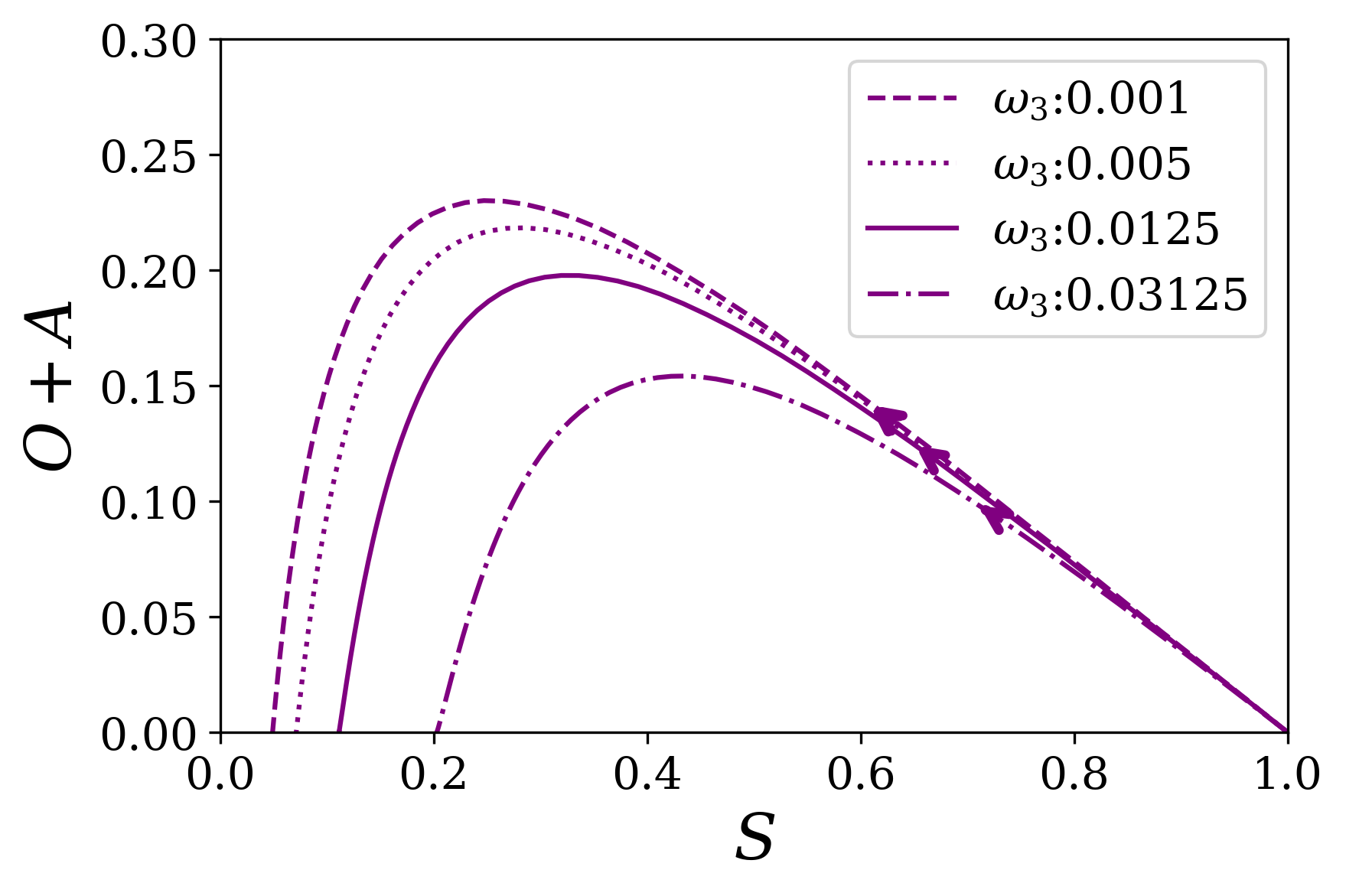}
    \end{subfigure}
    \hfill
    
    \begin{subfigure}[b]{0.32\textwidth}
    \caption{}
    \includegraphics[width=\textwidth]{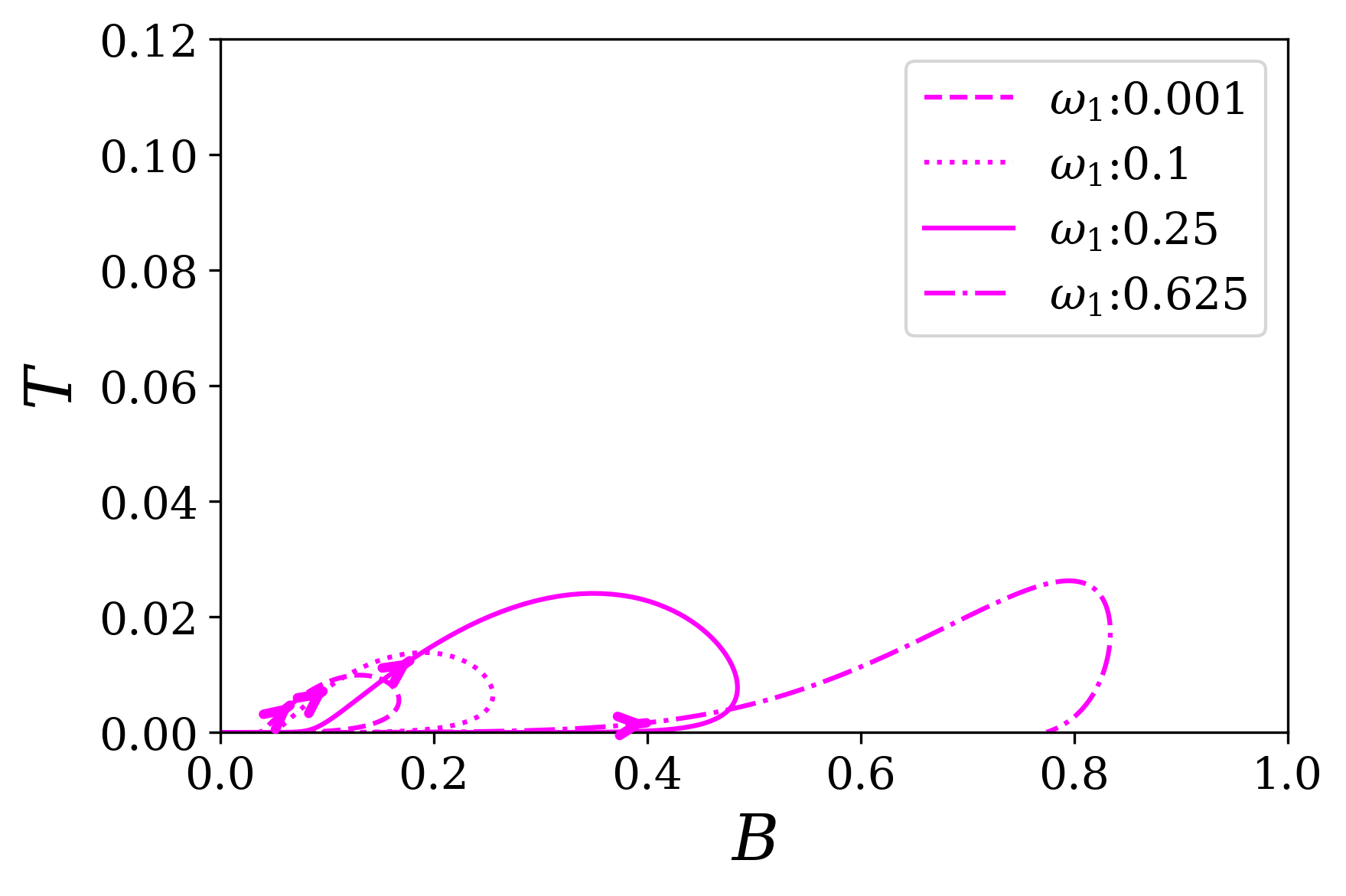}
    \end{subfigure}
    \hfill
    \begin{subfigure}[b]{0.32\textwidth}
    \caption{}
    \includegraphics[width=\textwidth]{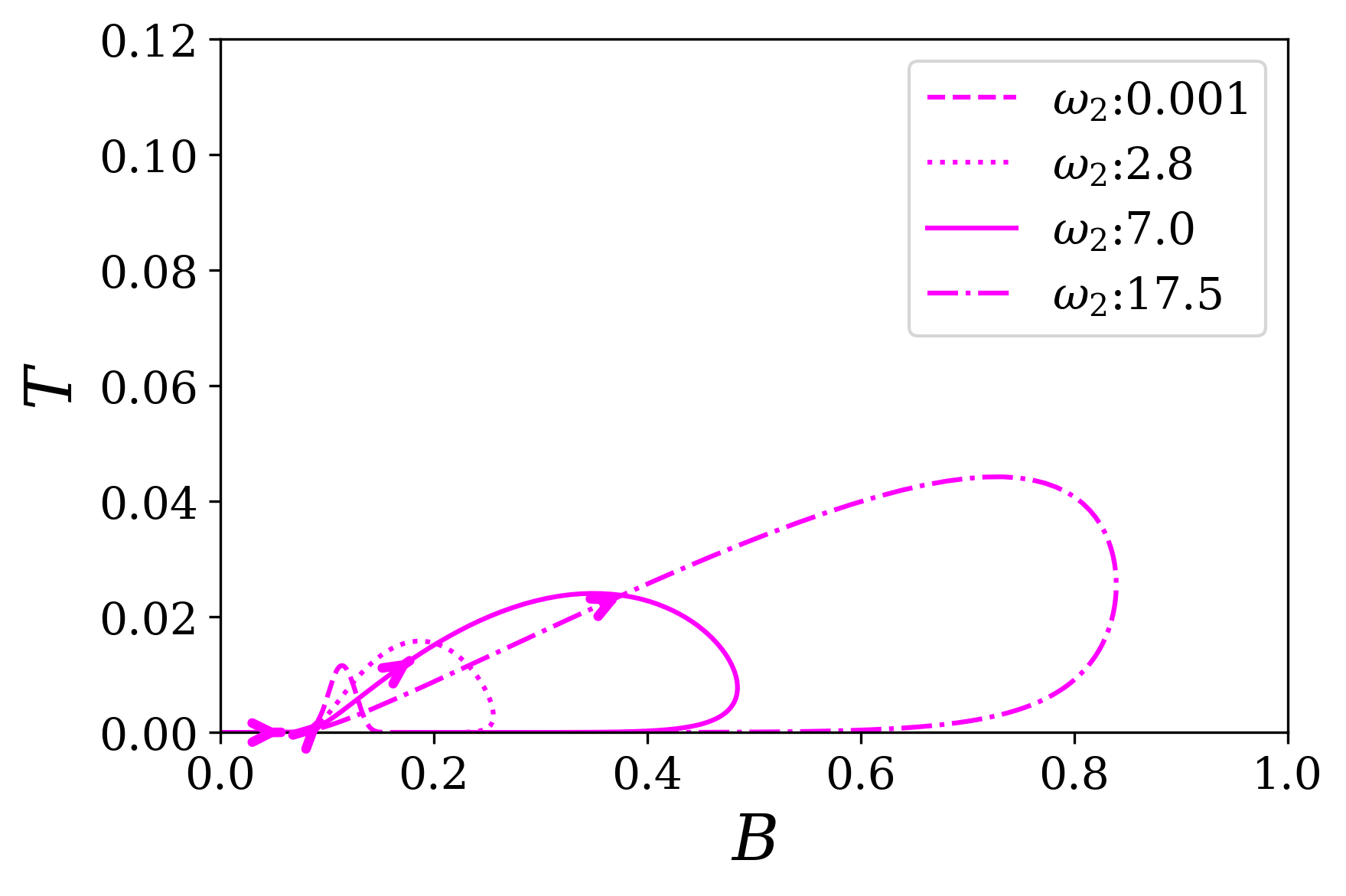}
    \end{subfigure}
    \hfill
    \begin{subfigure}[b]{0.32\textwidth}
    \caption{}
    \includegraphics[width=\textwidth]{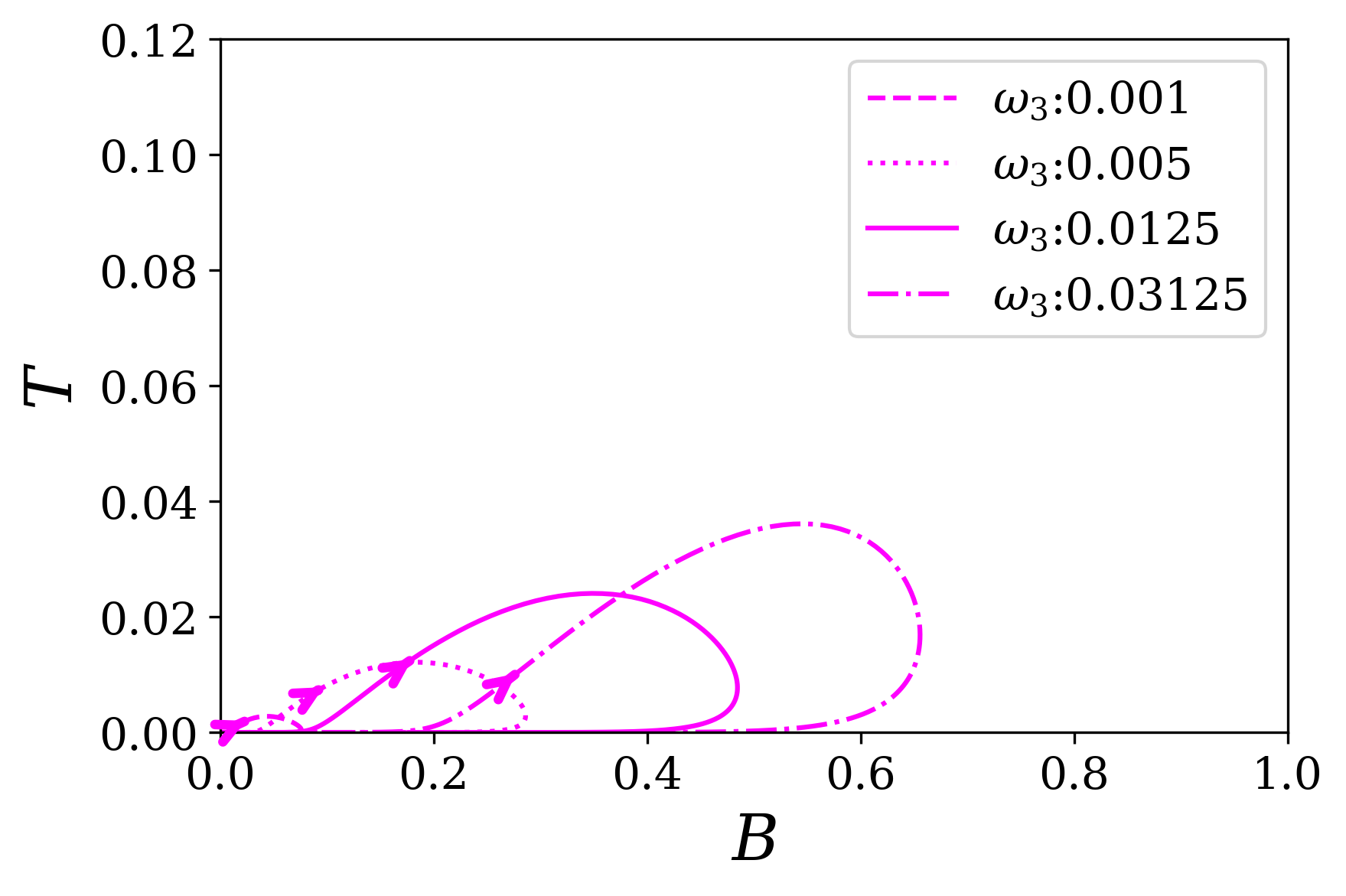}
    \end{subfigure}
    \caption{\textbf{Phase diagrams for different behavioural parameters fixing ${\Ro^D=3.28}$ in the epidemic regime.} Column one corresponds to changes in social influence ($\omega_1$), column two corresponds to changes in perception of illness threat ($\omega_2$), and column three corresponds to spontaneous uptake ($\omega_3$), respectively.  The rows show the susceptible versus observed epidemic ($S$ vs.\ $T$), the susceptible versus true epidemic ($S$ vs.\ $O +A$), and the behaviour versus observed epidemic ($B$ vs.\ $T$) phase planes.  The dashed line represents near removal of the behaviour construct, the dotted lines represent a reduction of the baseline value, the solid line represents the baseline values, and the dash-dot lines represent an increase in the baseline values.  The solid lines all represent the baseline values from Table \ref{tab:model_parameter}, ensuring the solid lines are comparable across each column.  Each simulation was run with the initial conditions $S_B(0) = 10^{-6}, I_N(0) = 10^{-6}, S_N = 1 - S_B - I_N$ and all other compartments empty: this ensured the early-stage dynamics of each model were approximately comparable. 
    }
    \label{fig:phase_epidemic_r0D}
\end{figure}

The transitory dynamics of the BaD testing and isolation model vary based on the behavioural uptake parameters (Figure \ref{fig:phase_epidemic_r0D}).  Across all three uptake parameters, the observed epidemic (Figure \ref{fig:phase_epidemic_r0D}, top row) is substantially smaller than the true epidemic in the population (Figure \ref{fig:phase_epidemic_r0D}, middle row). Two points of particular note stand out across all three behavioural constructs. First, increasing uptake of behaviour makes the epidemic look more severe through an increased reporting of cases ($T$) but the true effect of the epidemic is reduced ($O + A$). This is particularly evident when increasing the perception of illness threat (Figure \ref{fig:phase_epidemic_r0D} (b)), which is driven by the feedback mechanism linking $T$ and $B$.  Here, we observe a ``reporting epidemic'' that results in a larger behavioural response from the population, reducing the overall impact of the epidemic.  Second, removing any individual construct is sufficient to make sure the epidemic is hardly observed (Figure \ref{fig:phase_epidemic_r0D}, dashed lines).  This suggests an important interplay between the three behavioural constructs in ensuring that individuals in the population test and isolate in response to a new infection. However, each uptake mechanism has a different effect on the observed and true epidemic.

Social uptake of behaviour ($\omega_1$) can control the epidemic (Figure \ref{fig:phase_epidemic_r0D}, (a), (d)).  Increasing social pressure has the effect that we observe a larger epidemic, but the peak of the true epidemic is reduced due to increased public awareness and isolation.  When social pressure is increased sufficiently, we obtain a greatly reduced overall epidemic and observe almost every case.  This effect is driven by increasing $\Ro^{B} > 1$ so that the behaviour ``takes off'' socially, leading to a large and quick uptake of testing behaviour.  This social spread of behaviour also ensures that the disease-free equilibrium ($B_0$) is such that the infection cannot spread easily in the population (Figure \ref{fig:phase_epidemic_r0D} (g)).

Spontaneous uptake of behaviour ($\omega_3$) has a similar effect as social pressure (Figure \ref{fig:phase_epidemic_r0D}, (c), (f)).  Increasing spontaneous uptake in the model reduces the true epidemic size while increasing the peak observed number of cases. Removal of spontaneous uptake ensures that the epidemic is not observed in this model.  Due to the baseline parameter values (Table \ref{tab:model_parameter}), without a small influence of spontaneous behavioural uptake there is not enough awareness of testing in the population for testing to ``take off'', leading to a severe lack of testing and reporting of the epidemic.  In this setting, we would observe a silent epidemic where many individuals become unwell but do not seek testing and isolation. Increasing spontaneous uptake sufficiently has the same effect in controlling the epidemic as social pressure.
Like social pressure, this control of the epidemic is driven by the speed of behavioural uptake and the increase in $B(0)$ (Figure \ref{fig:phase_epidemic_r0D} (i)).

In contrast, the perception of illness threat ($\omega_2$) cannot control the epidemic (Figure \ref{fig:phase_epidemic_r0D}, (b), (e)).  When perception of illness threat is increased, we observe a ramp up of reported cases followed by a sharp decline and conclusion of the epidemic.  This is driven by the feedback mechanism in the model: the more cases reported to the population through the $T$ compartment, the sharper uptake of testing behaviour.  This feedback mechanism is the reason that perception of illness threat cannot control the epidemic.  Since $\omega_2$  does not influence the basic reproduction number or the disease free equilibrium (Figure \ref{fig:phase_epidemic_r0D} (h)), for the feedback mechanism to take effect we need to observe cases of infection in the population.

The effect of behavioural parameters on the transient dynamics are similar in an endemic regime, albeit with trajectories converging to steady states (Supplementary Figure \ref{fig:phase_endemic_r0D}).

\subsection{Endemic regime with waning immunity}
\label{sec:results_endemic_regime}

Here we detail our findings for an endemic regime with $\nu>0$ as detailed in \cref{tab:model_parameter}. We report on the impact of testing and isolation effects on endemic steady states, the impact of behaviour and disease contagiousness on endemic states, and the stability of steady states. 

\subsubsection{Testing and isolation effects on endemic steady states}

\begin{figure}[htbp]
    \centering
    \begin{subfigure}[b]{0.45\textwidth}
    \caption{}
    \includegraphics[width=\textwidth]{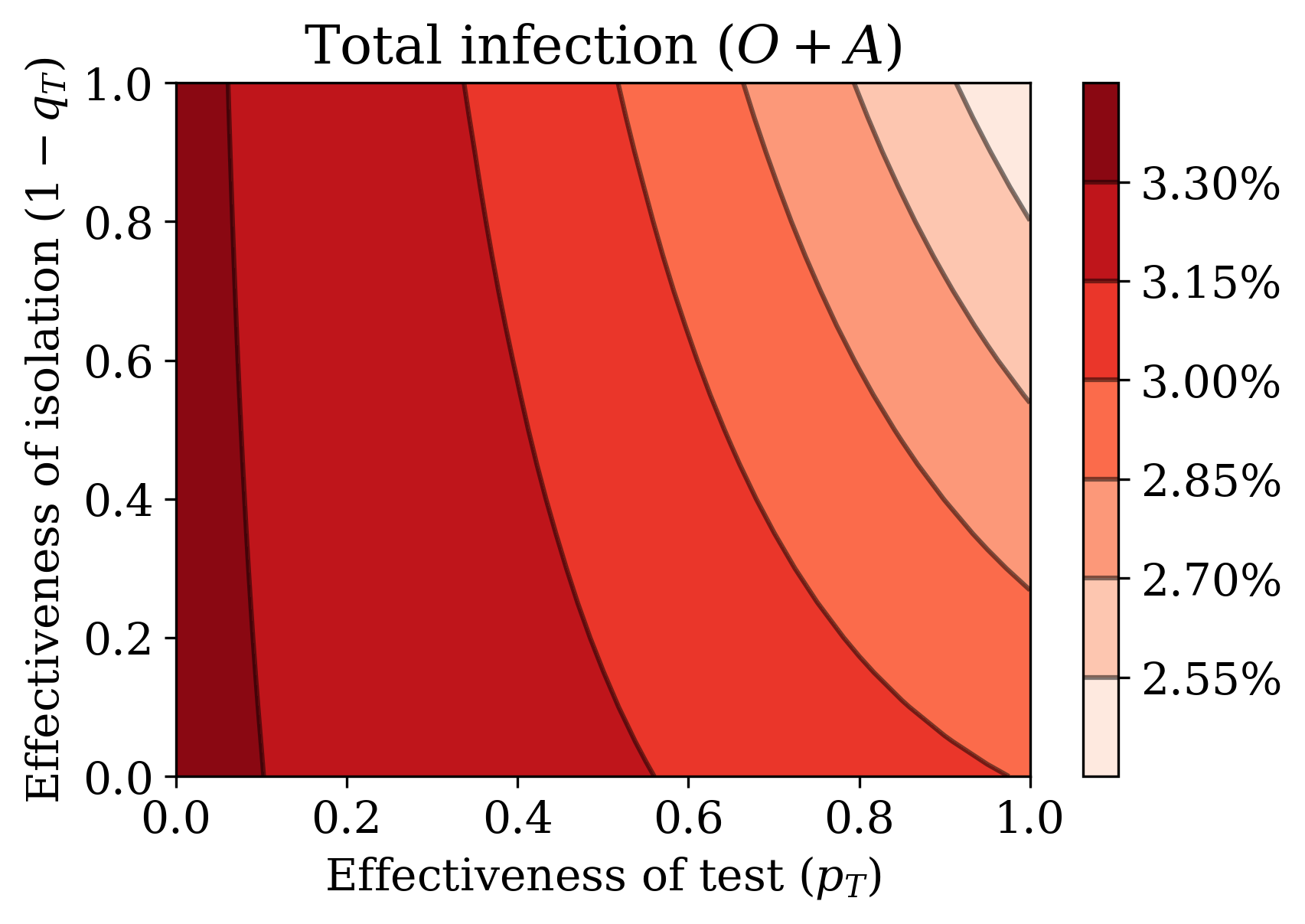}
    \end{subfigure}
    \hfill
    \begin{subfigure}[b]{0.45\textwidth}
    \caption{}
    \includegraphics[width=\textwidth]{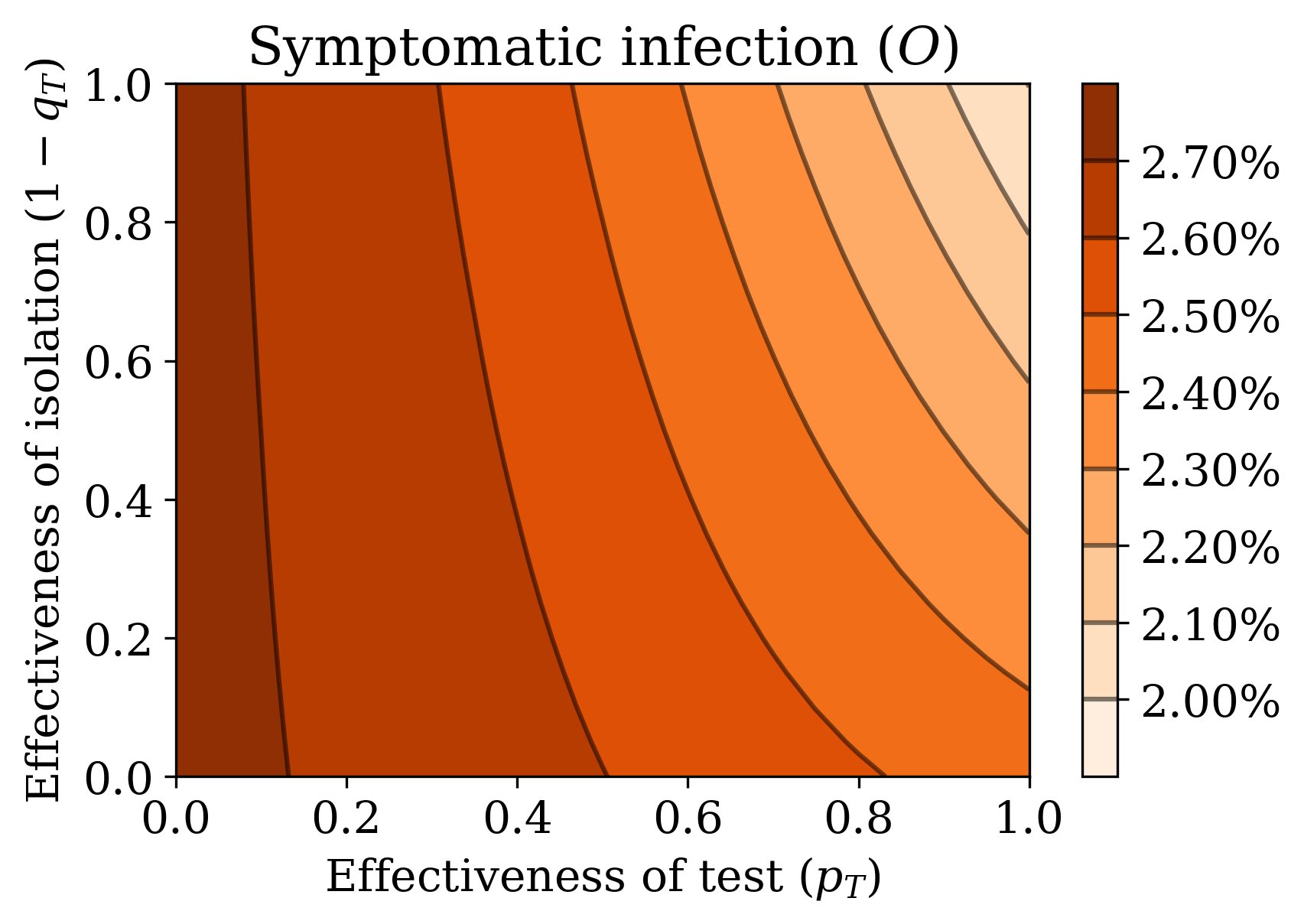}
    \end{subfigure}
    \hfill
    \begin{subfigure}[b]{0.45\textwidth}
    \caption{}
    \includegraphics[width=\textwidth]{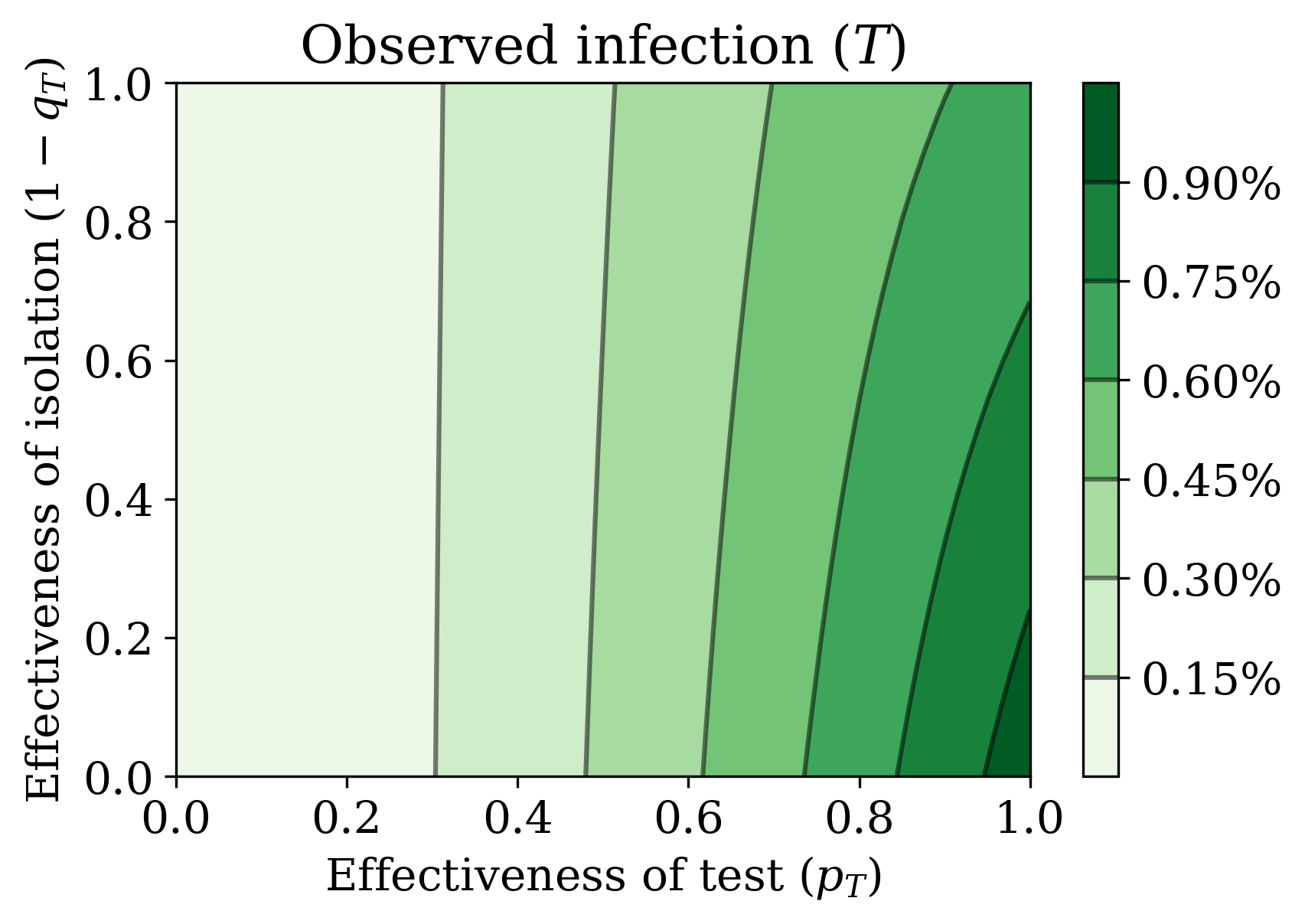}
    \end{subfigure}
    \hfill
    \begin{subfigure}[b]{0.45\textwidth}
    \caption{}
    \includegraphics[width=\textwidth]{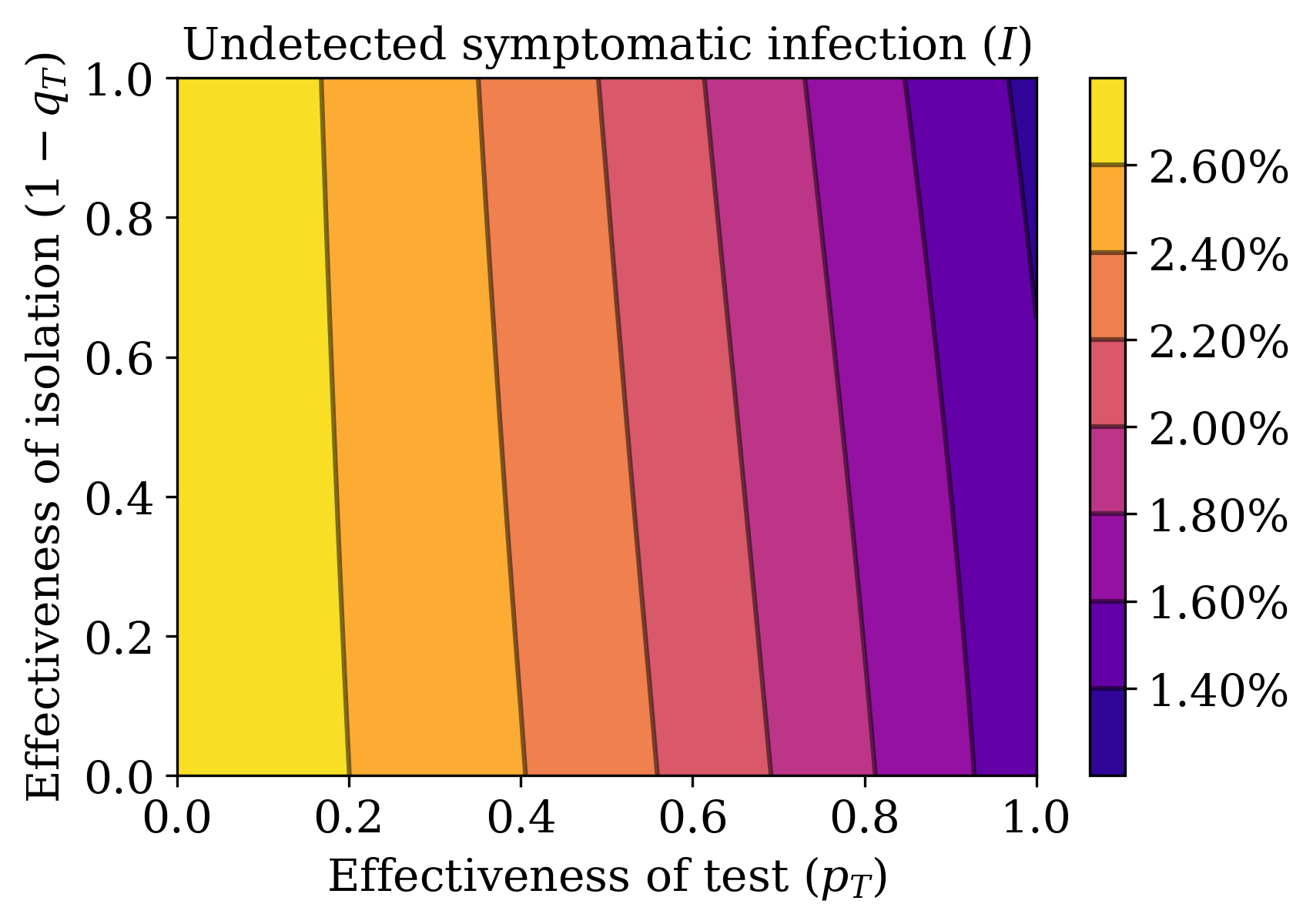}
    \end{subfigure}
    \hfill
    \begin{subfigure}[b]{0.45\textwidth}
    \caption{}
    \includegraphics[width=\textwidth]{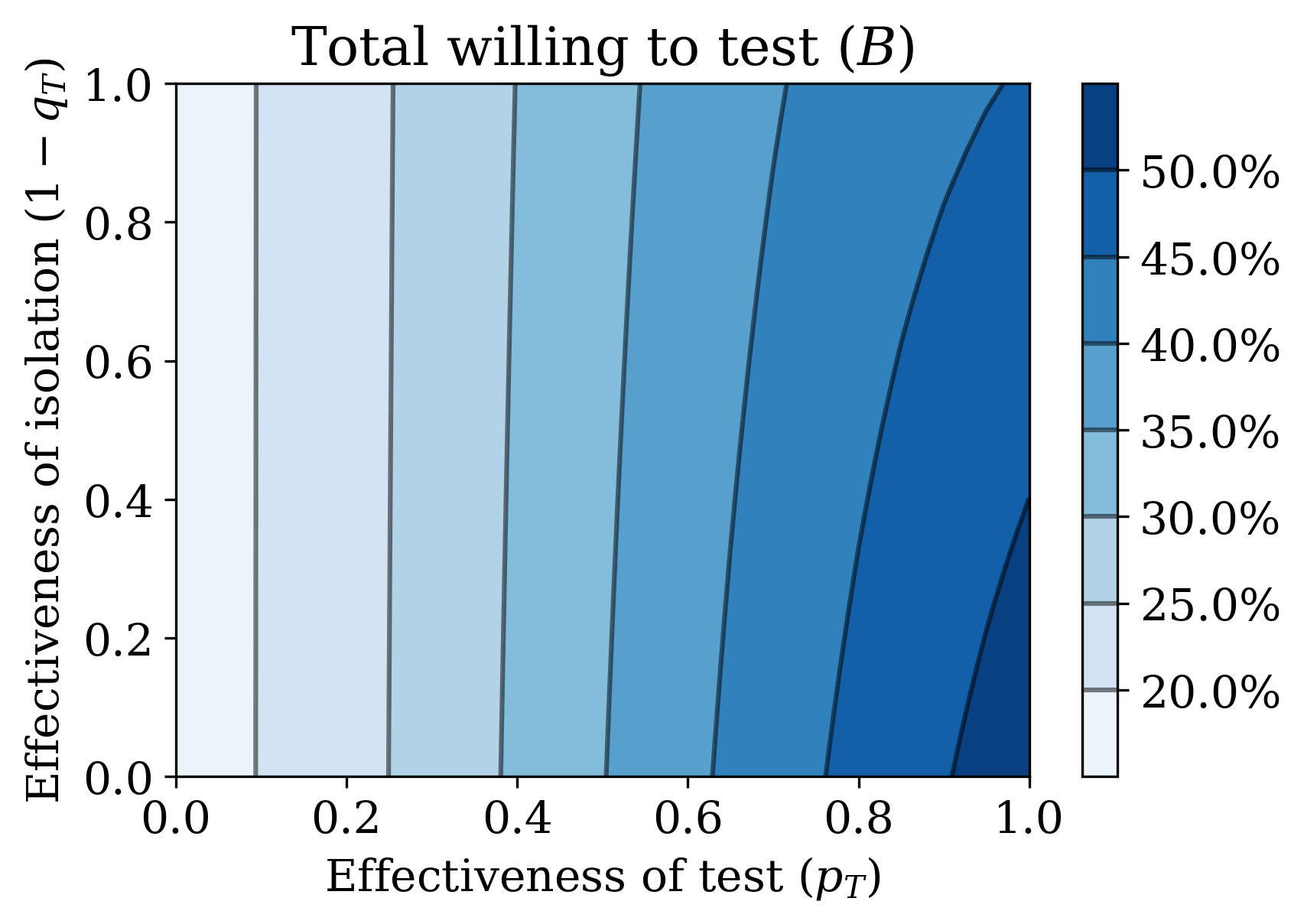}
    \end{subfigure}
    \hfill
    \begin{subfigure}[b]{0.45\textwidth}
    \caption{}
    \includegraphics[width=\textwidth]{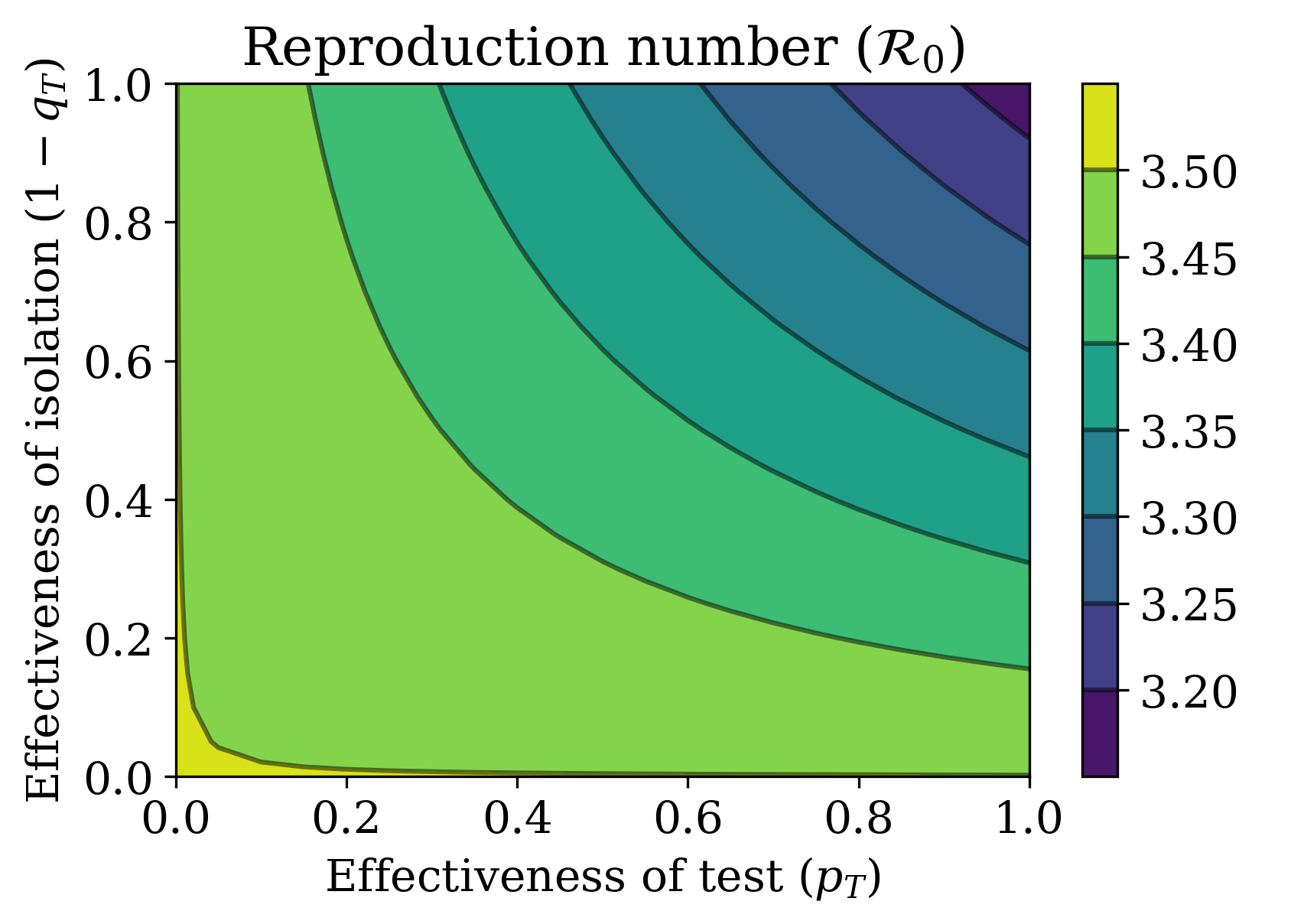}
    \end{subfigure}
    \caption{\textbf{The effect of isolation and test effectiveness on endemic states.} In each panel we display outputs for different combinations of isolation effectiveness ($1-q_T$, y-axis) and test effectiveness ($p_T$, x-axis). In all simulations we used a fixed behaviour-free reproduction number of $\Ro^D=3.28$.  \textbf{(a)} Total endemic infection state in the population ($O+A$); \textbf{(b)} Symptomatic endemic infection state in the population ($O$); \textbf{(c)} Observed symptomatic endemic infection state in the population ($T$); \textbf{(d)} Undetected symptomatic infection endemic state in the population ($I$); \textbf{(e)} Proportion of the population willing to test and isolate if symptomatic in the steady state of the population ($B$); \textbf{(f)} Basic reproduction number ($\Ro$).  For \textbf{(a)-(e)}, colour bars show the percentage of the population in each state.  For \textbf{(f)}, the colour bar shows the change in reproduction number.
    }
    \label{fig:pt_qt_effects}
\end{figure}

Testing effectiveness ($p_T$) and isolation effectiveness ($1-q_T$) can have a non-linear effect on the endemic steady states (Figure \ref{fig:pt_qt_effects}). However, similarly to the effects on the final size in the epidemic regime (Figure \ref{fig:sweep_finalSize}), for the steady state infection prevalence in the endemic regime improved testing effectiveness $p_T$ also has the more substantial effect. We observe the largest reduction in endemic infection prevalence (Figure \ref{fig:pt_qt_effects}, (a), (b)) for both high test effectiveness (large $p_T$) and effective isolation (low $q_T$; high $1-q_T$); this combination is also linked to the largest reduction in $\Ro$ (Figure \ref{fig:pt_qt_effects} (f)).  For lower levels of testing effectiveness, isolation has a negligible effect on the observed and unobserved symptomatic infection rates and the proportion of the population willing to test and isolate if symptomatic (Figure \ref{fig:pt_qt_effects}, (c), (d), (e)).  When testing effectiveness is high, we note a reduction in observed and unobserved symptomatic infection prevalences for improved isolation.  In these settings, the improved isolation is contributing to an overall reduction in infection, so observed cases are similarly reduced.  Due to the reduced number of cases observed for high values of $p_T$ and low values of $q_T$ (high $1-q_T$), we observe a corresponding reduction in the proportion of the population willing to test and isolate for the infection in the steady state of the system (Figure \ref{fig:pt_qt_effects} (e)).

\subsubsection{Behaviour and disease contagiousness on endemic states}

\begin{figure}[htbp]
    \centering
    \begin{subfigure}[b]{0.45\textwidth}
    \caption{}
    \includegraphics[width=\textwidth]{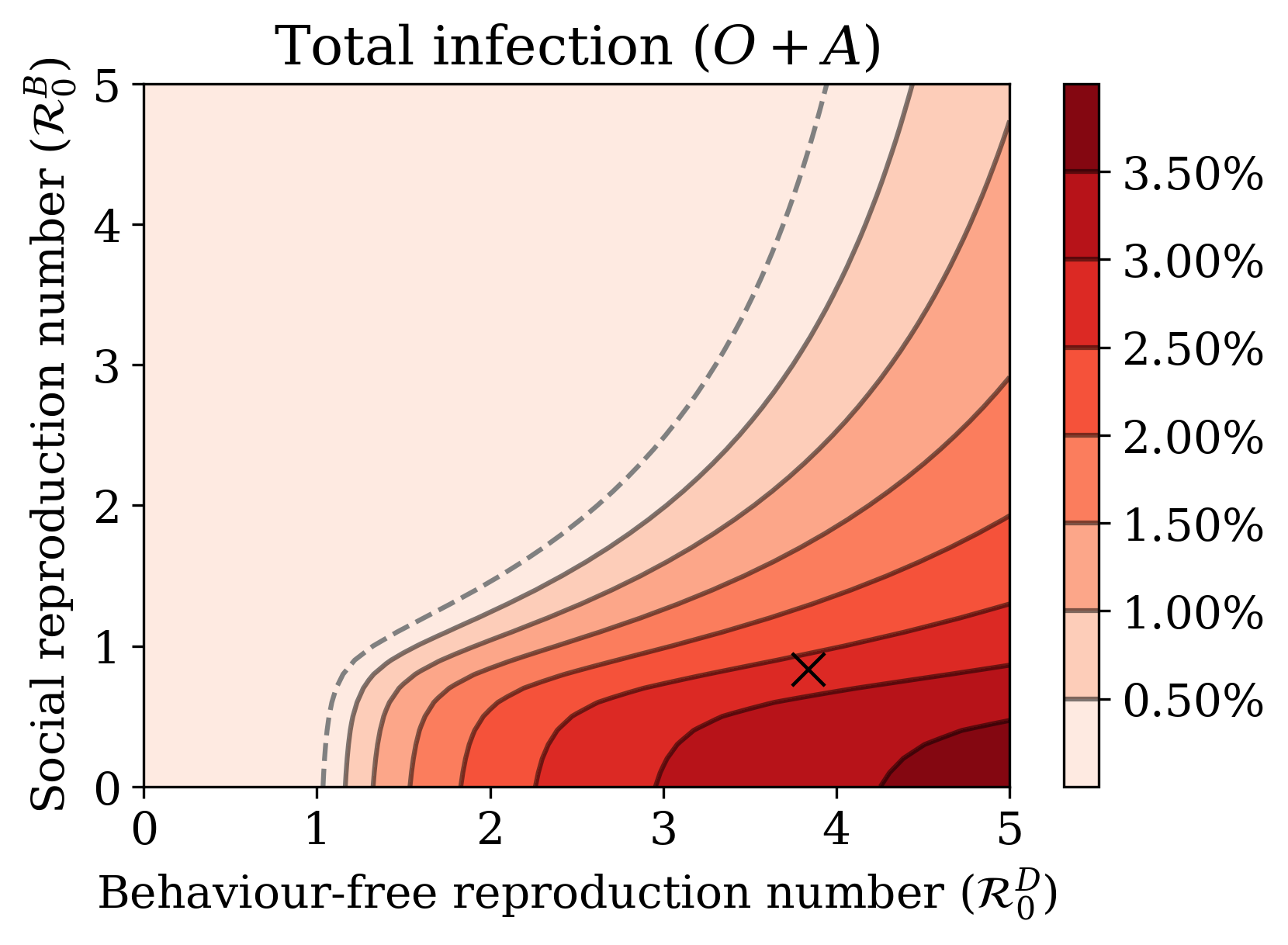}
    \end{subfigure}
    \hfill
    \begin{subfigure}[b]{0.45\textwidth}
    \caption{}
    \includegraphics[width=\textwidth]{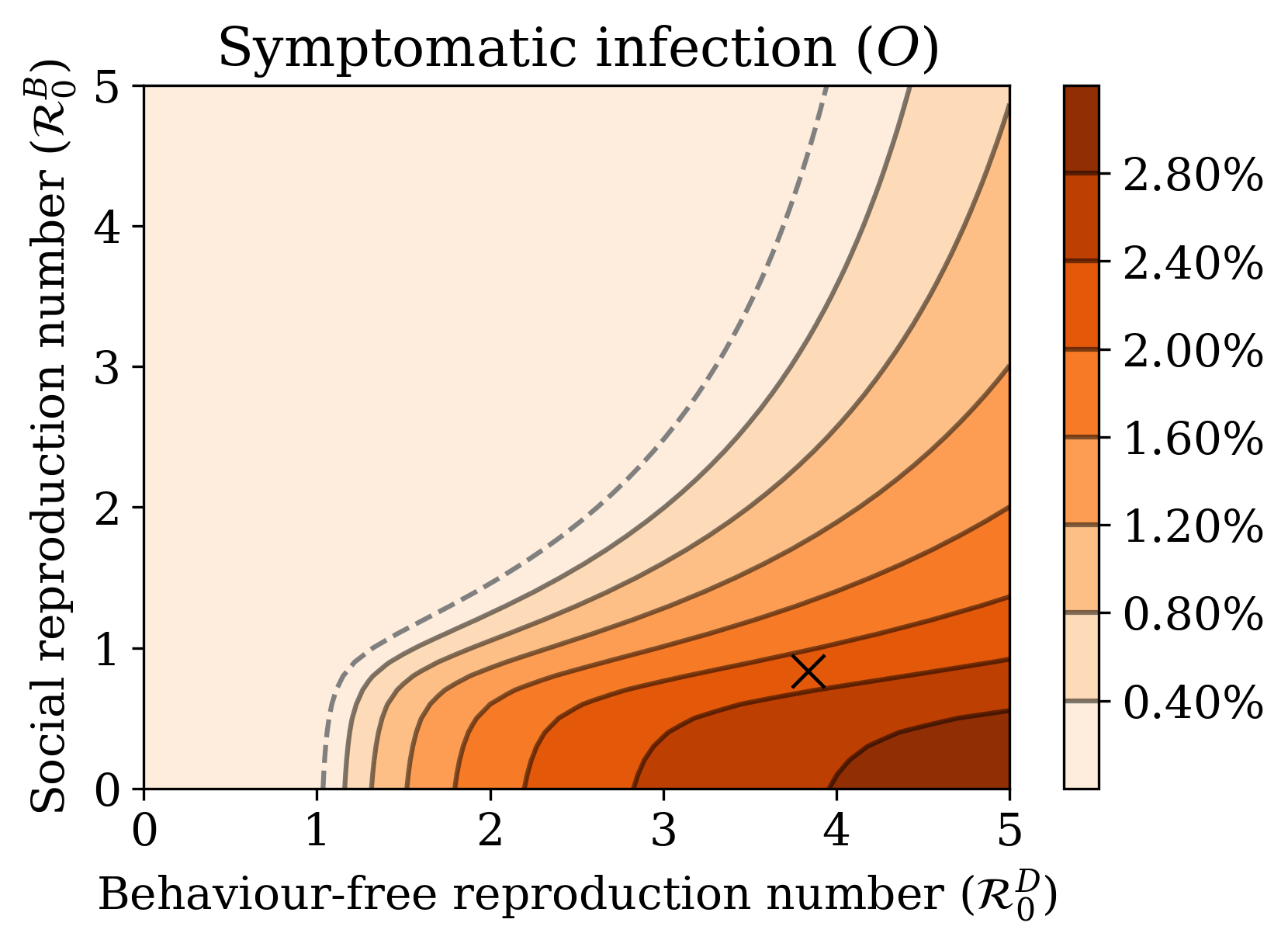}
    \end{subfigure}
    \hfill
    \begin{subfigure}[b]{0.45\textwidth}
    \caption{}
    \includegraphics[width=\textwidth]{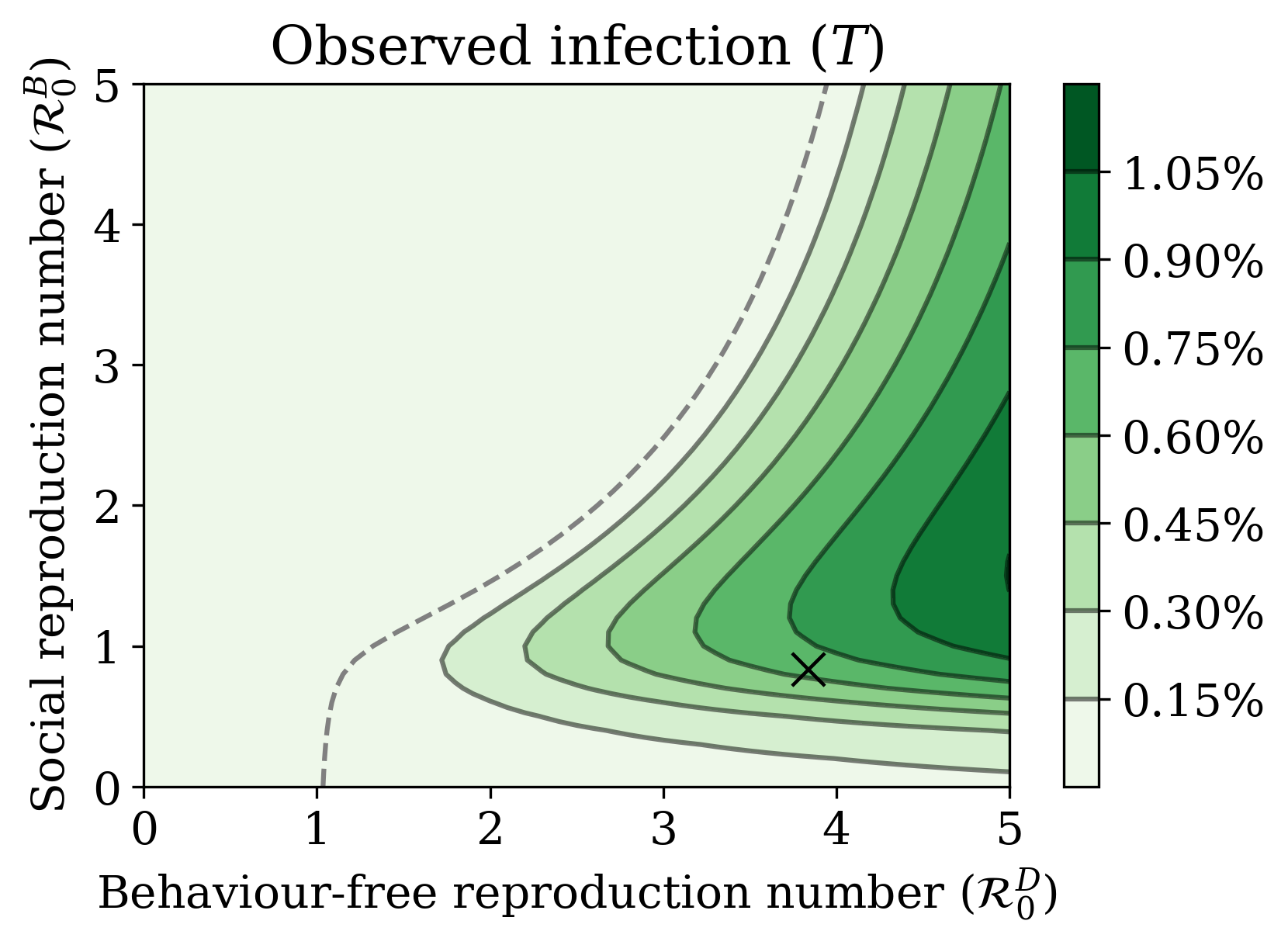}
    \end{subfigure}
    \hfill
    \begin{subfigure}[b]{0.45\textwidth}
    \caption{}
    \includegraphics[width=\textwidth]{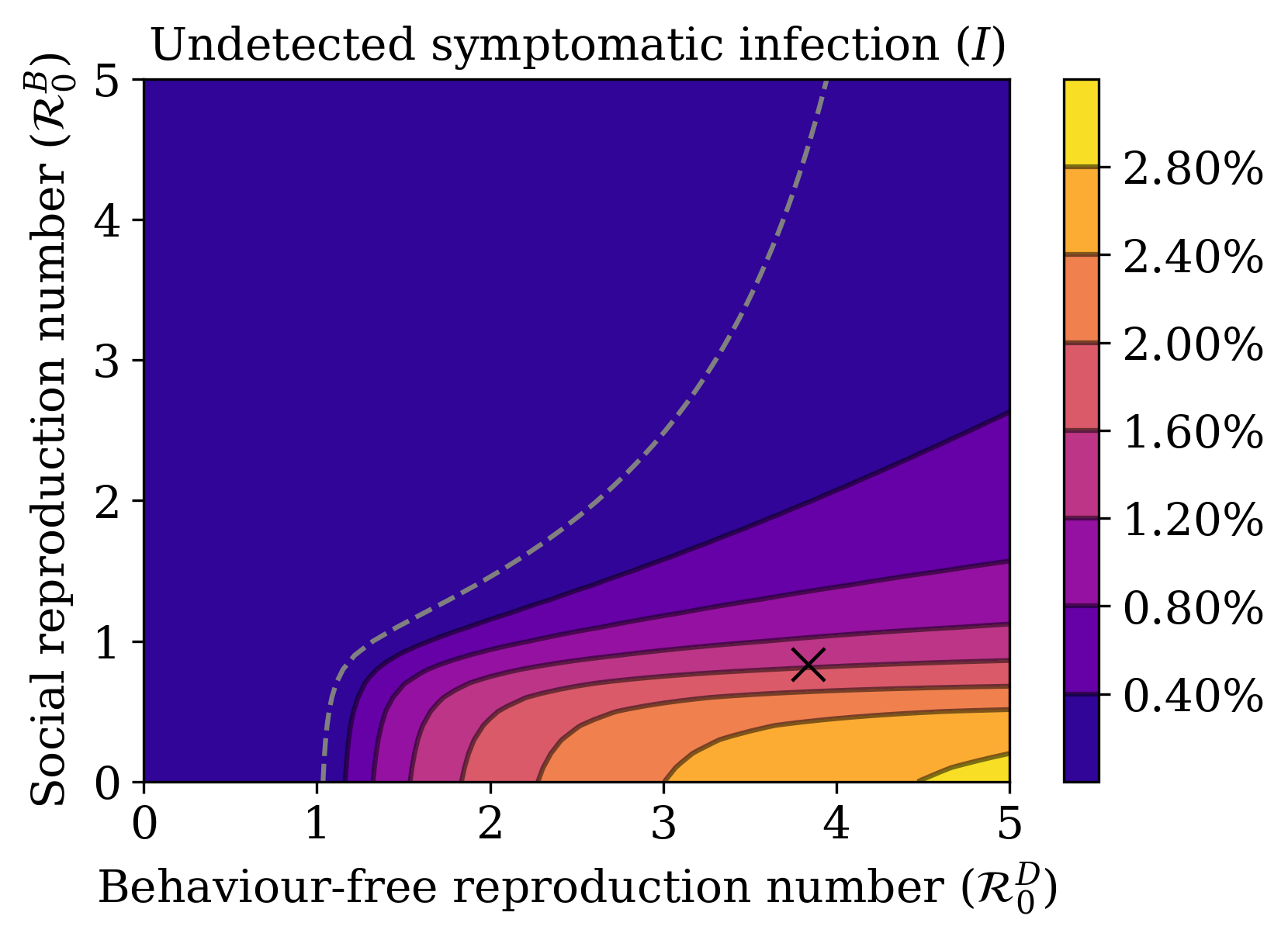}
    \end{subfigure}
    \hfill
    \begin{subfigure}[b]{0.45\textwidth}
    \caption{}
    \includegraphics[width=\textwidth]{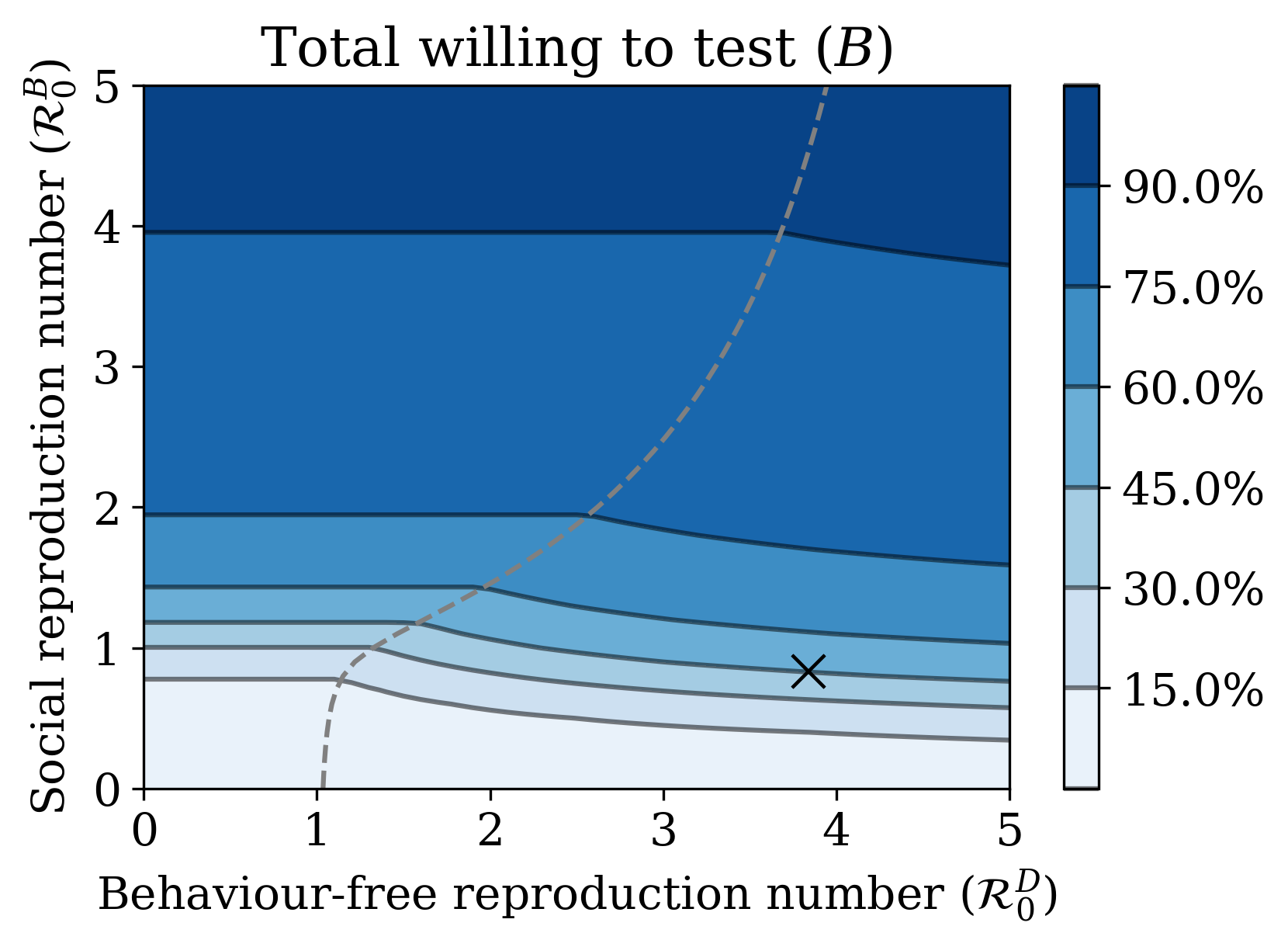}
    \end{subfigure}
    \hfill
    \begin{subfigure}[b]{0.45\textwidth}
    \caption{}
    \includegraphics[width=\textwidth]{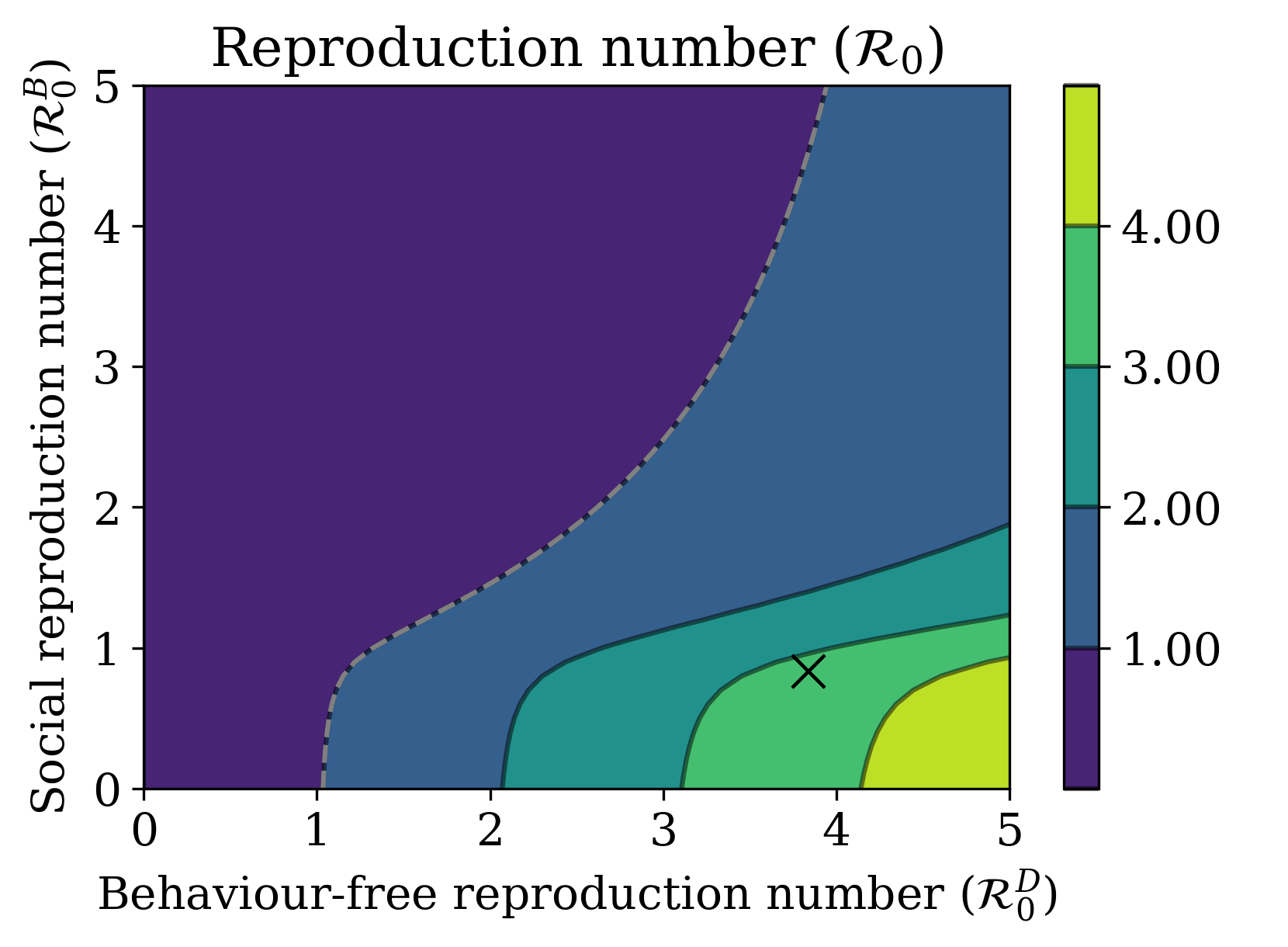}
    \end{subfigure}
    \caption{\textbf{The effect of the social reproduction number and the behaviour-free reproduction number on endemic states.} In each panel we display outputs for different combinations of the social reproduction number ($\Ro^{B}$, $y$-axis) and the behaviour-free reproduction number ($\Ro^{D}$, $x$-axis). The measures displayed are: \textbf{(a)} Total endemic infection state in the population ($O+A$); \textbf{(b)} Symptomatic endemic infection state in the population ($O$); \textbf{(c)} Observed symptomatic endemic infection state in the population ($T$); \textbf{(d)} Undetected symptomatic infection endemic state in the population ($I$); \textbf{(e)} Proportion of the population willing to test and isolate if symptomatic in the steady state of the population ($B$); \textbf{(f)} The basic reproduction number ($\Ro$).  For \textbf{(a)-(e)}, colour bars show the percentage of the population in each state.  For \textbf{(f)}, the colour bar shows the change in reproduction number. The dashed grey line shows $\Ro=1$ and the black cross shows the baseline parameters in Table \ref{tab:model_parameter}.
    }
    \label{fig:r0d_r0b_effects}
\end{figure}

Both the innate infectiousness of the pathogen ($\Ro^D$) and the social contagiousness of the testing behaviour ($\Ro^B$) have significant impacts on reported and unreported case numbers in the endemic steady state (Figure \ref{fig:r0d_r0b_effects}).  The boundary for elimination of the infection by reducing $\Ro<1$ is non-linear in the behaviour and diseases contagiousness measures (Figure \ref{fig:r0d_r0b_effects} (f)).  This non-linearity suggests that for a more infectious disease (larger $\Ro^D$) we require a substantial increase in the social influence of testing behaviour (larger $\Ro^B$) to eliminate the infection in the steady state.  This corresponds to a larger increase in the percentage of the population willing to test and isolate if symptomatic (Figure \ref{fig:r0d_r0b_effects} (e)).

When $\Ro > 1$, the prevalence of symptomatic and non-symptomatic infection increases with a more infectious disease (larger $\Ro^D$, Figure \ref{fig:r0d_r0b_effects}, (a), (b), (d)), although this increase is less noticeable when testing behaviour is strongly influenced by social interactions (larger $\Ro^B$).  However, the prevalence of reported infections does not necessarily increase with increased disease contagiousness (Figure \ref{fig:r0d_r0b_effects} (c)).  Indeed, when testing behaviour is negligibly influenced by social interactions, we observe near zero reported infections in the steady state, providing a message of false-elimination when in fact infection rates are at their highest steady state value.  When social contagiousness is increased for a fixed pathogen ($\Ro^D$), we observe an increase in reported cases followed by a decrease until elimination is reached ($\Ro<1$).  This non-linear hill structure in reported cases of symptomatic infections indicates that it is important to know which side of the hill a population is on, in order to understand how to interpret reported case numbers correctly.  For example, when testing rates are high and we observe a small proportion of the population testing positive for the infection, there is more confidence that the true case numbers throughout the population are reduced as well.  However, if both testing rates and case numbers are low, there is a possibility that the actual burden of the infection on the population is much larger than it appears.

\subsubsection{Stability of steady states}

\begin{figure}[htbp]
    \centering
    \begin{subfigure}[b]{0.32\textwidth}
    \caption{$\omega_1, \omega_2, \omega_3 > 0$}
    \includegraphics[width=\textwidth]{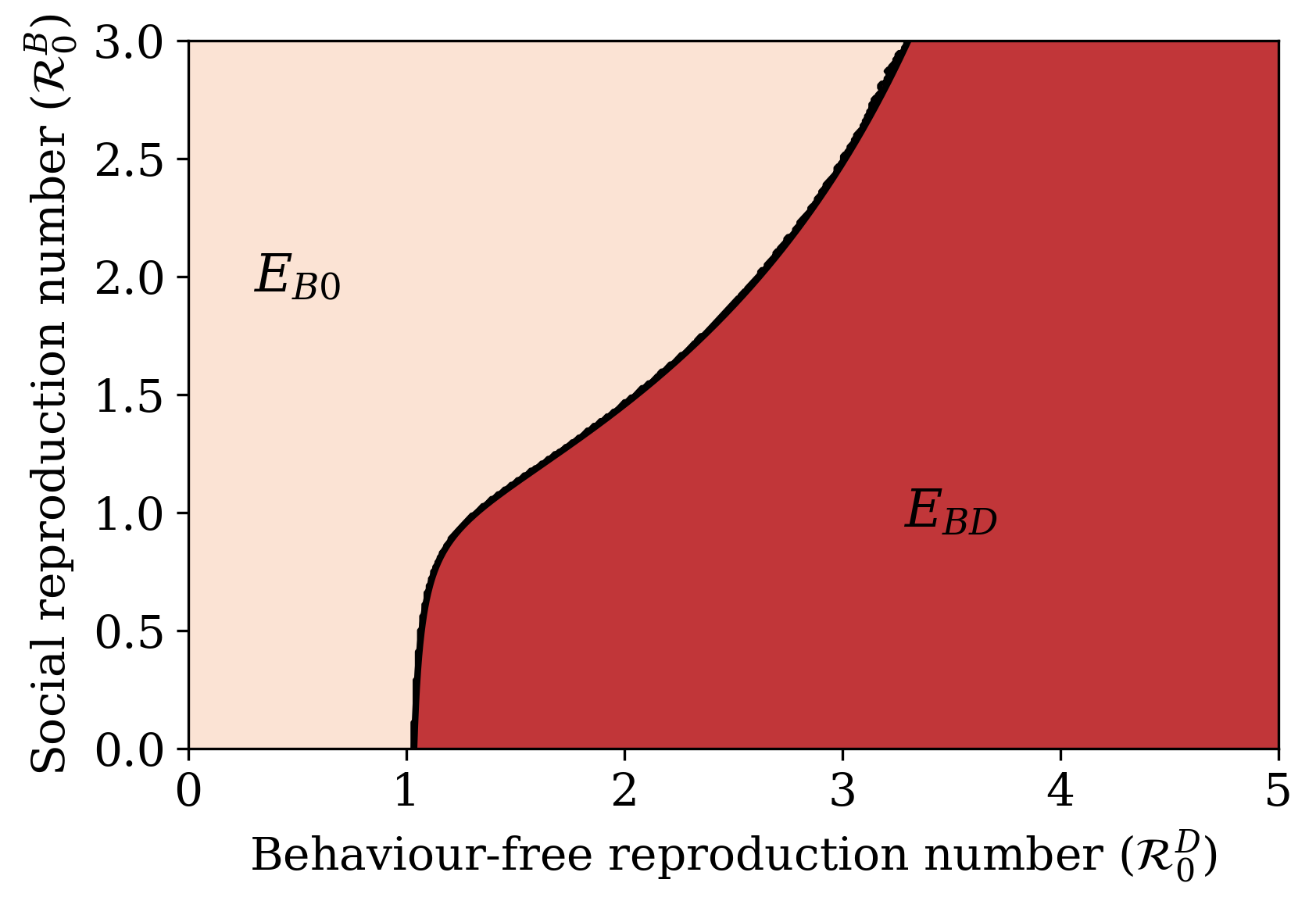}
    \end{subfigure}
    \hfill
    \begin{subfigure}[b]{0.32\textwidth}
    \caption{$\omega_3 = 0, \omega_1, \omega_2 > 0$}
    \includegraphics[width=\textwidth]{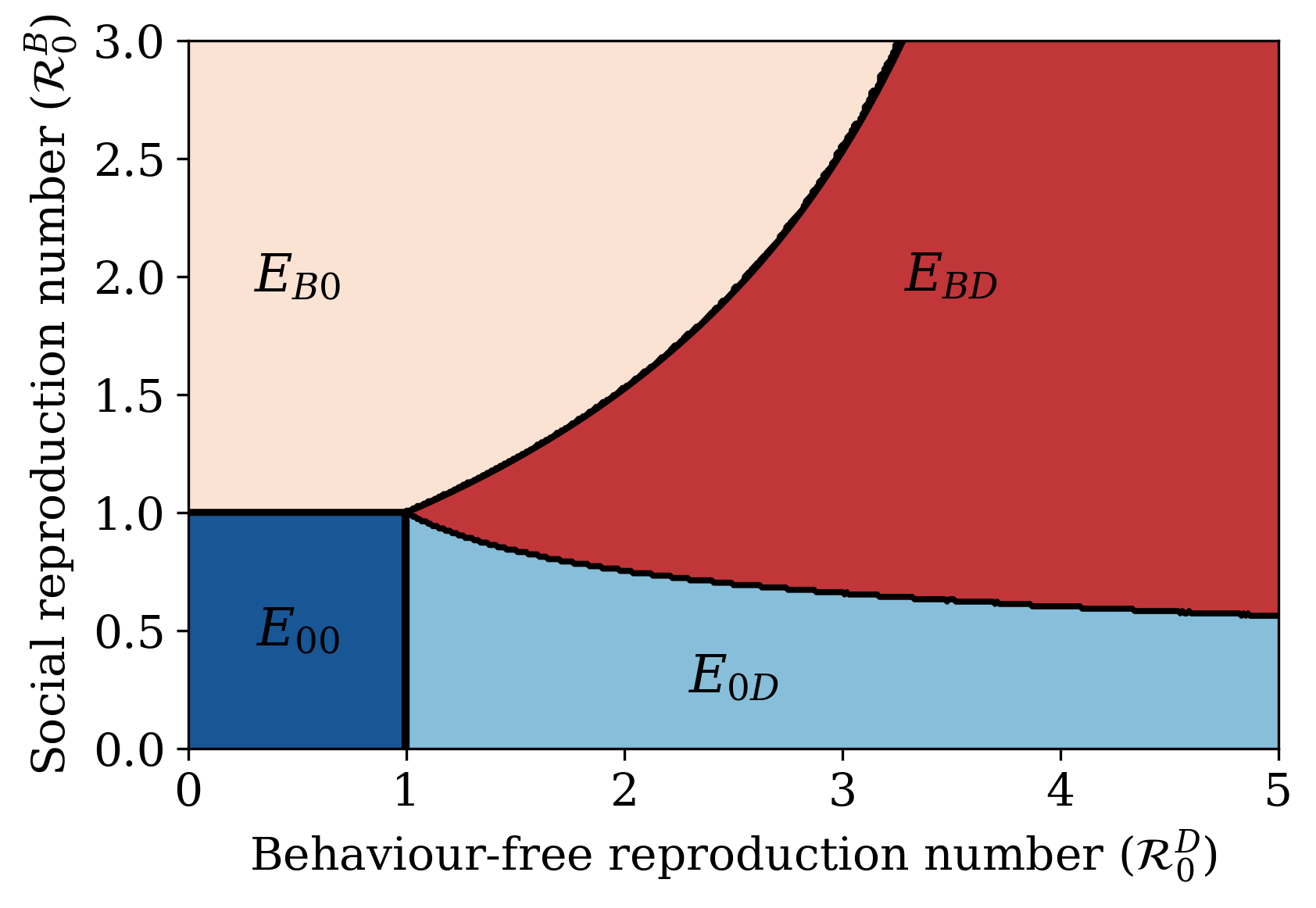}
    \end{subfigure}
    \hfill
    \begin{subfigure}[b]{0.32\textwidth}
    \caption{$\omega_2=\omega_3 = 0, \omega_1>0$}
    \includegraphics[width=\textwidth]{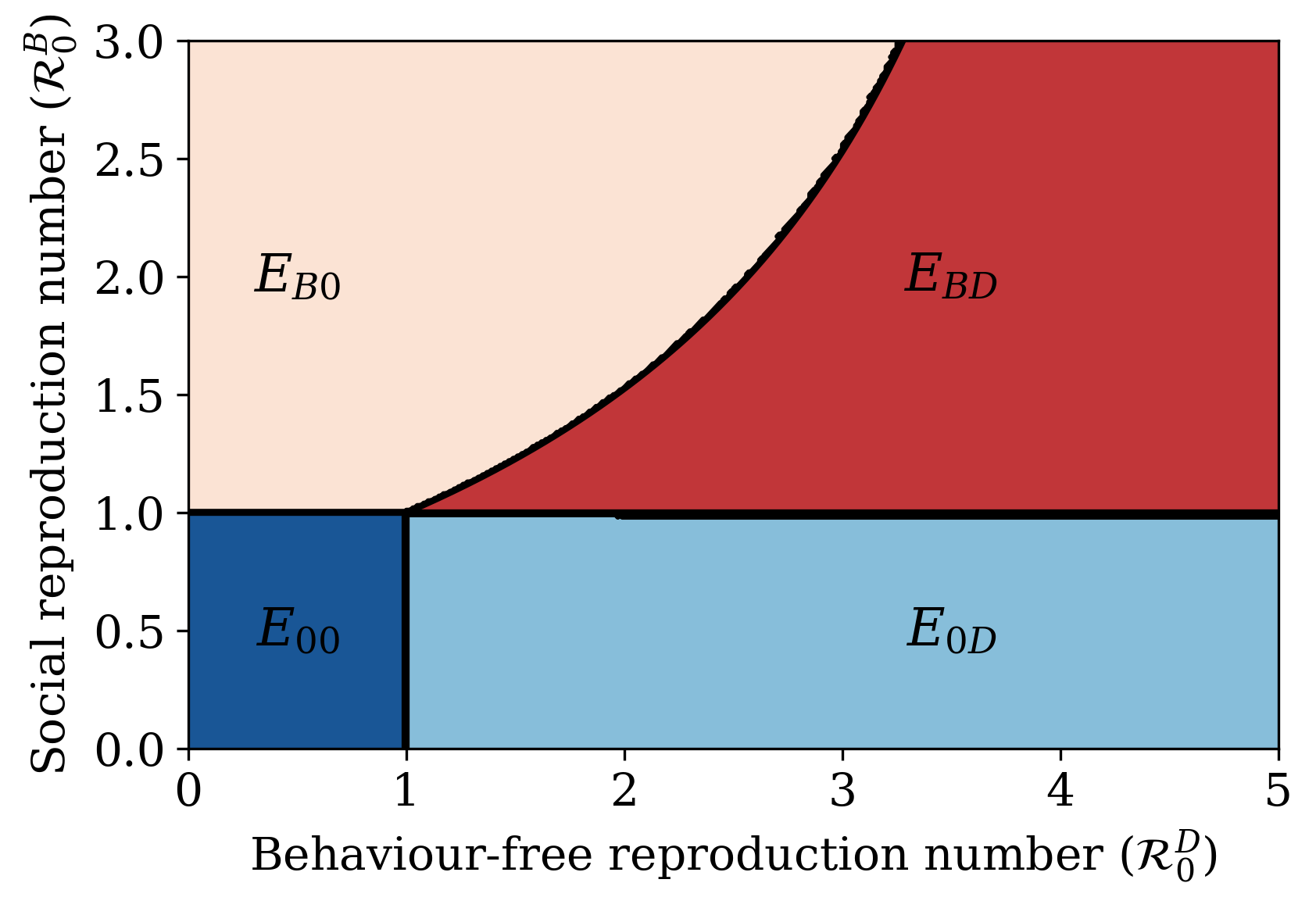}
    \end{subfigure}
    \hfill
    \caption{\textbf{Stable regions of the steady states for different combinations of behavioural parameters.} The $x$-axis shows the behaviour-free reproduction number ($\Ro^{D}$) varying through changes in $\beta$ and the $y$-axis shows the social reproduction number ($\Ro^B$) varying through changes in $\omega_1$. \textbf{(a)} Feasible steady states when all behavioural parameters are greater than zero ($\omega_1, \omega_2, \omega_3 > 0$).  \textbf{(b)} Feasible steady states when $\omega_1, \omega_2 > 0$ and $\omega_3 =0$. \textbf{(c)} Feasible steady states when $\omega_1 >0$ $\omega_2=\omega_3=0$.  In each figure: dark blue shows the behaviour and disease free state $E_{00}$; light blue shows the no behaviour, disease endemic state $E_{0D}$; light red shows the behaviour endemic, no disease state $E_{B0}$, and; dark red shows the behaviour and disease endemic state $E_{BD}$.}
    \label{fig:steadyStateRegions}
\end{figure}

There are four feasible steady state combinations for this BaD model for testing and isolation: no behaviour or disease (denoted $E_{00}$); no behaviour, disease endemic (denoted $E_{0D}$); behaviour endemic, no disease (denoted $E_{B0}$), and; behaviour and disease endemic (denoted $E_{BD}$).
The stable regions of the steady states agree with the regions previously identified by Ryan \textit{et al.} \cite{Ryan:2024} when $\omega_1,\omega_2,\omega_3>0$ (Figure \ref{fig:steadyStateRegions} (a)) and when $\omega_1>0, \omega_2=\omega_3=0$ (Figure \ref{fig:steadyStateRegions} (c)).  In contrast, when $\omega_3=0$ and $\omega_1,\omega_2>0$ we observe a non-linear boundary where the stable steady state switches from $E_{0D}$ to $E_{BD}$ (Figure \ref{fig:steadyStateRegions} (b)).  This boundary is driven by a non-linear relationship between social spread of behaviour ($\Ro^{B}$) and the perception of illness threat ($\omega_2$).  This suggests that there are parameter combinations where an infection can be endemic in a population without being detectable in the sense that negligibly few  tests are being conducted.  Since those who test positive ($T$) are explicitly a subset of those willing to test and isolate ($B$), the threshold observed in Figure \ref{fig:steadyStateRegions} (b) shows there is a need for enough people actively to seek testing and isolate if positive before the feedback mechanism of the perception of illness threat can take effect.

\section{Discussion}
\label{sec:discussion}

\subsection{Findings in context}
We have extended an existing behaviour and disease (BaD) model \cite{Ryan:2024} to investigate the epidemiological consequences of the behavioural dynamics of symptomatic testing and isolation. Using our model, through theoretical derivations we have computed the basic reproduction number and identified stability conditions for epidemiological-behavioural steady states. Numerical simulation of epidemic and endemic regimes have then shown a variety of relationships between an epidemiological outcome of interest and the testing associated behavioural parameter values.

Compared to the original BaD model \cite{Ryan:2024}, we observed similarities in the thresholds and steady states.  Despite the obvious complexity induced by a more complicated disease structure, we found here that the reproduction number can also be decomposed into four components describing the relative contribution for the mixing behavioural groups.  Further, Theorem \ref{thm:steady_states} shows a similar complexity arising in the steady states, with analytic expressions in terms of model states found numerically.  However, there are interesting differences in the dynamics and steady state properties.  In particular, the role of the perception of illness threat is different.  Here, we have modelled the perception of illness threat directly on the prevalence of positive tests in the population, as opposed to a less well-defined prevalence of cases in the original model.  This has caused an interesting threshold in the absence of spontaneous behavioural uptake (Figure \ref{fig:steadyStateRegions} (b)).  If testing behaviour is not socially contagious enough, we find that there exists a disease endemic state with no testing, which is not possible in the original BaD model.  The influence of perception of illness threat also creates the possibility of epidemics with false elimination (the hill structure in Figure \ref{fig:r0d_r0b_effects} (c)).

We have exemplified how BaD models enable us to explore the effects of interventions for reducing disease spread that are more targeted towards human behaviour. Previous individual-level transmission modelling studies have shown tracing and isolation alone are typically insufficient to control an epidemic \cite{Kucharski:2020}. Our results agree with these existing findings from a behavioural perspective,

Ultimately, the application and calibration of BaD transmission models to real-world problems is contingent on the availability of appropriate data sources \cite{Funk:2010,Bedson:2021}. Compliance with testing policy is an important data input. COVID-19 testing interventions have shown that the testing behaviours, in multiple settings, have demographic and socioeconomic heterogeneities. In 2020, an analysis of the demographic determinants of testing incidence and COVID-19 infections in New York City neighbourhoods reported that people residing in poor or immigrant neighbourhoods were less likely to be tested, but the likelihood that a test was positive was larger in those neighbourhoods \cite{Borjas:2020}. In Chile, test positivity and testing delays were higher in lower–socioeconomic status municipalities \cite{Mena:2021}. Mass testing data has also been used in conjunction with surveillance surveys to identify gaps in the uptake of public health interventions at both fine-scale levels and across sociodemographic groups in England \cite{Bajaj:2024} and Australia \cite{Guajardo:2025}. Improving our ability to onboard such data into BaD models, including real-time compliance estimates from community surveys \cite{Eales:2024}, can result in more impactful testing interventions for disease control.

\subsection{Study limitations and directions for further work}

We regard the model we have presented as an entry point for quantitative study of the feedbacks between symptomatic testing behaviour and infection dynamics. As with any model we acknowledge that it is, by necessity, a simplified representation of reality. It is important that we consider the modelling assumptions made and their potential limitations. Here we expand on the implications of our behavioural parameter assumptions, testing assumptions and reduction of model dimensionality via assuming the pathogen to have a single strain and there being a sole testing intervention deployed.

A core aspect of our presented model is parsimony. We therefore had fixed values for the behaviour effect parameters in the testing behavioural uptake rate ($\omega_{1},\omega_{2},\omega_{3}$) and testing behaviour abandonment rate ($\alpha_{1},\alpha_{2}$). Relaxing this assumption would permit temporal variation in these behavioural factors. One scenario of interest would be to have a decay rate on the perception of illness threat ($\omega_{2}$), corresponding to growing normality through time amongst the population for cases occurring and infection circulating in the community. Alternatively, we could vary the epidemiological metric to which the perception of illness threat is tied. We associated $\omega_{2}$ with those who tested positive. However, as there may be delays in reporting the count of positive tests to the wider population, different lags to this measure could be applied. Further, the case severity that the population responds to may alter during an outbreak (i.e.\ at different times the number of cases, hospitalisations and/or deaths may be attributed different importance); consideration of these behavioural feedbacks would require refinements to our model 
to include severe case outcomes (such as hospitalisations and deaths).

Another parsimony based assumption was that symptomatic `behaviour aware' individuals who tested negative (due to test sensitivity being below 100\%) had no change in their behaviour and were assumed to have the same infectiousness as those who were symptomatic and did not test. Removing this assumption would lead to more parameters being added to the model, increasing the parameter space to be explored beyond what we deem to be digestible in this single study. An ultimate ambition would be such processes being data-driven, meaning future analysis would benefit greatly from further behavioural data collection.
   
With regards to our testing assumptions, we implemented testing with idealised conditions of no delays in test accessibility and availability, or constraints on overall supply. In other words, we assumed that all those who wanted to test for infection could do so. Under circumstances where there were not sufficient tests both accessible and available, that would result in missed opportunities to identify infected individuals. If these individuals do not self-isolate during the infection episode (instead mixing with others in the population), we would anticipate higher peaks in infectious prevalence. These additional cases could then cascade into further onward transmission and new cases, putting further pressure on strained testing resources. Conceivably, issues with test accessibility and availability can have negative consequences on continued engagement with testing protocols for future infection episodes \cite{Embrett:2022,Zhang:2022}. These potential feedbacks between testing uptake and testing availability are an additional behavioural consideration that would benefit from data collection.

Associated with our idealised assumptions on testing capacity, another direction for enhancing the model is onboarding additional data steams for parameter estimation, such as for the test-seeking propensity. As an initial entry point, publicly available data at a coarse demographic-level may be provided by public health agencies. Using the COVID-19 pandemic as an example setting, in the UK statistics for testing were made available in weekly reports \cite{UKHSA:2022}. Yet, such data may not have the sufficient detail to capture the multiple potential influences on testing uptake, including employment conditions, geographical location, population density, beliefs about testing, health literacy, region, index of multiple deprivation (IMD) or other indicator of deprivation, age and/or gender. Pseudonymised individual-level data would be very beneficial for capturing such signals, which may then be used in data-driven models. Such individual-level data would likely not be publicly available for data protection reasons, but instead be held in secure data environments; for example, pseudonymised individual-level SARS-CoV-2 lateral flow testing data for Liverpool in the UK was used by Green \emph{et al.} \cite{Green:2021} to evaluate social and spatial inequalities. 

Lastly, we reduced the complexity of the modelled problem by treating the pathogen as only having a single strain and only considering one type of disease control (testing). The consideration of multiple strains/variants could be used to investigate the relationships to community survey data \cite{Eales:2024}. Furthermore, public health policies are often a combination of pharmaceutical (when available) and non-pharmaceutical interventions (NPIs). Multiple NPIs may also be used in conjunction with one another, such as social distancing and mask-usage in various geographic and social contexts during the COVID-19 pandemic~\cite{NitaBharti:2021, Bruin:2022}. Due attention should be given to the parameterisation of the behaviour and epidemiological feedbacks in the presence of a package of interventions -- the behavioural responses to testing may be influenced by the availability of other interventions -- which is presently an open question and merits further research.

\subsection{Conclusion}
Models of infectious disease dynamics have traditionally been bereft of data-driven and/or theoretical knowledge of outbreak behavioural dynamics. Our study has provided a structured modelling approach for the joint consideration of mechanistic behaviour and disease transmission processes to the dynamics of (symptomatic) testing and isolation. To choose the most appropriate behavioural model and structure (with reasonable assumptions on the components of the behavioural model, to fit the target behaviour, infection and questions of interest), we encourage interdisciplinary collaboration across the behavioural, biological, data and mathematical sciences. With these data and modelling advancements we would like to build public trust in modelling studies, which can have onward benefits for the public health, economic and social response to future infectious disease outbreaks.

\section*{Acknowledgements}

The authors would like to thank participants in the Isaac Newton Institute for Mathematical Sciences programme ``Modelling and inference for pandemic preparedness'' for helpful discussions on this work.
TH would like to thank Istvan Kiss for sharing observations on the history of behaviour and disease modelling.

\section*{Financial disclosure}

EMH is affiliated to the NIHR Health Protection Research Unit in Emerging and Zoonotic Infections (NIHR207393). The views expressed are those of the author(s) and not necessarily those of the NIHR or the Department of Health and Social Care. EMH is also funded by The Pandemic Institute, formed of seven founding partners: The University of Liverpool, Liverpool School of Tropical Medicine, Liverpool John Moores University, Liverpool City Council, Liverpool City Region Combined Authority, Liverpool University Hospital Foundation Trust, and Knowledge Quarter Liverpool (EMH is based at The University of Liverpool). The views expressed are those of the author(s) and not necessarily those of The Pandemic Institute. TH is supported by the Wellcome Trust (Grant number 227438/Z/23/Z). MGR is supported by the Marsden Fund under contract 20-MAU-50.
The authors would like to thank the Isaac Newton Institute for Mathematical Sciences for support and hospitality during the programme ``Modelling and inference for pandemic preparedness'', where work on this paper was initiated. This work was supported by EPSRC grant EP/Z000580/1.
The funders had no role in study design, data collection and analysis, decision to publish, or preparation of the manuscript. For the purpose of open access, the authors have applied a Creative Commons Attribution (CC BY) licence to any Author Accepted Manuscript version arising from this submission.

\section*{Conflicts of interest}

All authors declare they have no conflict of interest.

\section*{Code availability statement}

The code associated with this study is available at:\\ \url{https://github.com/Matthew-Ryan1995/BaD_testing_and_isolation}.

\newpage
\begin{appendices}
\section{Stability of the infection-free equilibrium}\label{append:stability}
We will first derive the Jacobian of the system and evaluate its eigenvalues at the infection-free equilibrium, thereby confirming Proposition \ref{thm:stability} that stability holds when $\Ro <1$.

For notational convenience, we define
$$\vecM_N=\left(\begin{array}{cccc}-\lambda-\nu & -\nu & -\nu & -\nu \\
\lambda & -\sigma & 0 & 0 \\0 & p_A\sigma & -\gamma & 0 \\0 & \left(1-p_A\right)\sigma & 0 & -\gamma\end{array}\right),
\qquad \vecn=\left(\begin{array}{c}S_N \\E_N \\A_N \\ I_N\end{array}\right)$$
and analogously for behavers: 
$$\vecM_B=\left(\begin{array}{cccc}-q_B\lambda-\nu & -\nu & -\nu & -\nu \\
q_B\lambda & -\sigma & 0 & 0 \\0 & p_A\sigma & -\gamma & 0 \\0 & \left(1-p_A\right)\left(1-p_T\right)\sigma & 0 & -\gamma\end{array}\right),
\qquad \vecb=\left(\begin{array}{c}S_B \\E_B \\A_B \\ I_B\end{array}\right)$$
with $$\lambda=\beta\left[I_N+I_B+q_A\left(A_N+A_B\right)+q_T T\right].$$ 

In this notation, the BaD model can be written in matrix form as
\begin{align}
\label{dM}
 \frac{\de \vecn}{\de t}    &= \vecM_N\vecn -\omega(B,T)\vecn+\alpha(N)\vecb +\nu N\vecu_1 \\
 \frac{\de \vecb}{\de t}    &= \vecM_B\vecb +\omega(B,T)\vecn-\alpha(N)\vecb +\nu \left(1-N-T\right)\vecu_1 \nonumber
\end{align}
where $\vecu_1$ is the first standard basis vector, having a
first element equal to $1$ and all other elements equal to $0$.

Next, the Jacobian $\vecJ$ of the full system \eqref{dM} has the block form
$$\vecJ=\left(
\begin{array}{c|cccc|cccc}
   & 0 & 0 & 0 & 0 & 0 & 0 & 0 & 0 \\
  \vecJ^{0}&   &  &  &  &  &  & & \\
   &   0 & 0 & 0 & 0 & 0 & \left(1-p_A\right) p_T \sigma & 0 & 0 \\
  \hline
    &   &  &  &  &  &  & & \\
  \vecJ^{n0} &    &  &  \vecJ^{nn}  &  &  &  \hspace{1cm}\vecJ^{nb} & & \\
  &   &  &  &  &  &  & & \\
   \hline
   &   &  &  &  &  &  & & \\
  \vecJ^{b0} &    &  &  \vecJ^{bn}  &  &  & \hspace{1cm} \vecJ^{bb} & & \\
  &   &  &  &  &  &  & & \\
  \end{array}
\right)$$
where $\vecJ^{0}$ is a $2 \times 2$ matrix given by
$$\vecJ^{0} =\left(\begin{array}{cc}\dfrac{\partial\omega}{\partial B}N+\dfrac{\partial\alpha}{\partial N}\left(1-N - T\right)-\omega-\alpha & -\dfrac{\partial\omega}{\partial T}N - \alpha \\[0.3cm]
0 & -\gamma\end{array}\right)$$ $\vecJ^{n0}$ and $\vecJ^{b0}$ are $4\times 2$ matrices given by
\begin{align*}
\vecJ^{n0} & =\left(\begin{array}{cc}\dfrac{\partial\omega}{\partial B}S_N+\dfrac{\partial\alpha}{\partial N}S_B+\nu & -q_T\beta S_N-\dfrac{\partial\omega}{\partial T}S_N \\ [0.3cm]
\dfrac{\partial\omega}{\partial B}E_N+\dfrac{\partial\alpha}{\partial N}E_B & q_T\beta S_N -\dfrac{\partial\omega}{\partial T}E_N \\ [0.3cm]
\dfrac{\partial\omega}{\partial B}A_N+\dfrac{\partial\alpha}{\partial N}A_B & -\dfrac{\partial\omega}{\partial T}A_N \\ [0.3cm]
\dfrac{\partial\omega}{\partial B}I_N+\dfrac{\partial\alpha}{\partial N}I_B & -\dfrac{\partial\omega}{\partial T}I_N\end{array}\right)
\\
\vecJ^{b0}&=\left(\begin{array}{cc}-\dfrac{\partial\omega}{\partial B}S_N-\dfrac{\partial\alpha}{\partial N}S_B-\nu & -q_Bq_T\beta S_B+\dfrac{\partial\omega}{\partial T}S_N-\nu \\ [0.3cm]
-\dfrac{\partial\omega}{\partial B}E_N-\dfrac{\partial\alpha}{\partial N}E_B & q_Bq_T\beta S_B +\dfrac{\partial\omega}{\partial T}E_N \\ [0.3cm]
-\dfrac{\partial\omega}{\partial B}A_N-\dfrac{\partial\alpha}{\partial N}A_B & \dfrac{\partial\omega}{\partial T}A_N \\ [0.3cm]
-\dfrac{\partial\omega}{\partial B}I_N-\dfrac{\partial\alpha}{\partial N}I_B & \dfrac{\partial\omega}{\partial T}I_N\end{array}\right) 
\end{align*}
and the other four sub-matrices $\vecJ^{nn}$, $\vecJ^{nb}$, $\vecJ^{bn}$ and $\vecJ^{bb}$ have dimension $4\times 4$, and are given by
\begin{align*}
\vecJ^{nn} & =\left(\begin{array}{cccc}-\lambda-\omega-\nu & -\nu & -q_A\beta S_N -\nu & -\beta S_N -\nu\\
\lambda & -\sigma-\omega & q_A\beta S_N & \beta S_N \\
0 & p_A\sigma &  -\gamma-\omega & 0 \\
0 & \left(1-p_A\right)\sigma & 0 & -\gamma-\omega\end{array}\right)
\\
\vecJ^{nb} & =\left(\begin{array}{cccc}\alpha & 0 & -q_A\beta S_N & -\beta S_N \\0 & \alpha & q_A\beta S_N & \beta S_N \\0 & 0 & \alpha & 0 \\0 & 0 & 0 & \alpha\end{array}\right) \\
\vecJ^{bn} & =\left(\begin{array}{cccc}\omega & 0 & -q_Aq_B\beta S_B & -q_B\beta S_B \\0 & \omega & q_Aq_B\beta S_B & q_B\beta S_B \\0 & 0 & \omega & 0 \\0 & 0 & 0 & \omega\end{array}\right) \\
\vecJ^{bb} & =\left(\begin{array}{cccc}-q_B\lambda-\alpha-\nu & -\nu & -q_Aq_B\beta S_B -\nu & -q_B\beta S_B -\nu\\ q_B\lambda & -\sigma-\alpha & q_Aq_B\beta S_B & q_B\beta S_B \\0 & p_A\sigma &  -\gamma-\alpha & 0 \\0 & \left(1-p_A\right)\left(1-p_T\right)\sigma & 0 & -\gamma-\alpha\end{array}\right)\, .
\end{align*}

At the infection-free steady state, we have $S_N=N_0$, $S_B=B_0$, with $B_0=1-N_0.$ It follows that 
$$\omega(B_0,0) = \omega_1B_0 + \omega_3, \quad \alpha(N_0)=\alpha_1 N_0+\alpha_2.$$ We also have 
$$\dfrac{\partial\omega}{\partial B} = \omega_1, \quad
\dfrac{\partial\omega}{\partial T} = \omega_2, \quad
\dfrac{\partial\alpha}{\partial N} = \alpha_1.$$

Substituting into the Jacobian blocks gives, we obtain
\begin{align*}
\vecJ^{n0} & =\left(\begin{array}{cc}\omega_1N_0+\alpha_1B_0+\nu & -q_T\beta N_0-\omega_2N_0 \\ 
0 & q_T\beta N_0  \\ 
0 & 0 \\ 
0 & 0\end{array}\right) \\
\vecJ^{b0} & =\left(\begin{array}{cc}-\omega_1N_0-\alpha_1B_0-\nu & -q_Bq_T\beta B_0+\omega_2N_0-\nu \\ 
0 & q_Bq_T\beta B_0  \\ 
0 & 0 \\ 
0 & 0\end{array}\right)\\
\vecJ^{nn} & =\left(\begin{array}{cccc}-\omega(B_0,0)-\nu & -\nu & -q_A\beta N_0 -\nu & -\beta N_0 -\nu\\ 0 & -\sigma-\omega(B_0,0) & q_A\beta N_0 & \beta N_0 \\0 & p_A\sigma &  -\gamma-\omega(B_0,0) & 0 \\0 & \left(1-p_A\right)\sigma & 0 & -\gamma-\omega(B_0,0)\end{array}\right)\\
\vecJ^{nb} & =\left(\begin{array}{cccc}\alpha(N_0) & 0 & -q_A\beta N_0 & -\beta N_0 \\0 & \alpha(N_0) & q_A\beta N_0 & \beta N_0 \\0 & 0 & \alpha(N_0) & 0 \\0 & 0 & 0 & \alpha(N_0)\end{array}\right)\\
\vecJ^{bn} & =\left(\begin{array}{cccc}\omega(B_0,0) & 0 & -q_Aq_B\beta B_0 & -q_B\beta B_0 \\0 & \omega(B_0,0) & q_Aq_B\beta B_0 & q_B\beta B_0 \\0 & 0 & \omega(B_0,0) & 0 \\0 & 0 & 0 & \omega(B_0,0)\end{array}\right)\\
\text{and}\quad
\vecJ^{bb}& =\left(\begin{array}{cccc}-\alpha(N_0)-\nu & -\nu & -q_Aq_B\beta B_0 -\nu & -q_B\beta B_0 -\nu\\ 0 & -\sigma-\alpha(N_0) & q_Aq_B\beta B_0 & q_B\beta B_0 \\0 & p_A\sigma &  -\gamma-\alpha(N_0) & 0 \\0 & \left(1-p_A\right)\left(1-p_T\right)\sigma & 0 & -\gamma-\alpha(N_0)\end{array}\right).
\end{align*}

The infection-free steady state where $S_N=N_0$ and $S_B=B_0$ follows
standard results for disease dynamics without a behavioural component and so is stable when $\Ro<1$.

\section{Derivation of the endemic steady states}\label{append:steady_states}
Here we present details on derivations underlying Proposition~\ref{thm:steady_states}. When infection is present and ignoring higher order terms, the dynamics of behaviour are governed by
\begin{align}
	 \frac{\de N}{\de t} & = -\omega(B,T) N + \alpha(N) (B - T) \nonumber\\
     & = (\omega_1 - \alpha_1)N^2 - (\omega_2 T + \omega_1 + \omega_3 + \alpha_2 - \alpha_1(1-T))N + \alpha_2(1-T) .\label{eqn:N_ode}
\end{align}
At equilibrium, $N^*$ satisfies $G(N^*)=0$ (Equation~\eqref{eqn:G_N}). Since $G(0) = \alpha_2(1-T^*) > 0$ and $G(1) = -[(\omega_2+\alpha_1+\alpha_2)T^* + \omega_3] < 0$, there exists a unique solution $N^* \in (0,1)$  depending on $T^*$. The behaviour prevalence is then $B^*=1-N^*$.

For $X\in\{S,E,A,I,R\}$ write $X=X_N+X_B$ and let the total symptomatic be $O=I+T$. Summing Equations \eqref{eqn:In}, \eqref{eqn:Ib} and \eqref{eqn:T} together gives
\begin{align*}
\frac{\de O}{\de t} = (1-p_{A})\sigma(E_N + E_B) - \gamma O.
\end{align*}
Similarly, adding Equations \eqref{eqn:An} and \eqref{eqn:Ab}, Equations \eqref{eqn:Rn} and \eqref{eqn:Rb} gives 
\begin{align*}
\frac{\de A}{\de t} & = p_{A}\sigma E - \gamma A \\
\frac{\de R}{\de t} & =  \gamma(I+A+T) - \nu R.
\end{align*}

Hence, fixing $O^* = I^*+T^*$, setting the equations above to zero yields
\begin{align}
\label{eqn:Estar}
	E^* & = E_N^* + E_B^*  = \frac{\gamma}{(1-p_A)\sigma} O^*\\
\label{eqn:Astar}
	A^* & = A_N^* + A_B^*  = \frac{p_A}{(1-p_A)}O^*\\
\label{eqn:Rstar}
	R^* & = R_N^* + R_B^* = \frac{\gamma}{\nu(1-p_A)} O^* \\
\label{eqn:Sstar}
    S^* & = S_N^* + S_B^*   = 1 - (E^* + A^* + O^*  + R^*).
\end{align}

Further, knowing $T^*$ (and hence $N^*$ and $B^* = 1 - N^*$), the equilibrium rates $\lambda^*, \omega^*$ and $\alpha^*$ are in Equations \eqref{eqn:lambdastar}-\eqref{eqn:alphastar}. 

Next, we derive the behavioural steady states $S_B^*, E_B^*, A_B^*, I_B^*$ and $R_B^*$.

Setting Equations \eqref{eqn:En} and \eqref{eqn:Eb} to zero gives
\begin{align}
	\label{eqn:SN1}
	\lambda^* S_N^* & = (\omega^* + \sigma) E^* - (\alpha^* + \omega^* + \sigma) E_B^* \\
	\label{eqn:SB1}
	\lambda^* S_B^* & = \frac{-\omega^* E^* + (\alpha^* + \omega^* + \sigma)E_B^*}{q_B} .
\end{align}
Adding Equations \eqref{eqn:SN1} and \eqref{eqn:SB1} gives
\begin{equation*}
	%\label{eqn:Ebstar}
	E_B^* = \frac{q_B \lambda^* S^* + (\omega^* - q_B(\omega^* + \sigma))E^*}{(1-q_B)(\alpha^* + \omega^* + \sigma)} .
\end{equation*}
Setting Equations \eqref{eqn:Ab} and \eqref{eqn:Ib} to zero respectively gives
\begin{align*}
	%\label{eqn:Abstar}
	A_B^* &= \frac{p_A \sigma E_B^* + \omega^* A^*}{\gamma + \alpha^* + \omega^*}\\
	%\label{eqn:Ibstar}
	I_B^* &= \frac{(1-p_A)(1-p_T) \sigma E_B^* + \omega^* I^*}{\gamma + \alpha^* + \omega^*} .
\end{align*}
Further, setting Equation \eqref{eqn:Rb} to zero gives
\begin{equation*}
	%\label{eqn:Rbstar}
	R_B^* = \frac{\gamma(A_B^* + I_B^* + T^*) + \omega^* R^*}{\alpha^* + \omega^* + \nu}\,.
\end{equation*}
It follows that
\begin{equation*}%\label{eqn:Sbstar}
     S_B^* = B^* - (E_B^* + A_B^* + I_B^* + T^* + R_B^*).
\end{equation*}
Given $T^*$ and $O^*$, the non-behaver states $S_N^*, E_N^*, A_N^*, I_N^*$ and $R_N^*$ are obtained by subtraction:
\begin{align*}
	S_N^* & = S^* - S_B^* \\
	E_N^* & = E^* - E_B^* \\
	A_N^* & = A^* - A_B^* \\
	R_N^* & = R^* - R_B^*\\
    I_N^* & = N^* - (S_N^* + E_N^* + A_N^* + R_N^*) .
\end{align*}

Finally, setting Equations \eqref{eqn:Sn} and \eqref{eqn:T} to zero gives the simultaneous Equations \eqref{eqn:constraint_1} and \eqref{eqn:constraint_2} which fixes $I^*$ and $T^*$ (and hence $O^*$).

\end{appendices}

\clearpage

\vspace*{2cm}
\begin{center}
\textsc{\LARGE Supplementary Figures}
\end{center}
\vspace*{3cm}

\renewcommand{\figurename}{Supplementary Figure}

\begin{figure}[htbp]
    \centering
    \begin{subfigure}[b]{0.45\textwidth}
    \caption{}
    \includegraphics[width=\textwidth]{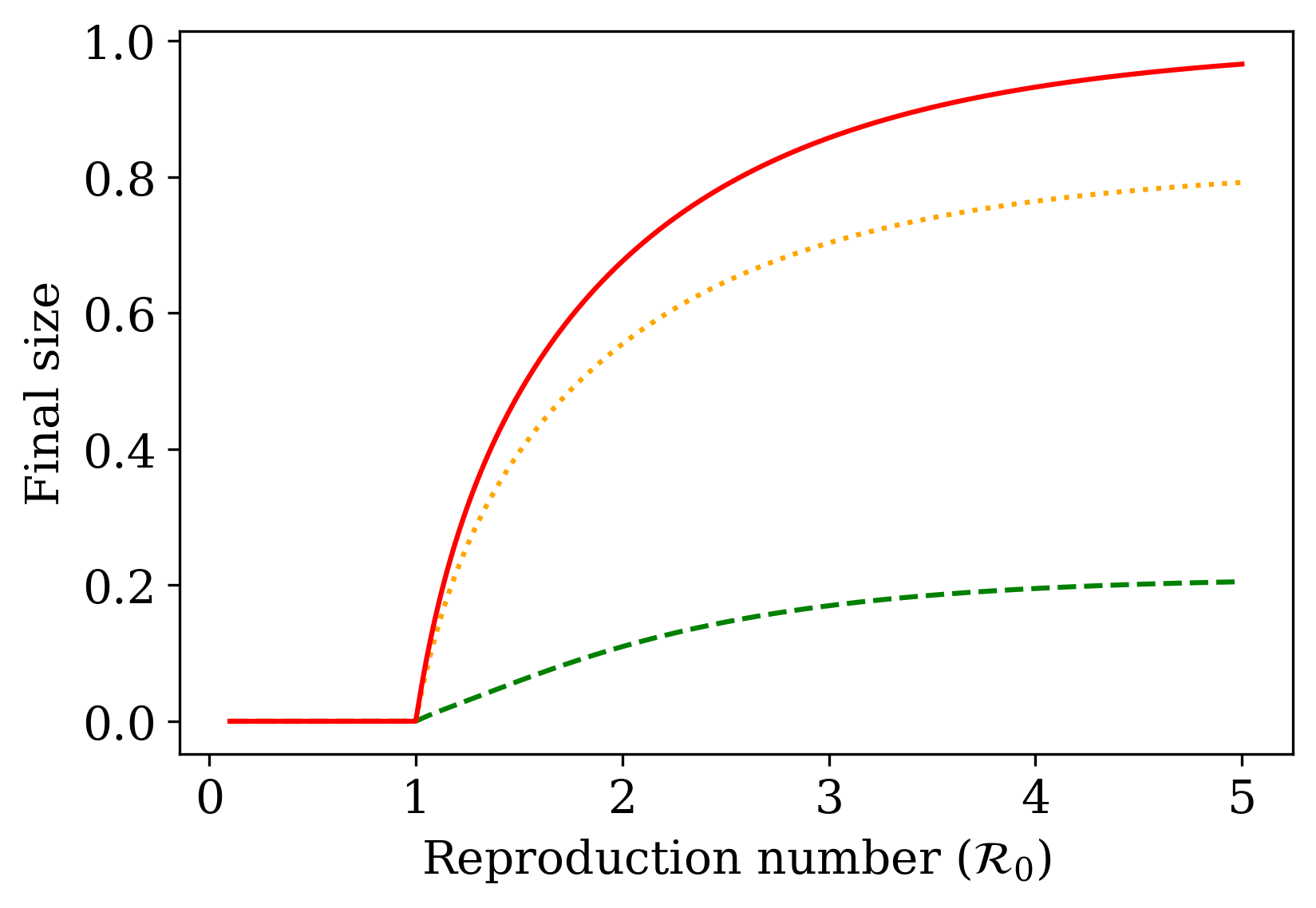}
    \end{subfigure}
    \hfill
    \begin{subfigure}[b]{0.45\textwidth}
    \caption{}
    \includegraphics[width=\textwidth]{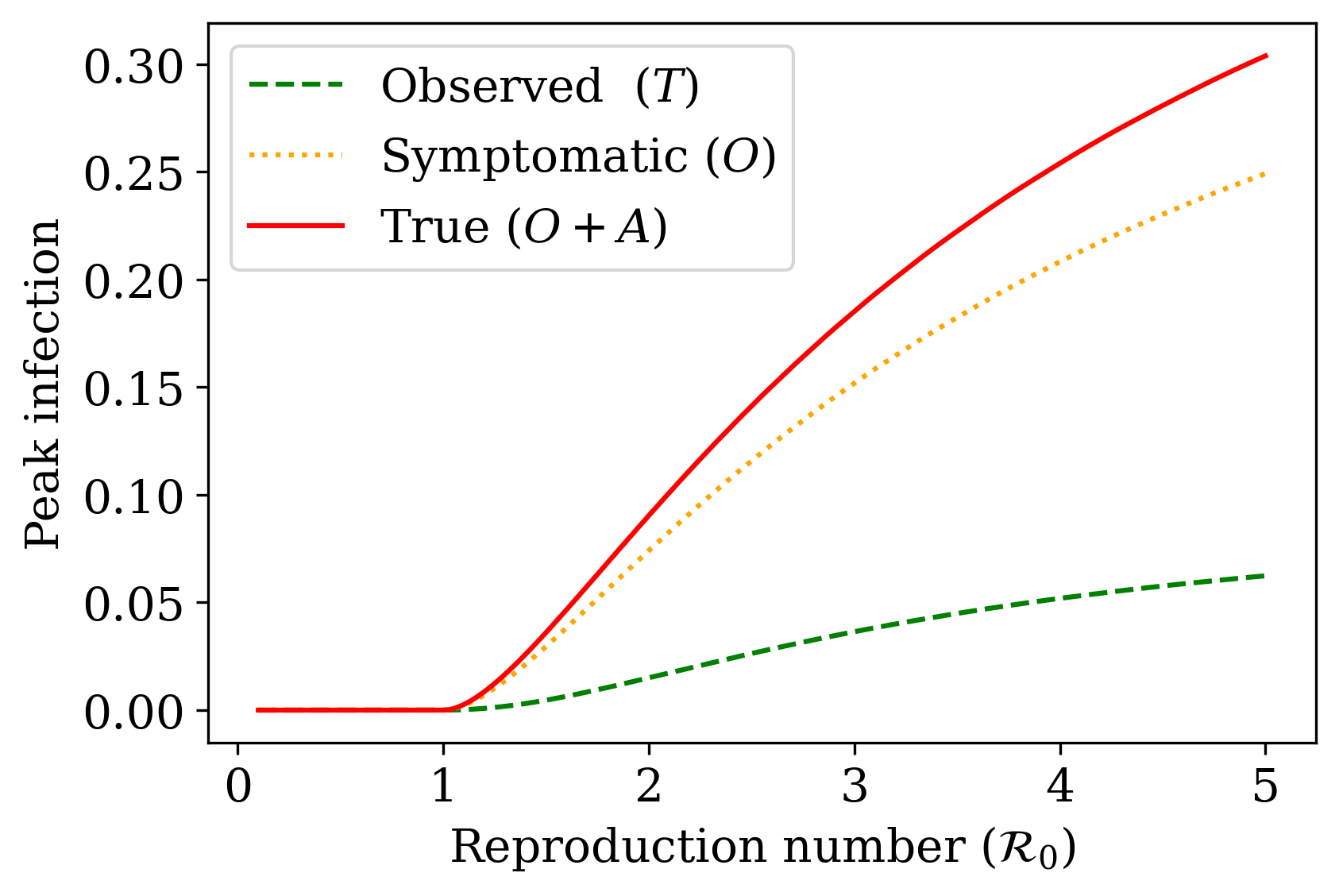}
    \end{subfigure}
    \caption{\textbf{Change in (a) final size and (b) peaks for different ${\Ro}$ values.}  The dashed, green line represented the observed epidemic ($T$), the dotted, orange line represents the symptomatic epidemic ($O = I+T$), and the solid, red line represents the total epidemic ($O+A$).
    }
    \label{fig:peak_fs_by_R0}
\end{figure}

\begin{figure}[htbp]
    \centering
    \begin{subfigure}[b]{0.45\textwidth}
    \caption{}
    \includegraphics[width=\textwidth]{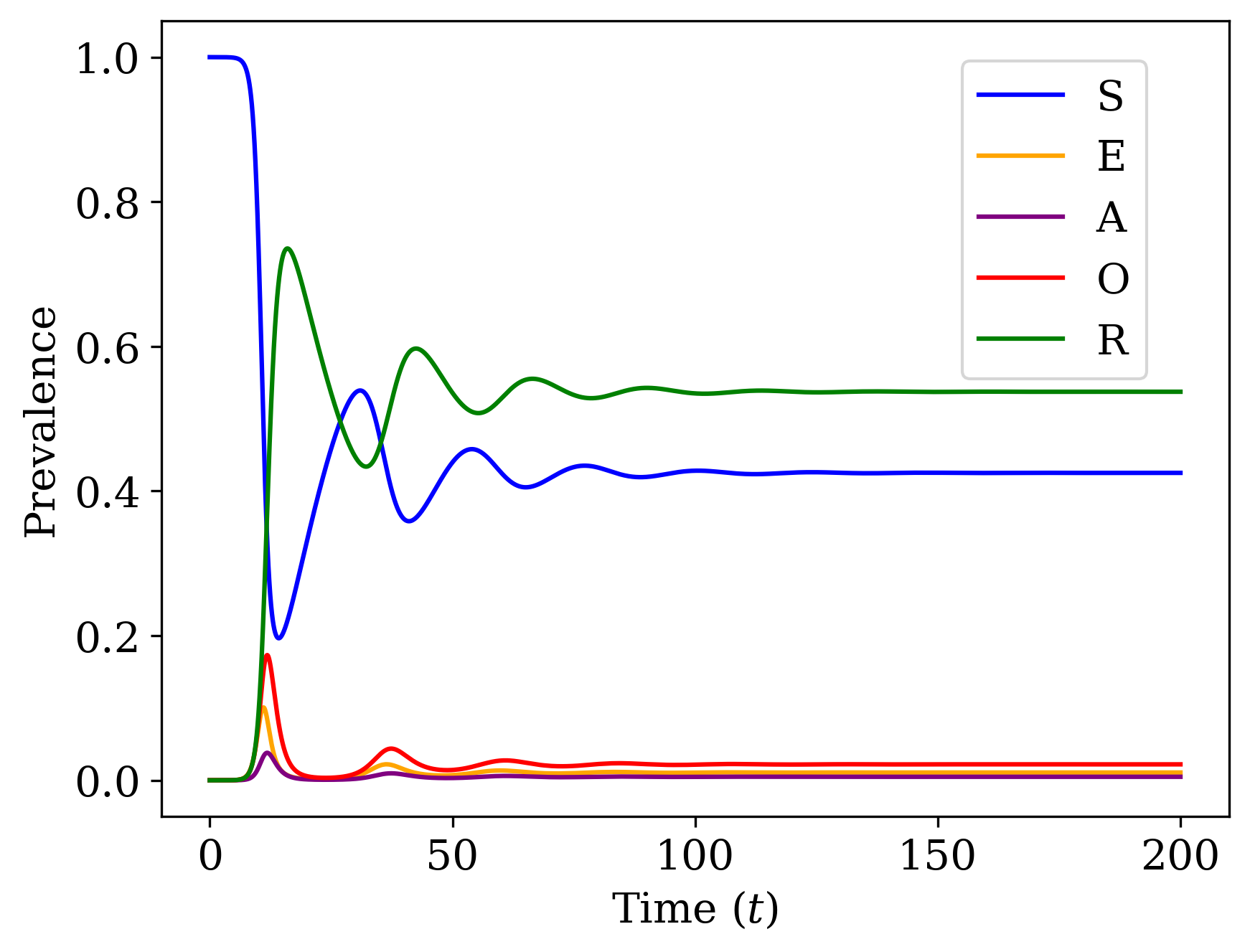}
    \end{subfigure}
    \hfill
    \begin{subfigure}[b]{0.45\textwidth}
    \caption{}
    \includegraphics[width=\textwidth]{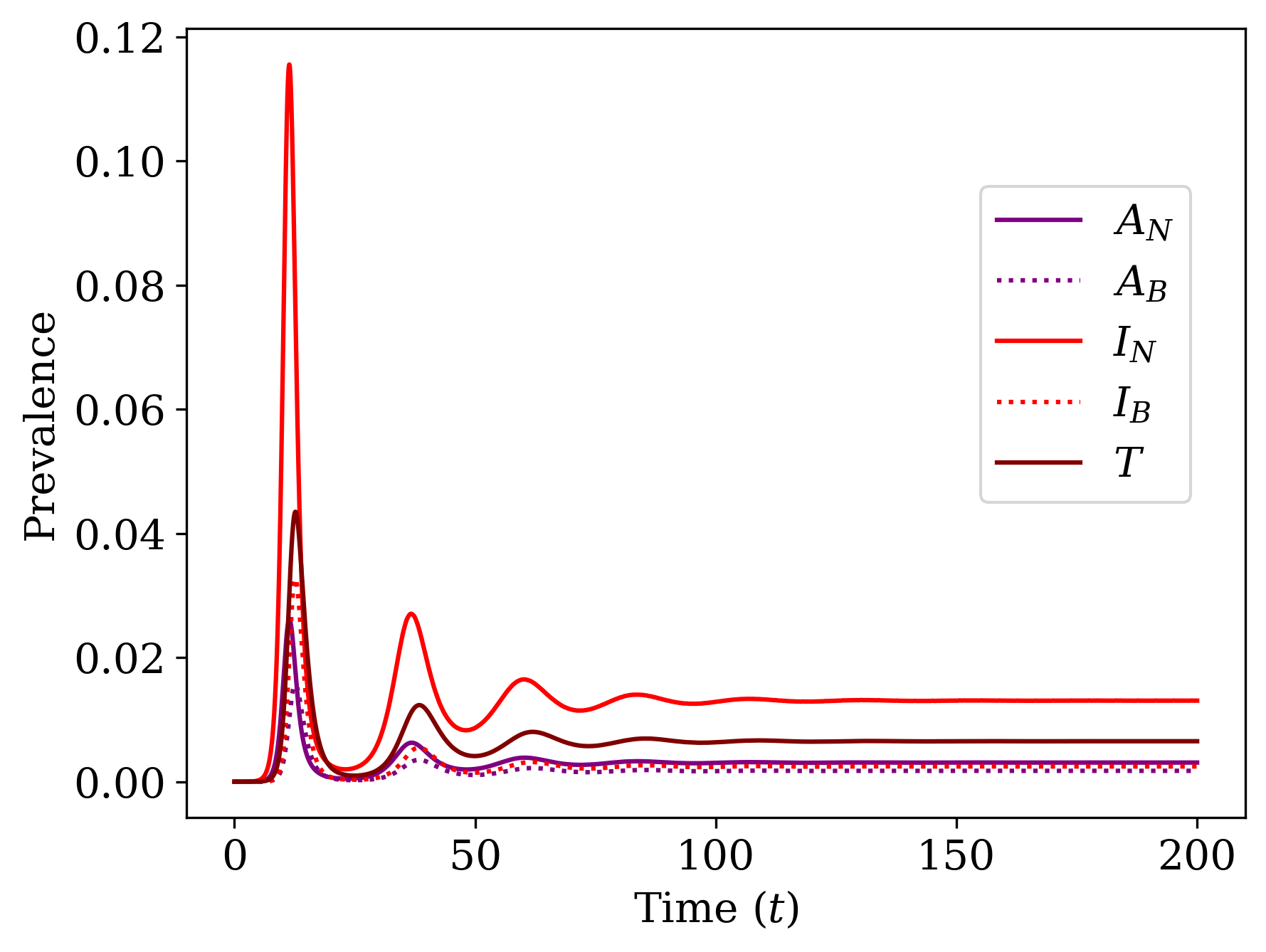}
    \end{subfigure}
    \hfill
    \begin{subfigure}[b]{0.45\textwidth}
    \caption{}
    \includegraphics[width=\textwidth]{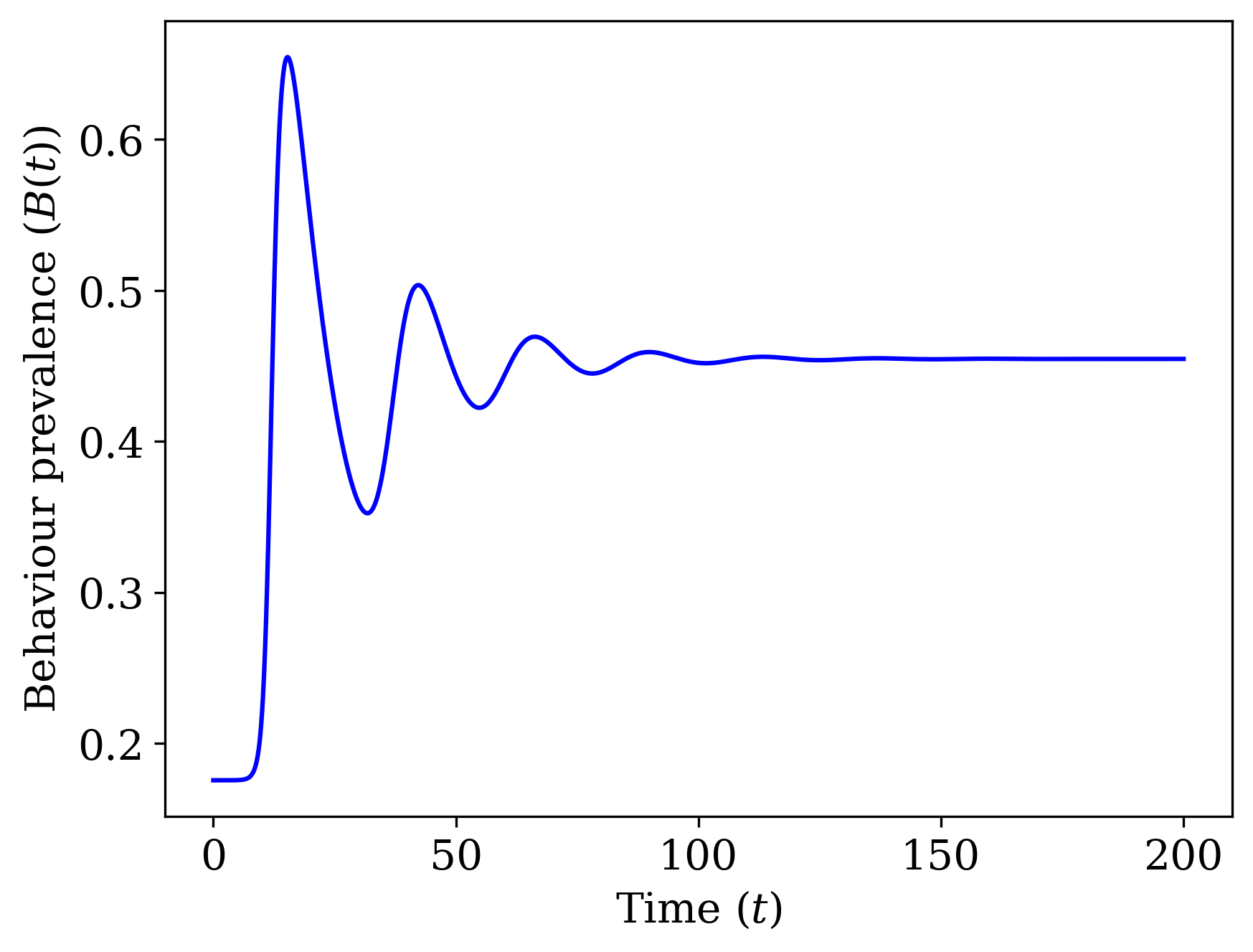}
    \end{subfigure}
    \hfill
    \begin{subfigure}[b]{0.45\textwidth}
    \caption{}
    \includegraphics[width=\textwidth]{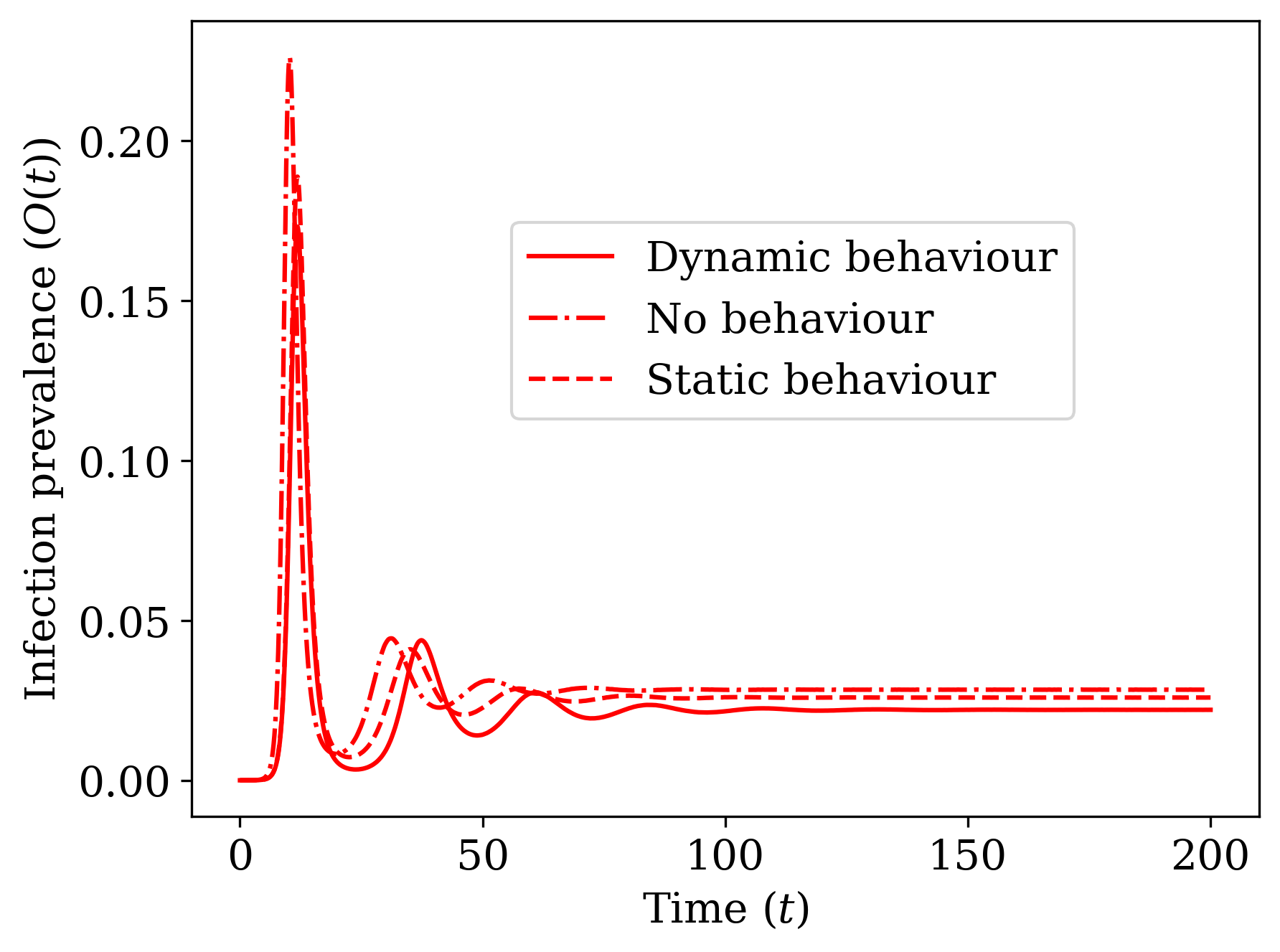}
    \end{subfigure}
    \caption{\textbf{Transient dynamics of the behaviour and disease model for testing and isolation with baseline parameters from Table 1 in the main text.}  (a) The transient dynamics for each epidemiological compartment. (b) The transient dynamics for each of the five infectious classes, where dotted lines present behaviours.  (c) The transient dynamics of behaviours ($B(t)$).  (d) A comparison of the symptomatic infection prevalence $O(t)$ across: 1) the baseline parameters with dynamic behaviour, 2) A model with no behaviour at all, and 3) a model where behaviour is held static throughout the whole simulation.
    }
    \label{fig:transient}
\end{figure}

\begin{figure}[htbp]
    \centering
    \begin{subfigure}[b]{0.3\textwidth}
    \caption{}
    \includegraphics[width=\textwidth]{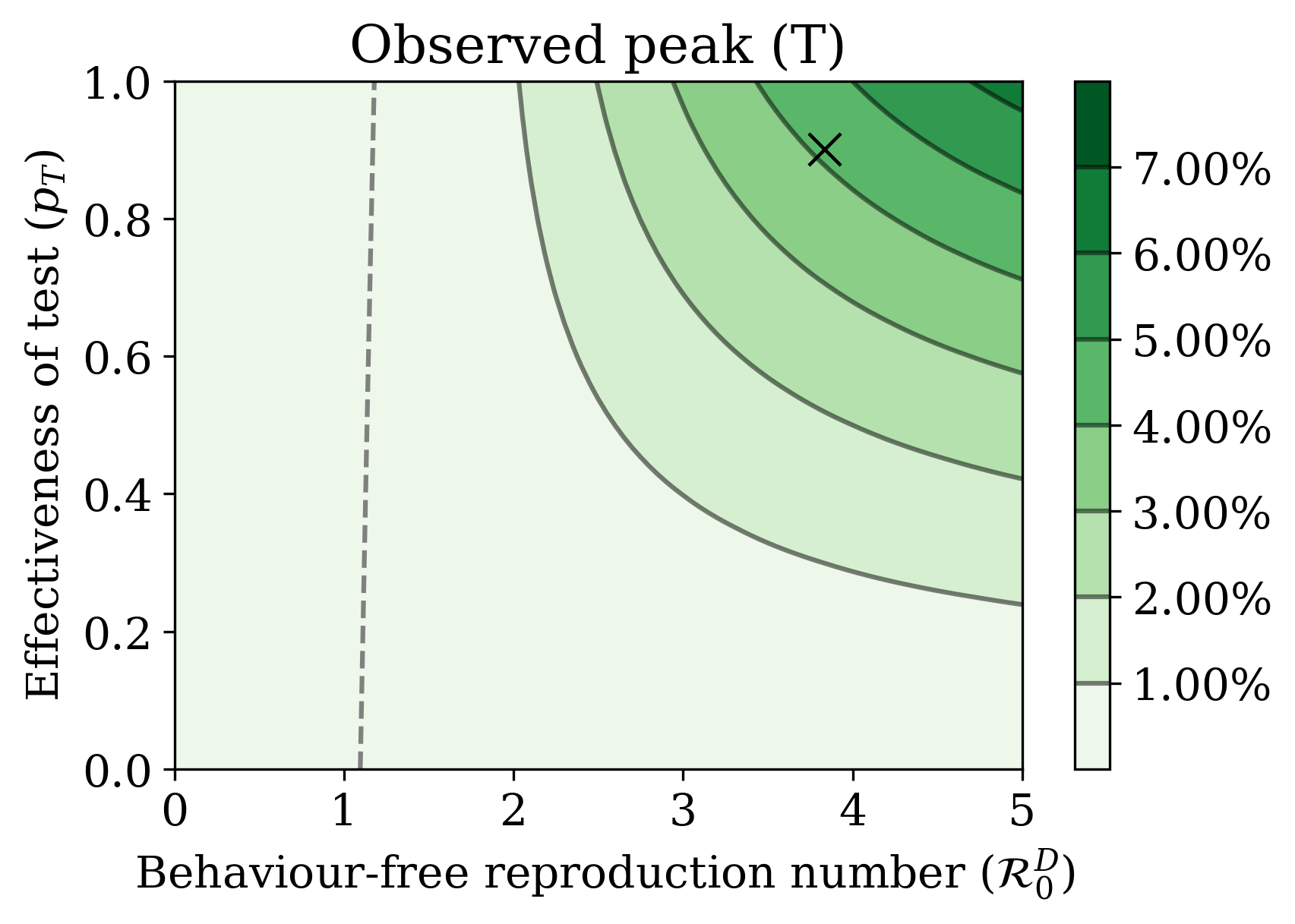}
    \end{subfigure}
    \hfill
    \begin{subfigure}[b]{0.3\textwidth}
    \caption{}
    \includegraphics[width=\textwidth]{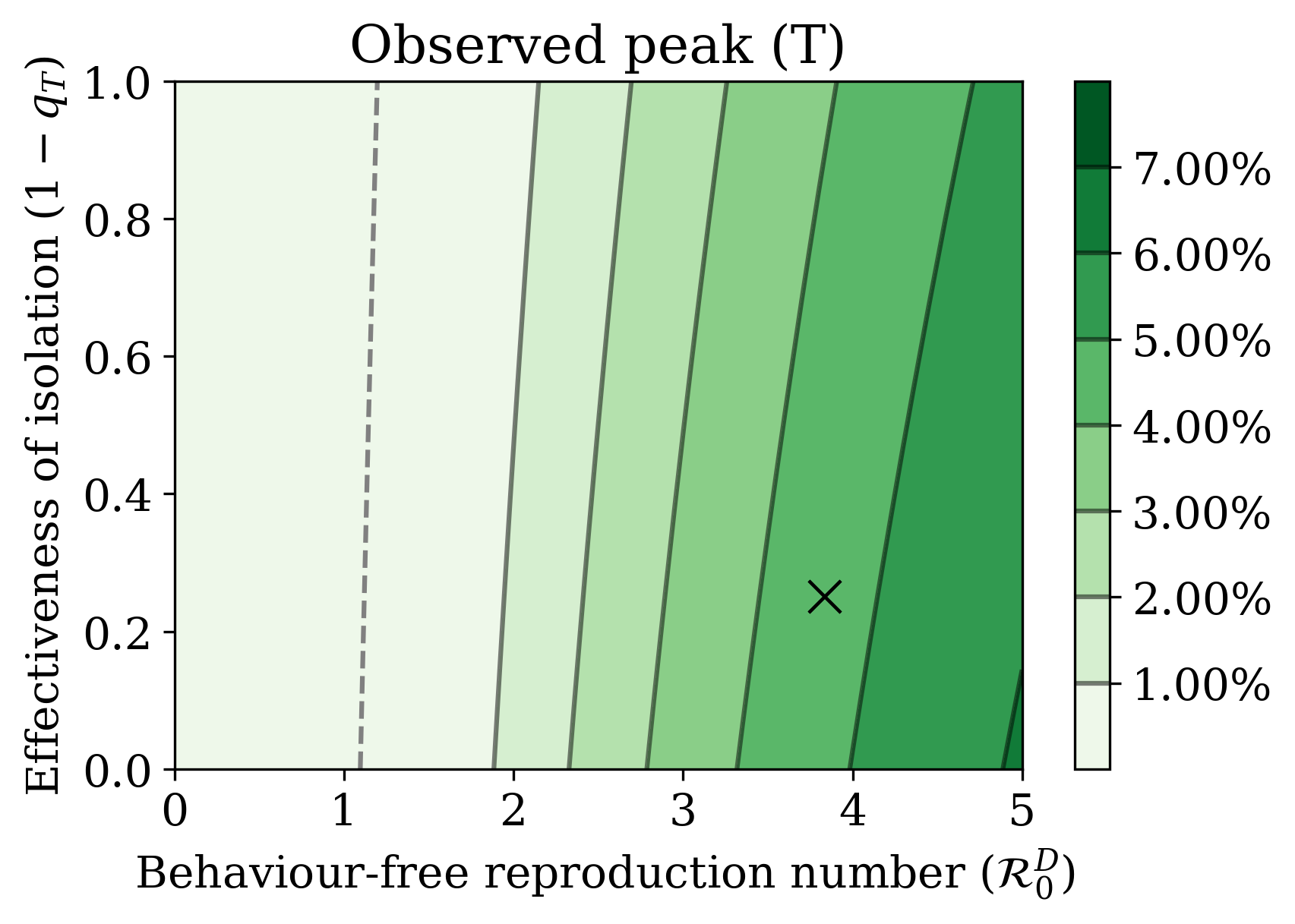}
    \end{subfigure}
    \hfill
    \begin{subfigure}[b]{0.3\textwidth}
    \caption{}
    \includegraphics[width=\textwidth]{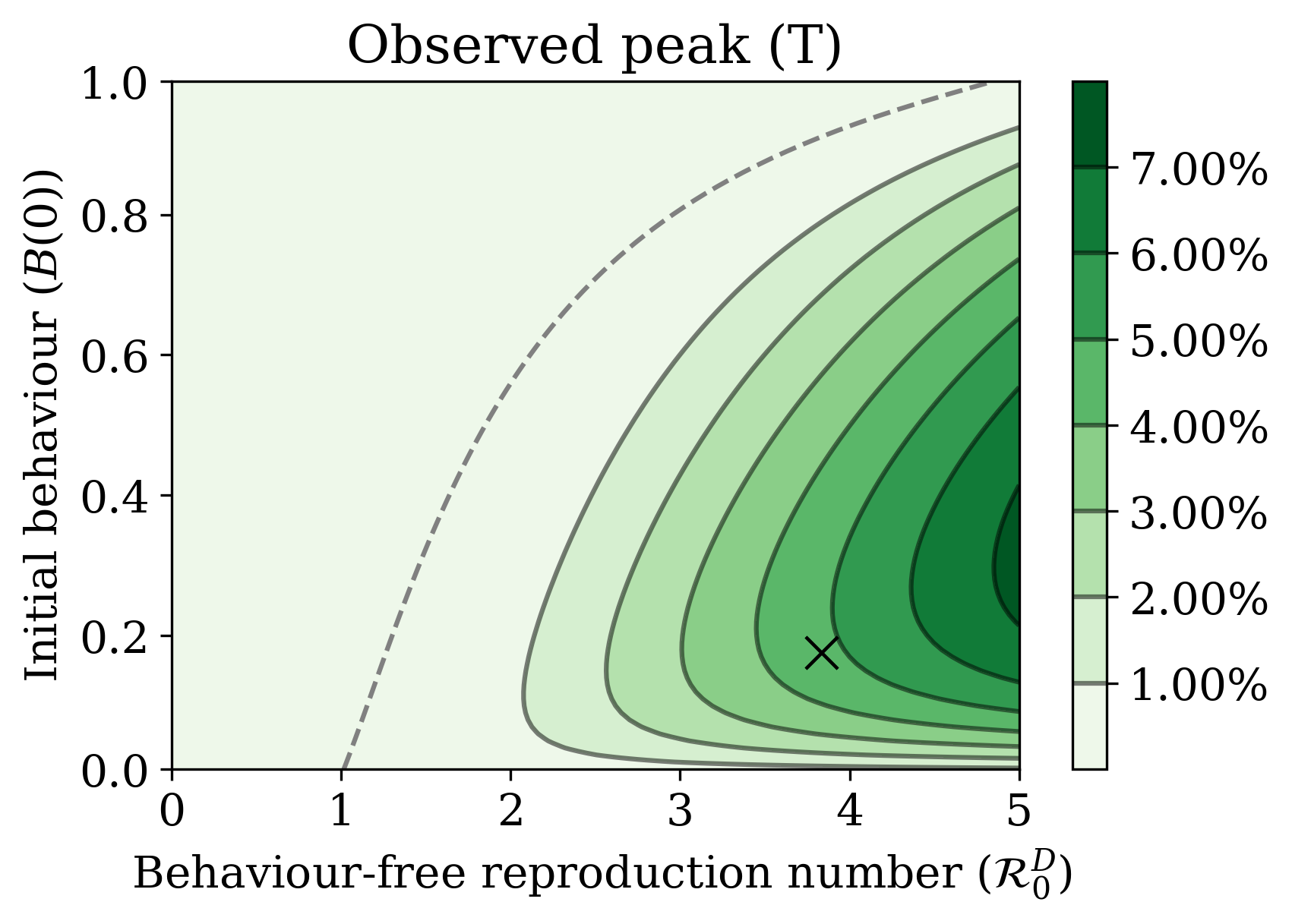}
    \end{subfigure}
    \hfill

    \begin{subfigure}[b]{0.3\textwidth}
    \caption{}
    \includegraphics[width=\textwidth]{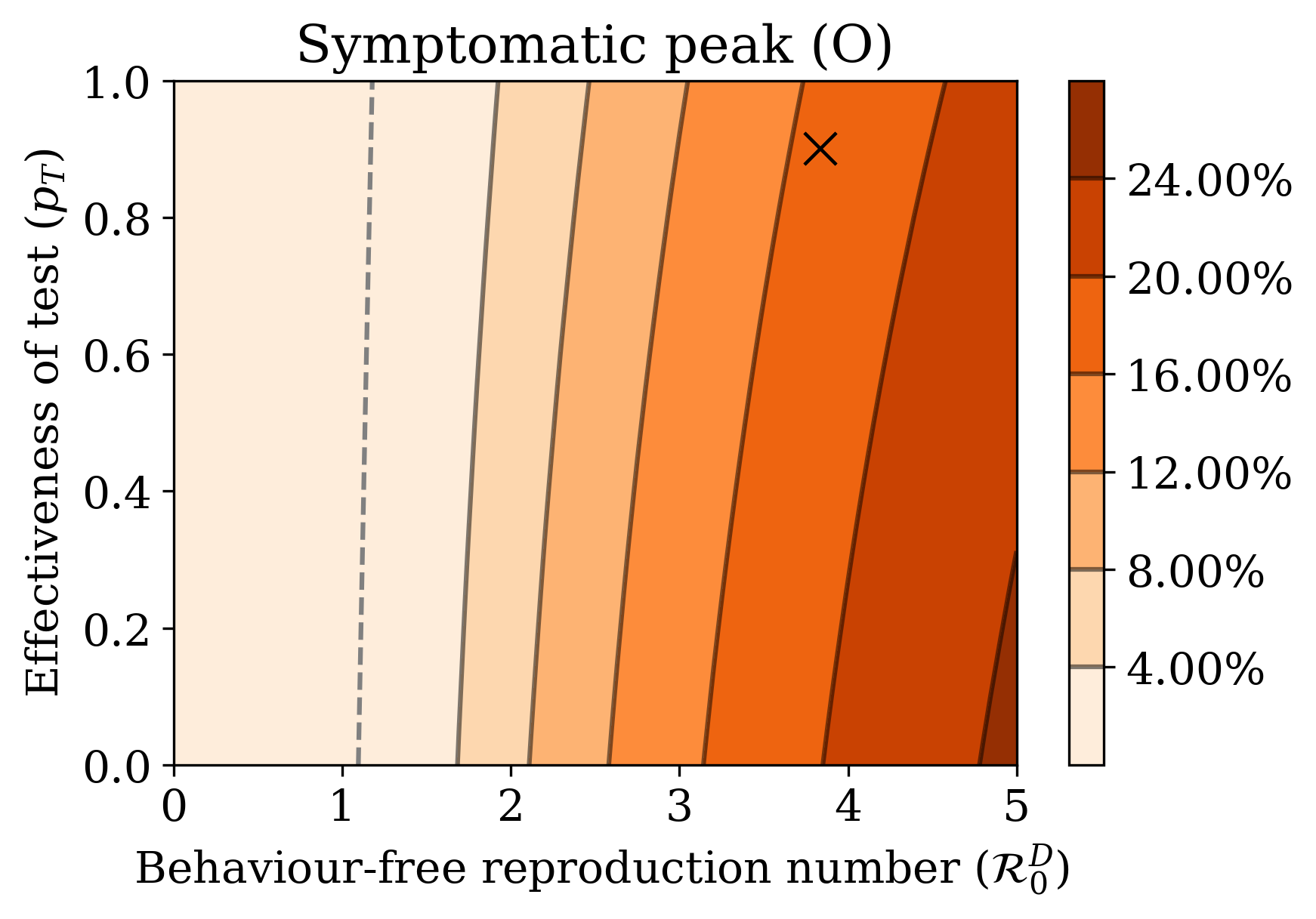}
    \end{subfigure}
    \hfill
    \begin{subfigure}[b]{0.3\textwidth}
    \caption{}
    \includegraphics[width=\textwidth]{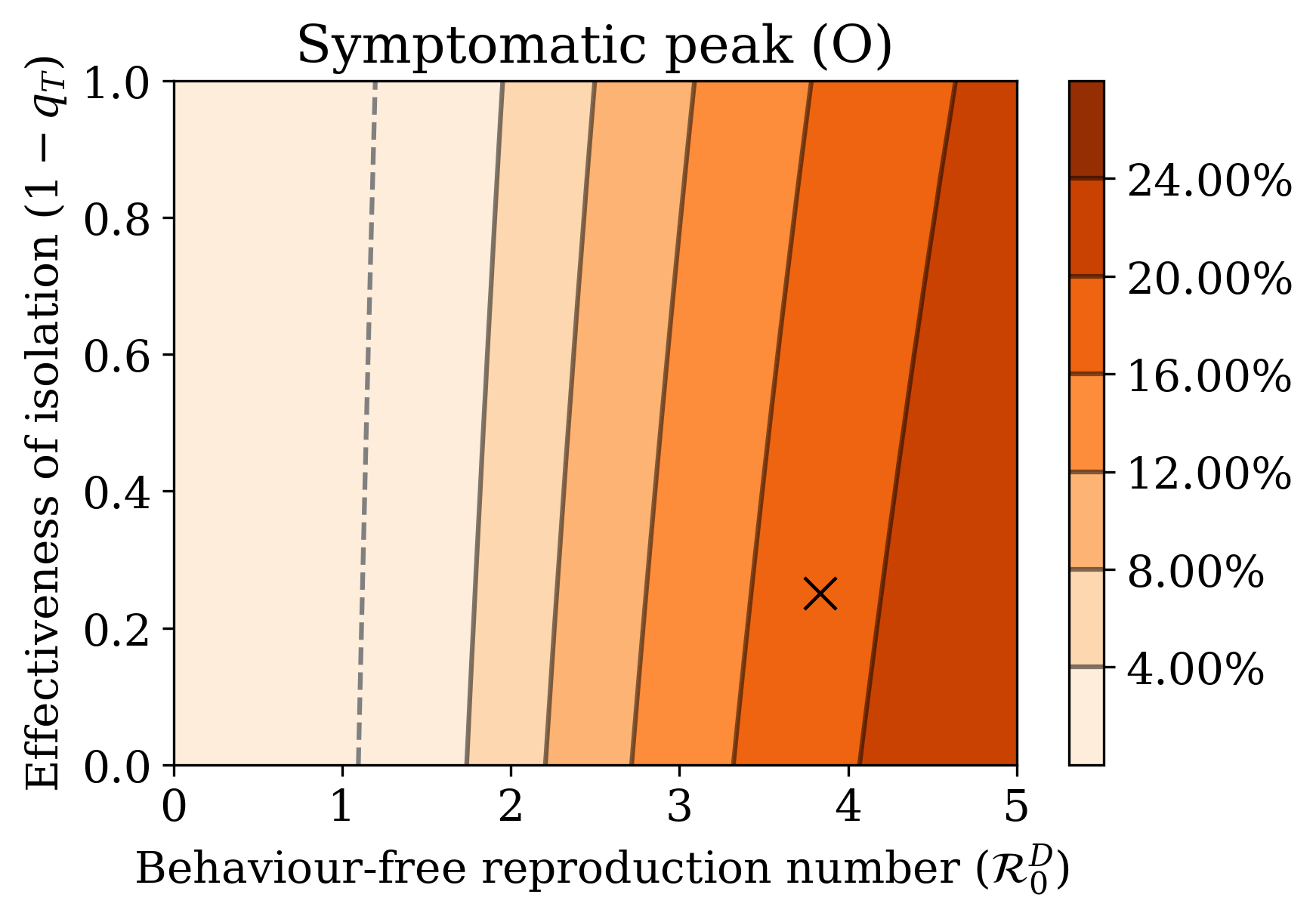}
    \end{subfigure}
    \hfill
    \begin{subfigure}[b]{0.3\textwidth}
    \caption{}
    \includegraphics[width=\textwidth]{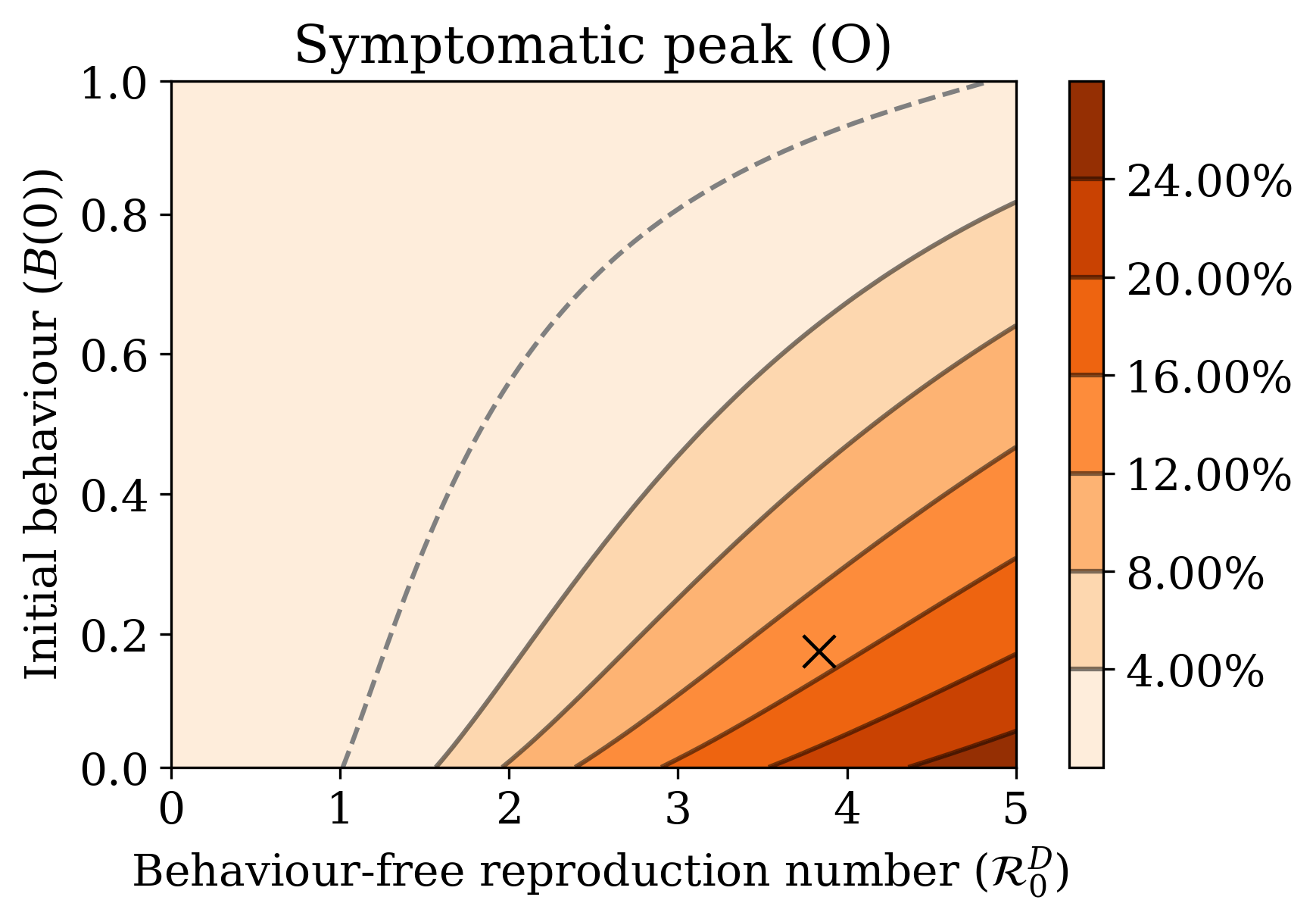}
    \end{subfigure}
    \hfill

    \begin{subfigure}[b]{0.3\textwidth}
    \caption{}
    \includegraphics[width=\textwidth]{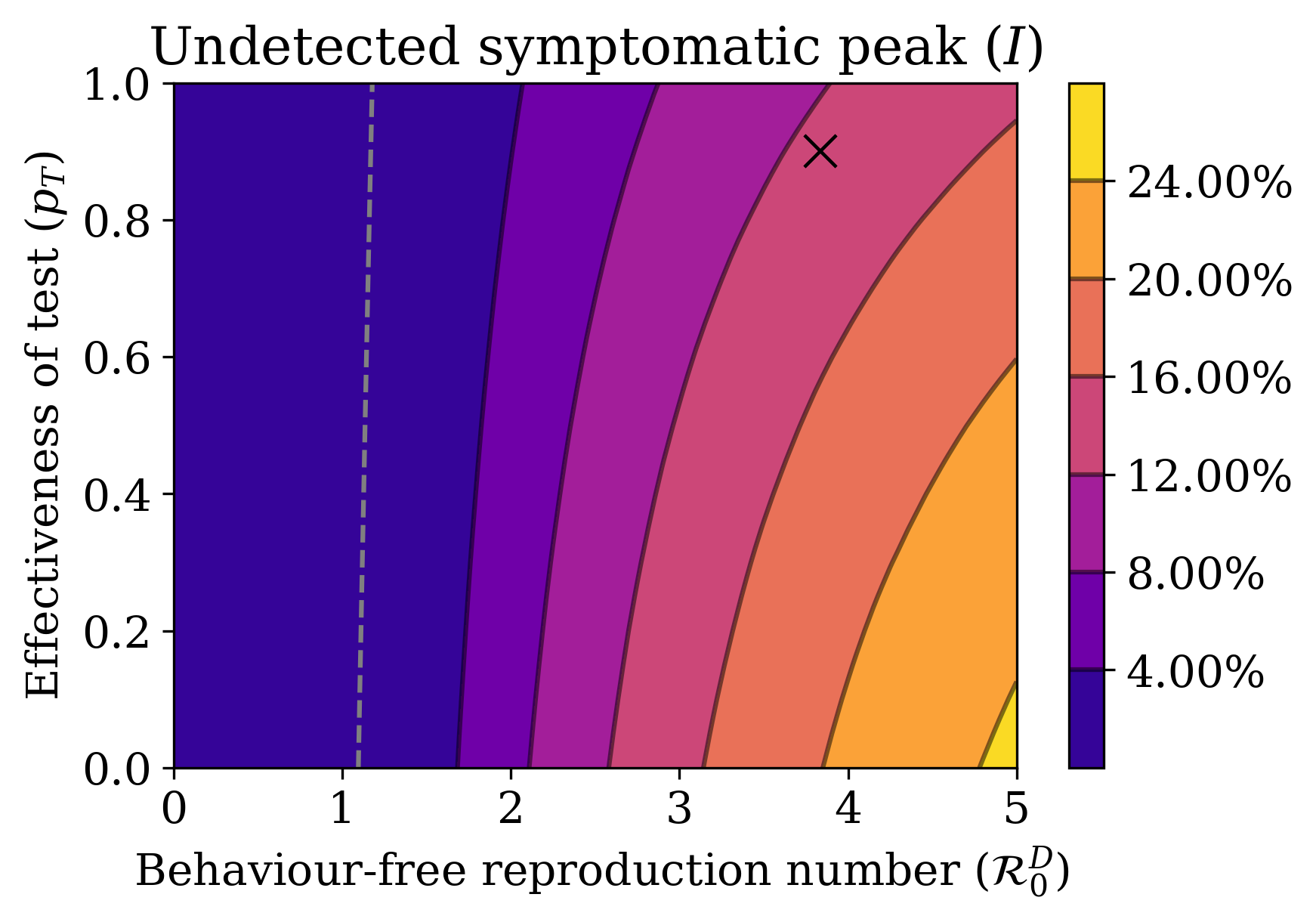}
    \end{subfigure}
    \hfill
    \begin{subfigure}[b]{0.3\textwidth}
    \caption{}
    \includegraphics[width=\textwidth]{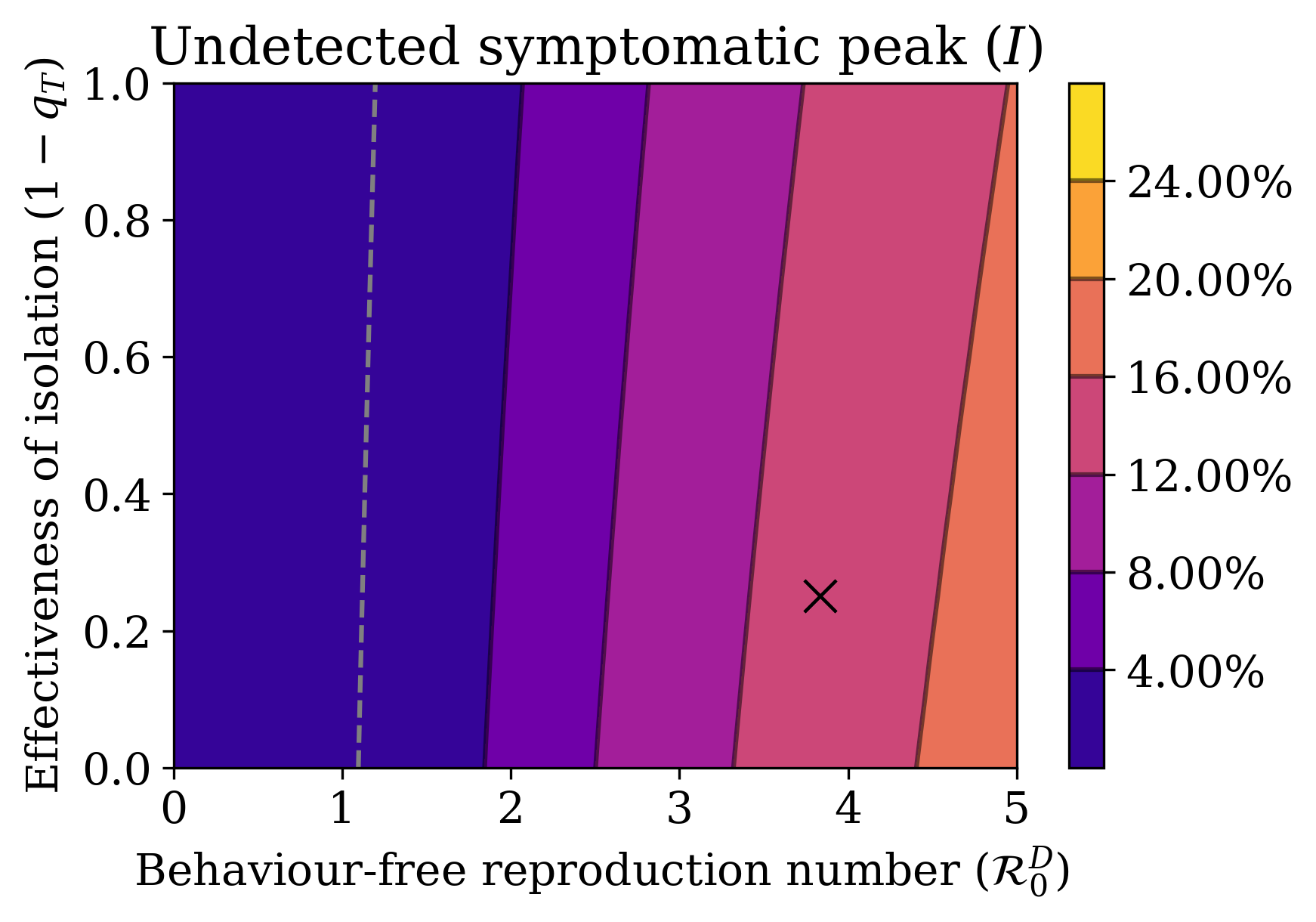}
    \end{subfigure}
    \hfill
    \begin{subfigure}[b]{0.3\textwidth}
    \caption{}
    \includegraphics[width=\textwidth]{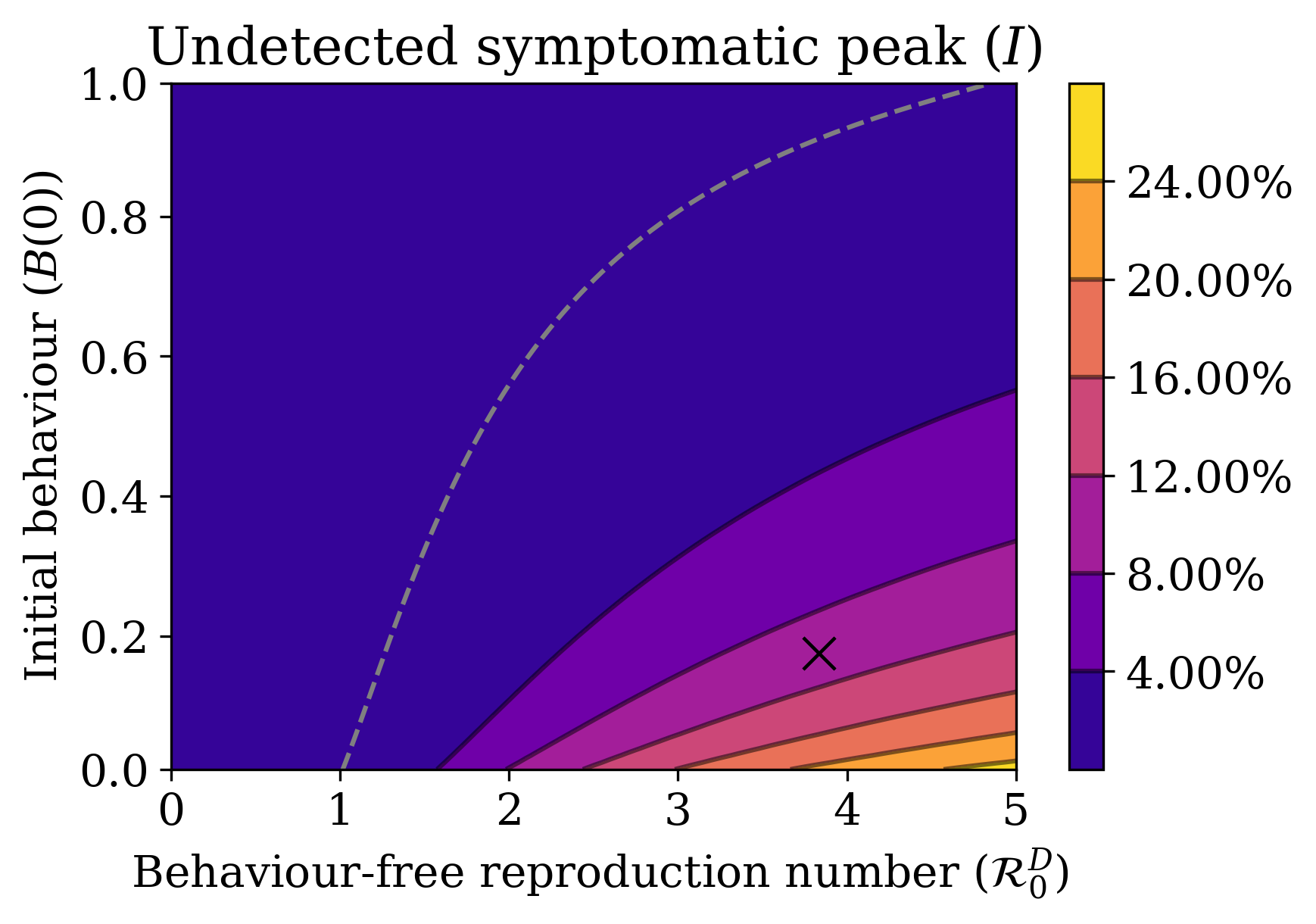}
    \end{subfigure}

    \caption{\textbf{Impact of test effectiveness ($p_T$), isolation effectiveness ($1-q_T$), initial condition of behaviour (${B(0)}$) and the behaviour-free reproduction number (${\Ro^{D}}$) on observed symptomatic (${T}$), undetected symptomatic (${I}$), and total symptomatic (${O=I+T}$) peaks. } The colour bars represent the peak infection prevalence as a percentage of the population and are consistent across each row.  Note, we vary the behaviour-free reproduction number instead of $\Ro$ due to the dependence of $\Ro$ on the vertical axes.  The dashed grey lines show $\Ro=1$ and the black cross indicates the baseline parameter values from Table 1 in the manuscript.  
    }
    \label{fig:sweep_peak}
\end{figure}

\begin{figure}[htbp]
    \centering
    \begin{subfigure}[b]{0.32\textwidth}
    \caption{}
    \includegraphics[width=\textwidth]{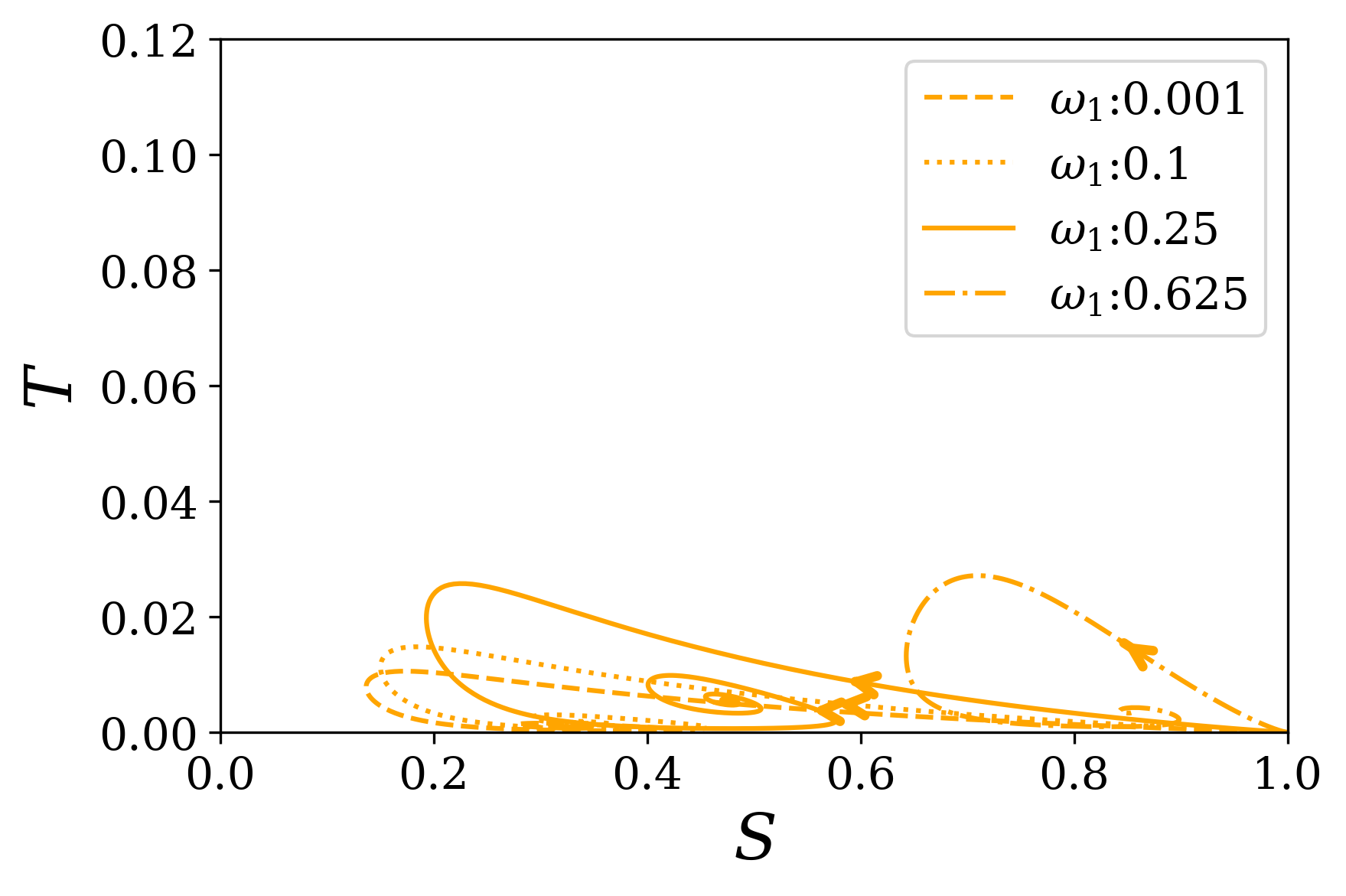}
    \end{subfigure}
    \hfill
    \begin{subfigure}[b]{0.32\textwidth}
    \caption{}
    \includegraphics[width=\textwidth]{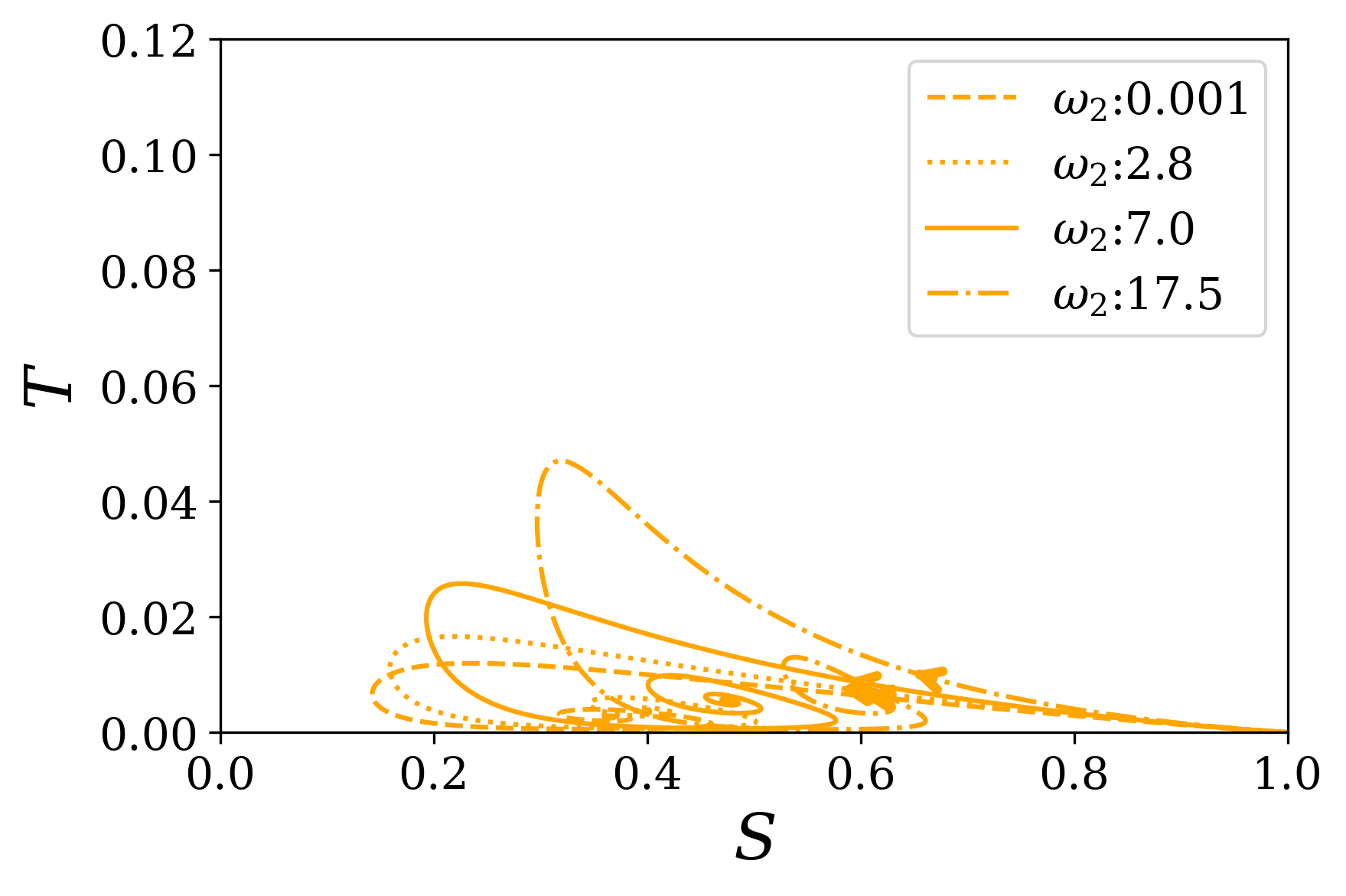}
    \end{subfigure}
    \hfill
    \begin{subfigure}[b]{0.32\textwidth}
    \caption{}
    \includegraphics[width=\textwidth]{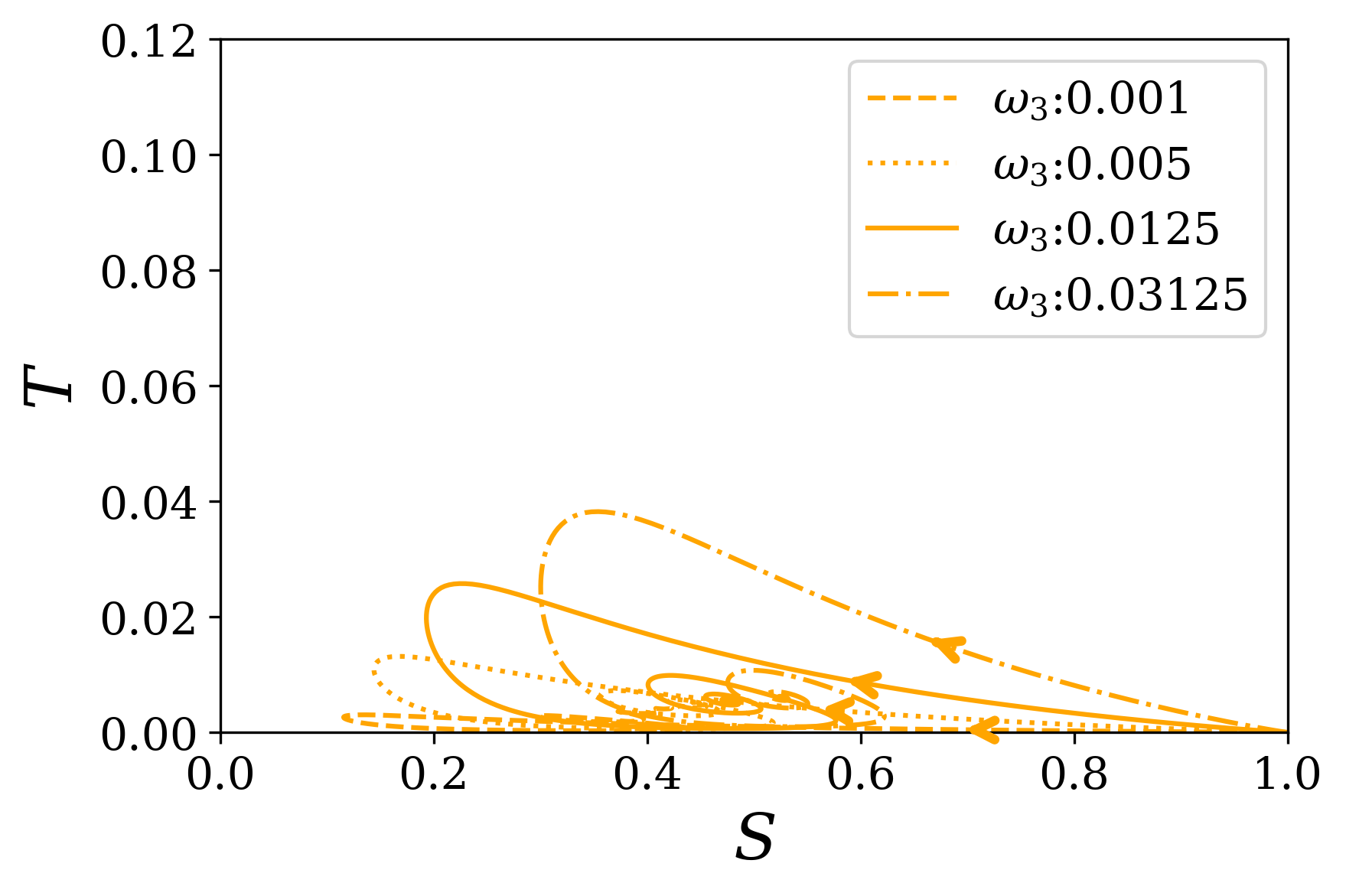}
    \end{subfigure}
    \hfill
    
    \begin{subfigure}[b]{0.32\textwidth}
    \caption{}
    \includegraphics[width=\textwidth]{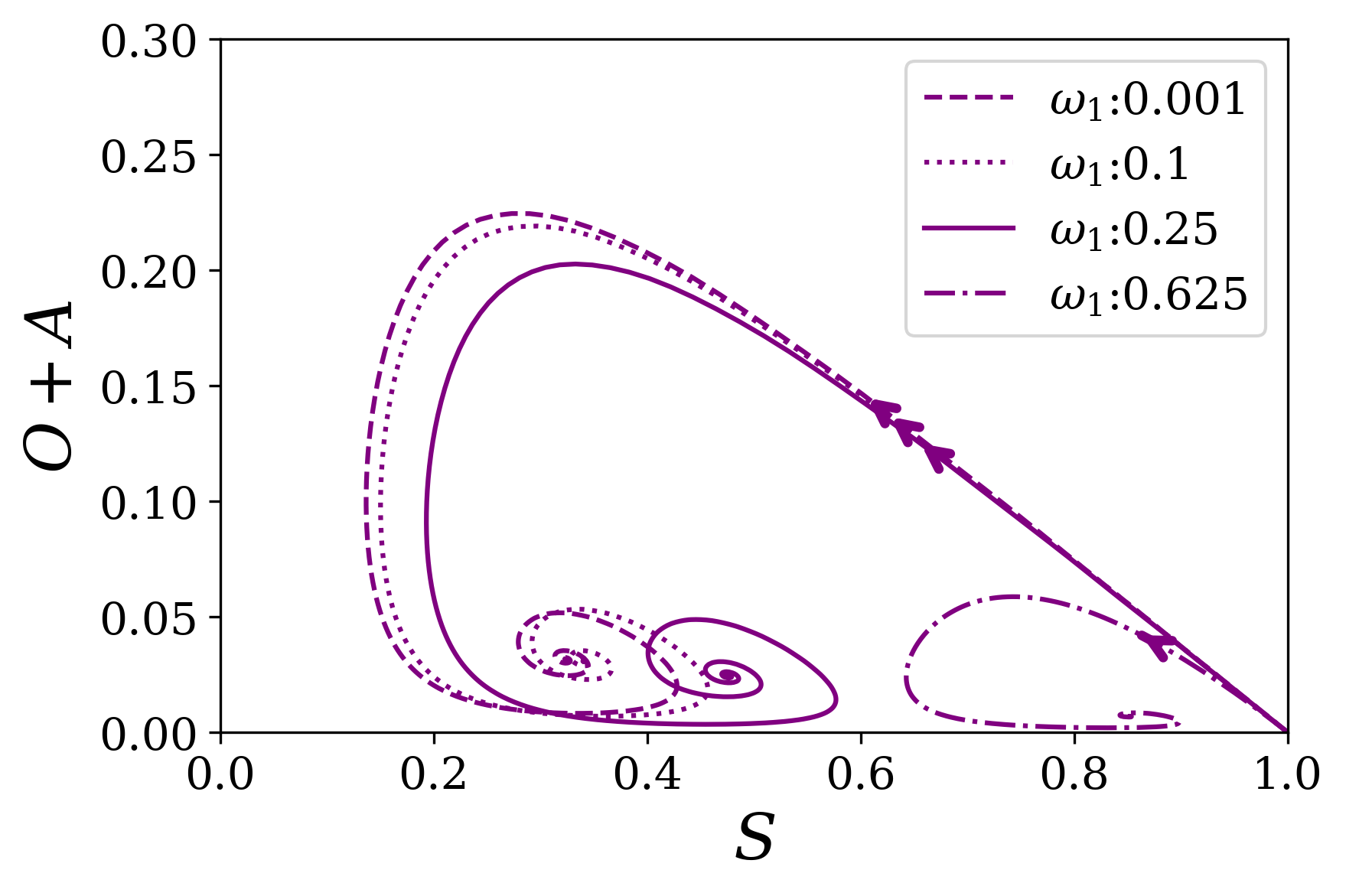}
    \end{subfigure}
    \hfill
    \begin{subfigure}[b]{0.32\textwidth}
    \caption{}
    \includegraphics[width=\textwidth]{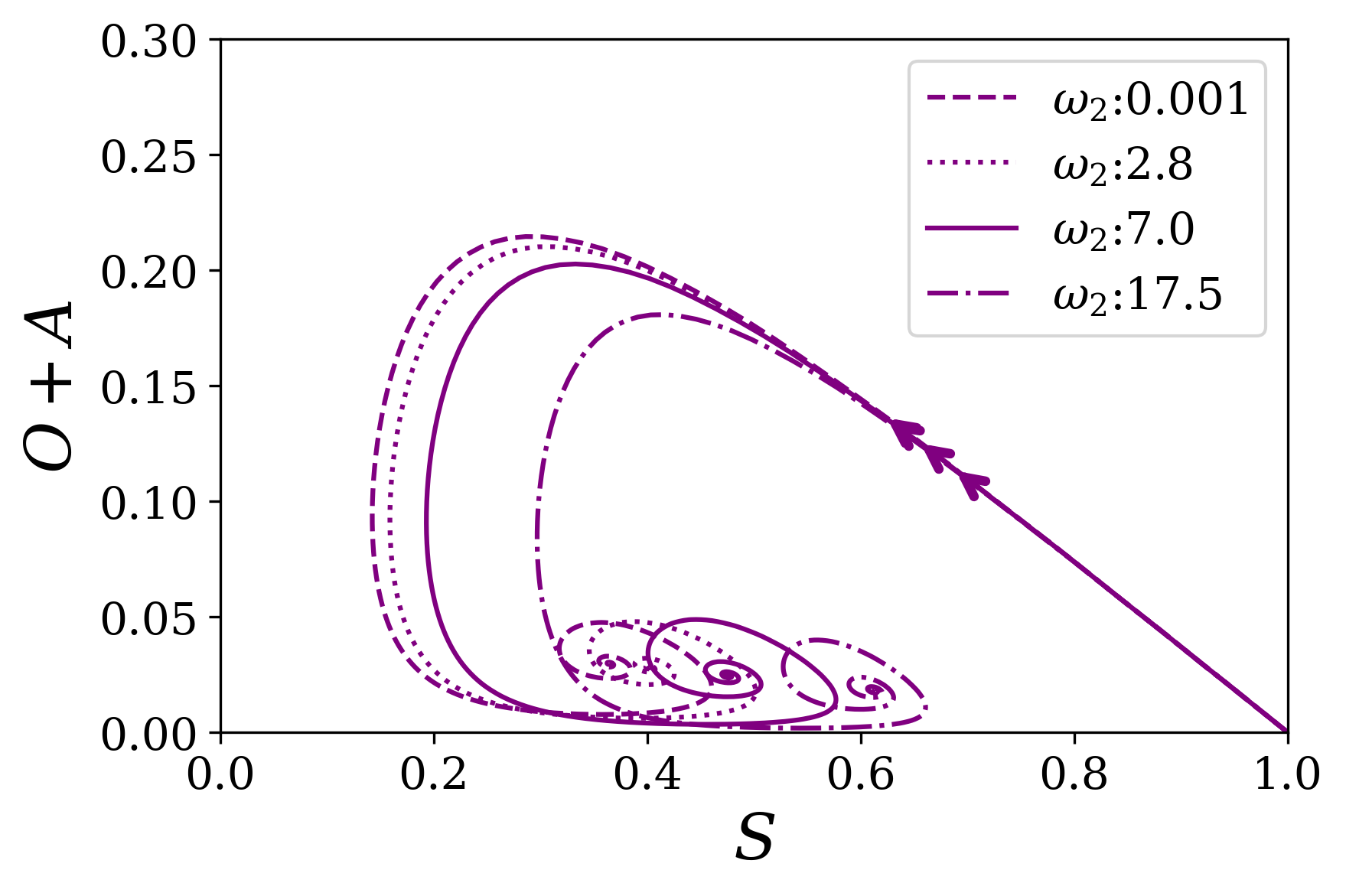}
    \end{subfigure}
    \hfill
    \begin{subfigure}[b]{0.32\textwidth}
    \caption{}
    \includegraphics[width=\textwidth]{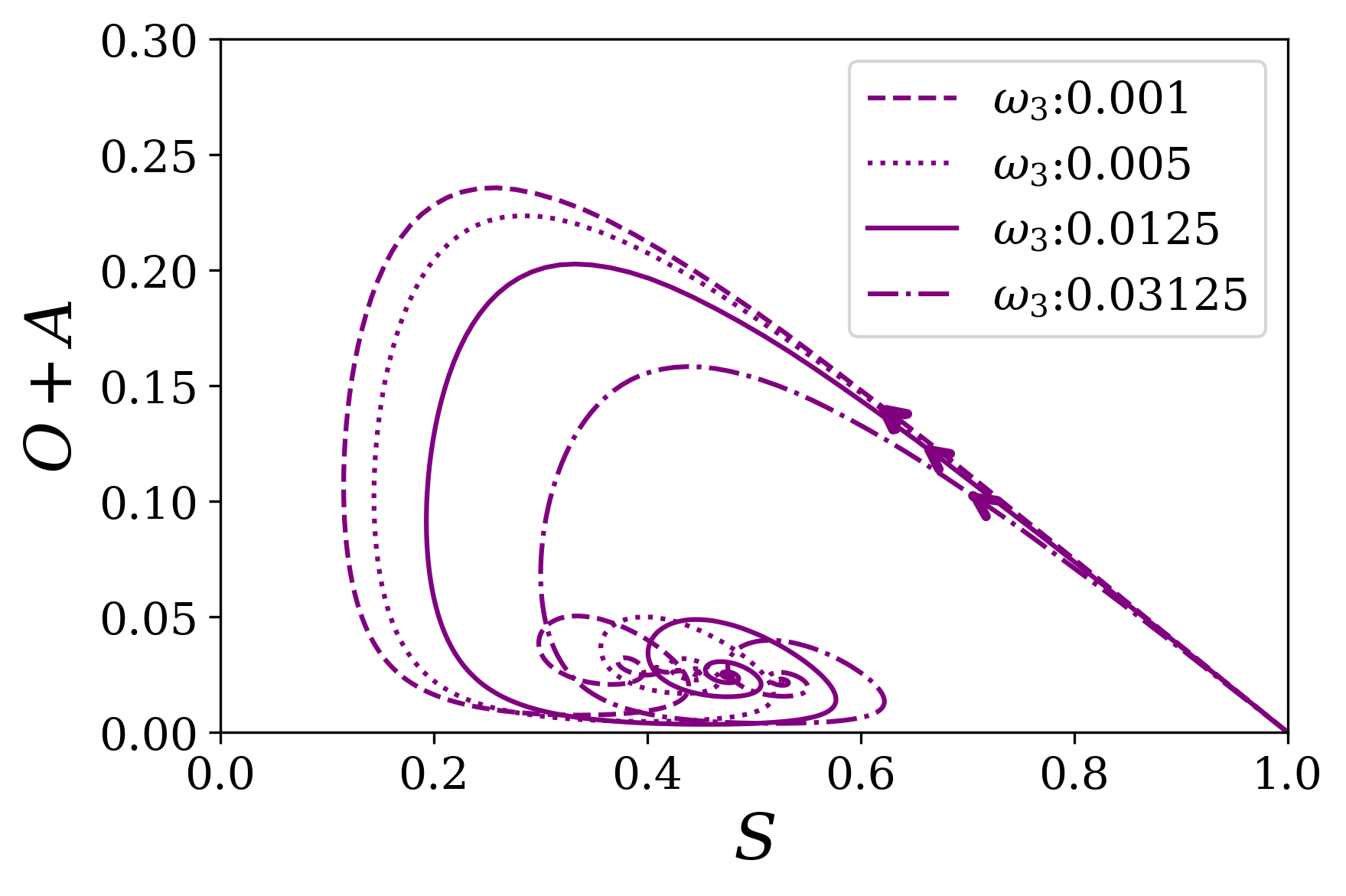}
    \end{subfigure}
    \hfill
    
    \begin{subfigure}[b]{0.32\textwidth}
    \caption{}
    \includegraphics[width=\textwidth]{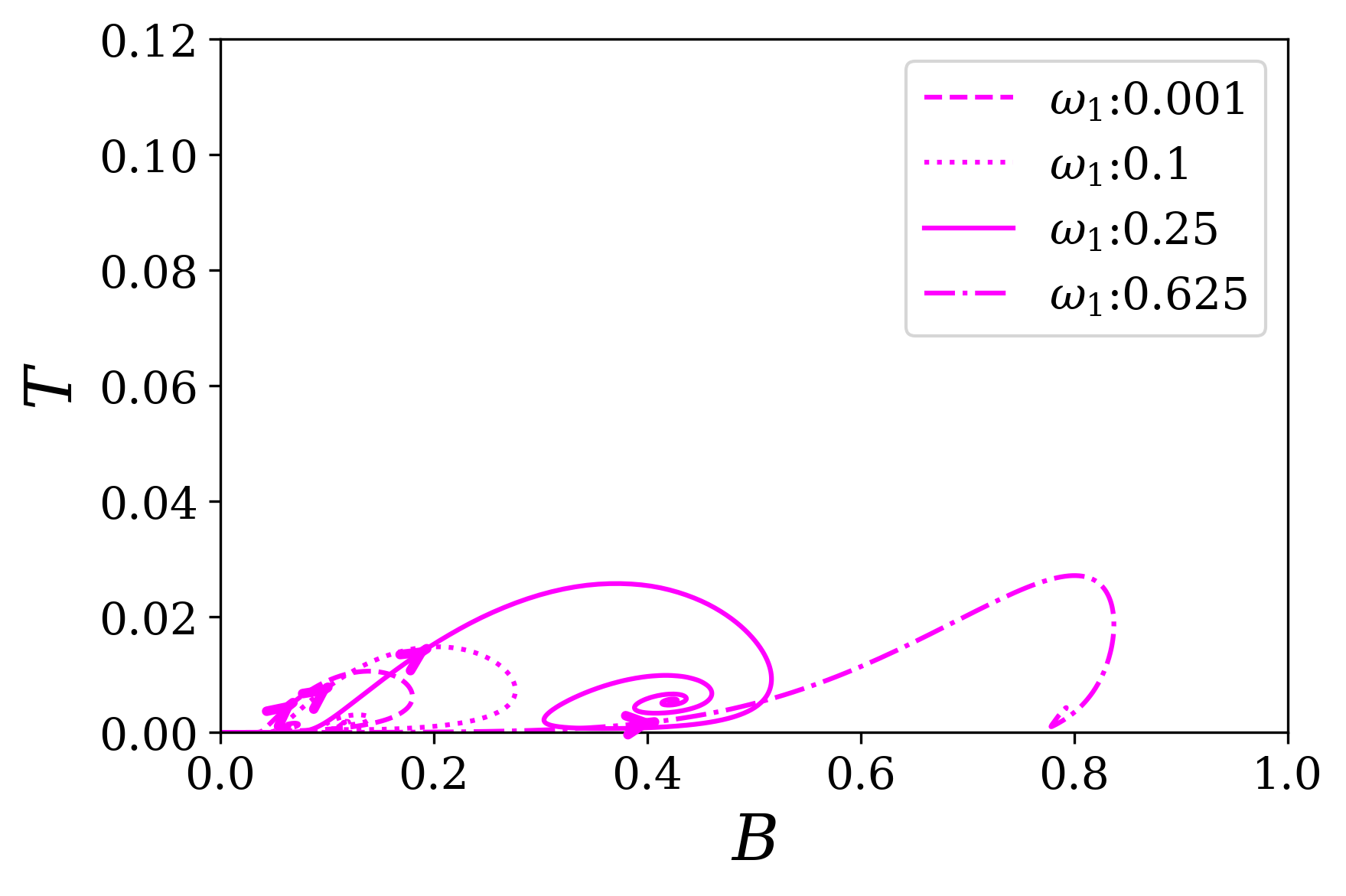}
    \end{subfigure}
    \hfill
    \begin{subfigure}[b]{0.32\textwidth}
    \caption{}
    \includegraphics[width=\textwidth]{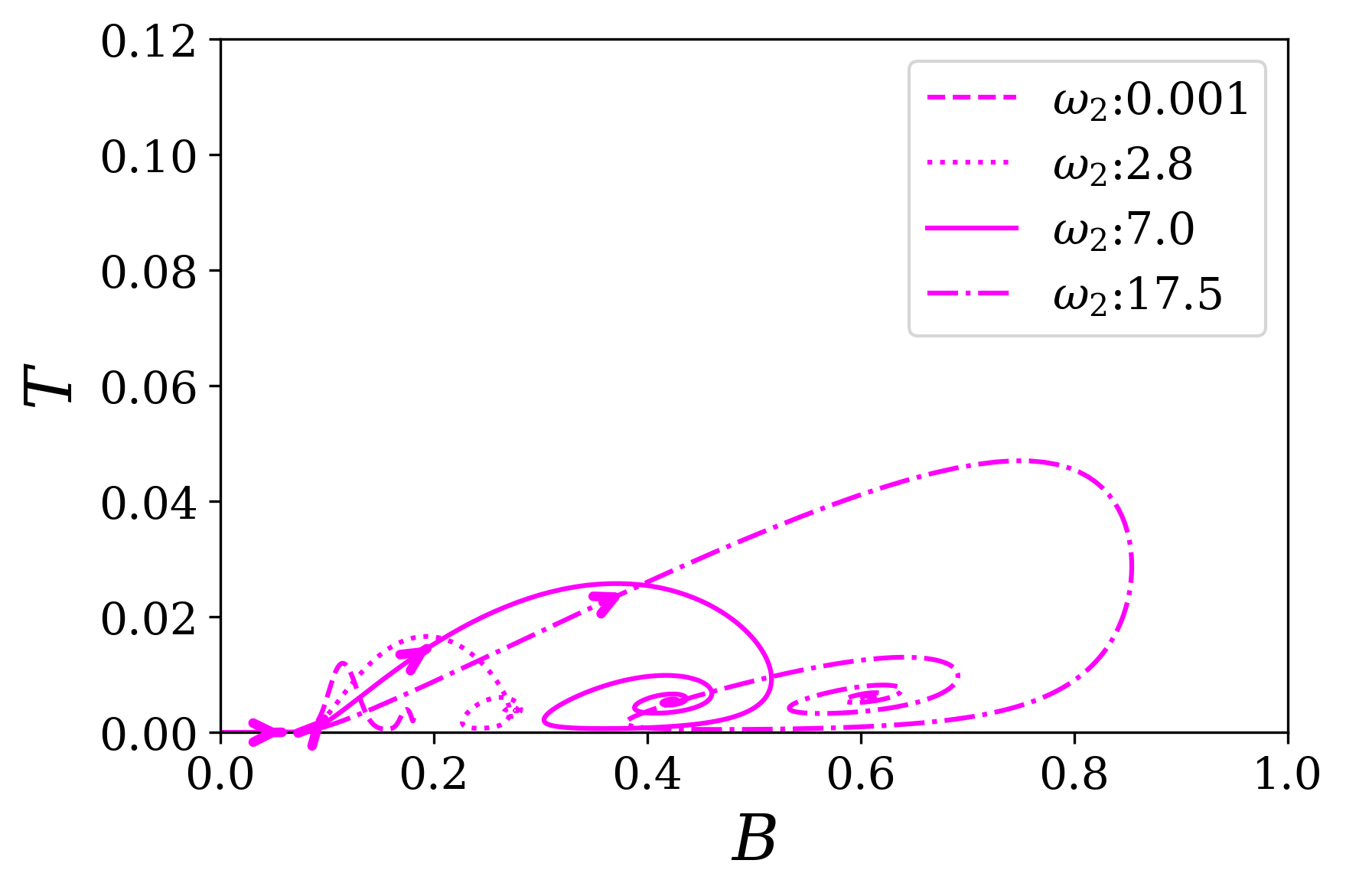}
    \end{subfigure}
    \hfill
    \begin{subfigure}[b]{0.32\textwidth}
    \caption{}
    \includegraphics[width=\textwidth]{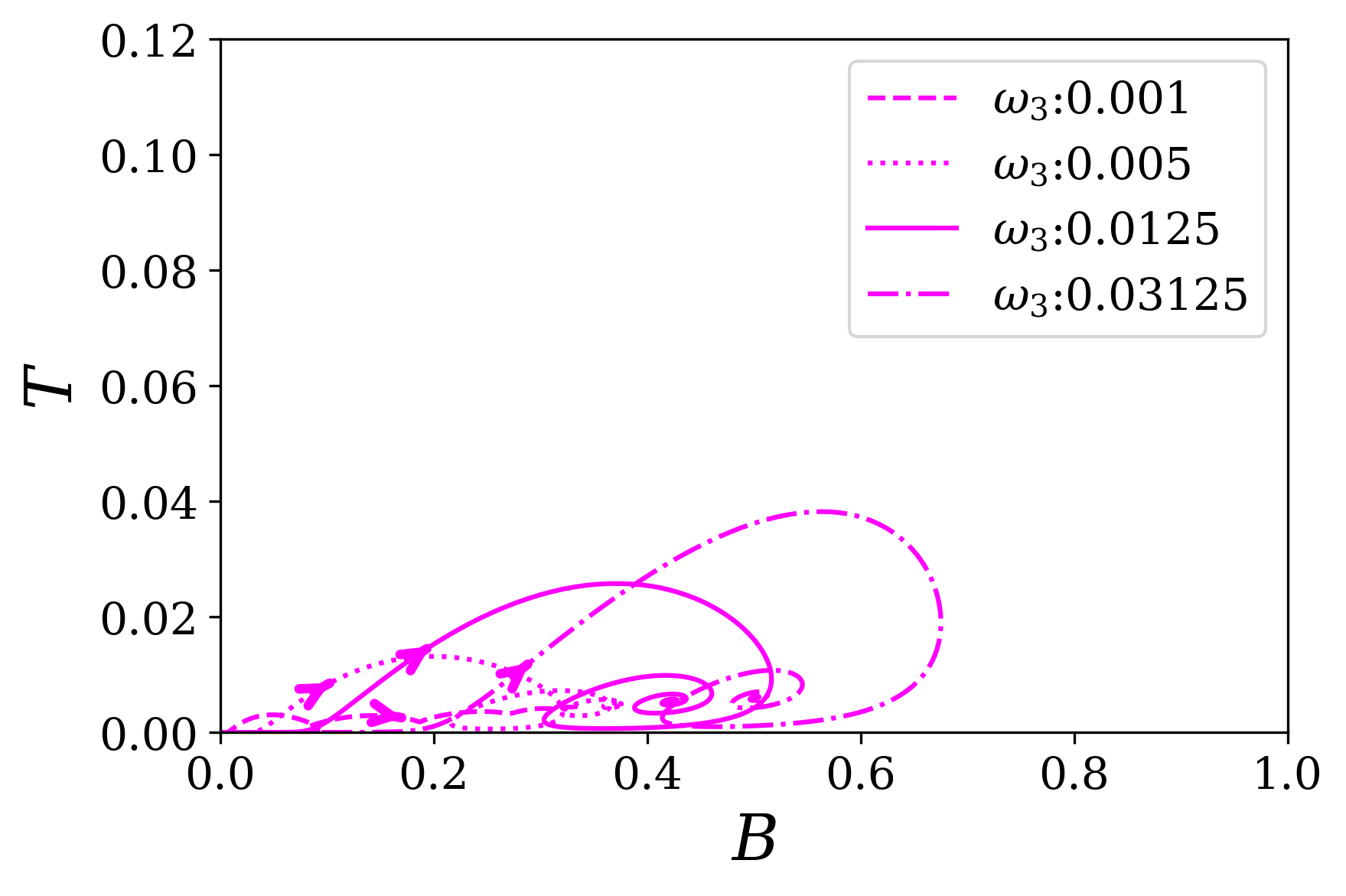}
    \end{subfigure}
    
    \caption{
    \textbf{Phase diagrams for different behavioural parameters fixing ${\Ro^D=3.28}$ in the endemic regime.} Column one corresponds to changes in social influence ($\omega_1$), column two corresponds to changes in perception of illness threat ($\omega_2$), and column three corresponds to spontaneous uptake ($\omega_3$), respectively.  The rows show the susceptible versus observed infection ($S$ vs $T$), the susceptible versus true infection ($S$ vs $O +A$), and the behaviour versus observed infection ($B$ vs $T$) phase planes.  The dashed line represents near removal of the behaviour construct, the dotted lines represent a reduction of the base line value, the solid line represents the baseline values, and the dash-dot lines represent an increase in the baseline values.  The solid lines all represent the baseline values from Table 1 in the manuscript, ensuring the solid lines are comparable across each column.  Each simulation was run with the initial conditions $S_B(0) = 10^{-6}, I_N(0) = 10^{-6}, S_N = 1 - S_B - I_N$ and all other compartments empty: this ensured the early-stage dynamics of each model were approximately comparable.
    }
    \label{fig:phase_endemic_r0D}
\end{figure}


\begin{thebibliography}{10}

\bibitem{Daley:1965}
Daley DJ, Kendall DG.
\newblock {Stochastic Rumours}.
\newblock IMA Journal of Applied Mathematics. 1965;1(1):42-55.

\bibitem{Centola:2010}
Centola D.
\newblock {The Spread of Behavior in an Online Social Network Experiment}.
\newblock Science. 2010;329(5996):1194-7.

\bibitem{Cori:2009}
Cori A, Boëlle PY, Thomas G, Leung GM, Valleron AJ.
\newblock {Temporal Variability and Social Heterogeneity in Disease
  Transmission: The Case of {SARS in Hong Kong}}.
\newblock {PLOS Computational Biology}. 2009;5(8):e1000471.

\bibitem{Funk:2009}
Funk S, Gilad E, Watkins C, Jansen VAA.
\newblock {The spread of awareness and its impact on epidemic outbreaks}.
\newblock PNAS. 2009;106(16):6872-7.

\bibitem{Poletti:2009}
Poletti P, Caprile B, Ajelli M, Pugliese A, Merler S.
\newblock {Spontaneous behavioural changes in response to epidemics}.
\newblock Journal of theoretical biology. 2009;260(1):31-40.

\bibitem{Kiss:2010}
Kiss IZ, Cassell J, Recker M, Simon PL.
\newblock {The impact of information transmission on epidemic outbreaks}.
\newblock Mathematical Biosciences. 2010;225(1):1-10.

\bibitem{Kamp:2010}
Kamp C.
\newblock {Demographic and behavioural change during epidemics}.
\newblock Procedia Computer Science. 2010;1(1):2253-9.

\bibitem{Perra:2011}
Perra N, Balcan D, Gonçalves B, Vespignani A.
\newblock {Towards a Characterization of Behavior-Disease Models}.
\newblock PLOS ONE. 2011;6(8):e23084.

\bibitem{Funk:2010}
Funk S, Salathé M, Jansen VAA.
\newblock {Modelling the influence of human behaviour on the spread of
  infectious diseases: a review}.
\newblock Journal of The Royal Society Interface. 2010;7(50):1247-56.

\bibitem{Manfredi:2013}
Manfredi P, D'Onofrio A.
\newblock {Modeling the Interplay Between Human Behavior and the Spread of
  Infectious Diseases}.
\newblock New York: Springer; 2013.

\bibitem{Wang:2015}
Wang Z, Andrews MA, Wu ZX, Wang L, Bauch CT.
\newblock {Coupled disease–behavior dynamics on complex networks: A review}.
\newblock Physics of Life Reviews. 2015;15:1-29.

\bibitem{House:2011}
House T.
\newblock {Modelling behavioural contagion}.
\newblock Journal of The Royal Society Interface. 2011;8(59):909-12.

\bibitem{daSilva:2019}
da~Silva PCV, Vel\'asquez-Rojas F, Connaughton C, Vazquez F, Moreno Y,
  Rodrigues FA.
\newblock {Epidemic spreading with awareness and different timescales in
  multiplex networks}.
\newblock Physical Review E. 2019;100:032313.

\bibitem{Kolok:2025}
Kolok CB, \'Odor G, Keliger D, Karsai M.
\newblock {Epidemic paradox induced by awareness driven network dynamics}.
\newblock Physical Review Research. 2025;7:L012061.

\bibitem{Gozzi:2024}
Gozzi N, Perra N, Vespignani A.
\newblock {Comparative Evaluation of Behavioral-Epidemic Models Using
  {COVID-19} Data}.
\newblock [medRxiv:2024110824316998]. 2024.
\newblock Available from:
  \url{https://www.medrxiv.org/content/early/2024/11/10/2024.11.08.24316998}.

\bibitem{Hill:2024}
Hill EM, Ryan M, Haw D, Lynch MP, McCabe R, Milne AE, et~al.
\newblock {Integrating human behaviour and epidemiological modelling: unlocking
  the remaining challenges}.
\newblock Mathematics in Medical and Life Sciences. 2024;1(1):2429479.

\bibitem{Marshall:2022}
Marshall GC, Skeva R, Jay C, Silva MEP, Fyles M, House T, et~al.
\newblock {Public perceptions and interactions with {UK COVID-19} Test, Trace
  and Isolate policies, and implications for pandemic infectious disease
  modelling [version 1; peer review: awaiting peer review] }.
\newblock F1000Research. 2022;11(1005).

\bibitem{Shearer:2024}
Shearer FM, McCaw JM, Ryan GE, Hao T, Tierney NJ, Lydeamore MJ, et~al.
\newblock {Estimating the impact of test–trace–isolate–quarantine systems
  on {SARS-CoV-2} transmission in {A}ustralia}.
\newblock Epidemics. 2024;47:100764.

\bibitem{Ryan:2024}
Ryan M, Brindal E, Roberts M, Hickson RI.
\newblock {A behaviour and disease transmission model: incorporating the Health
  Belief Model for human behaviour into a simple transmission model}.
\newblock Journal of The Royal Society Interface. 2024;21(215):20240038.

\bibitem{Ryan:2025}
Ryan M, Brindal E, Hickson RI.
\newblock {Behaviour and infection feedback loops inform early-stage behaviour
  emergence and the efficacy of interventions}.
\newblock Mathematics in Medical and Life Sciences. 2025;2(1):2452444.

\bibitem{Eales:2024}
Eales O, Teo M, Price DJ, Hao T, Ryan GE, Senior KL, et~al.
\newblock Temporal trends in test-seeking behaviour during the {COVID-19}
  pandemic.
\newblock medRxiv. 2024.
\newblock Available from:
  \url{https://www.medrxiv.org/content/early/2024/06/07/2024.06.06.24308566}.

\bibitem{Sayyar:2023}
Sayyar G, R{\"o}st G.
\newblock Epidemic Patterns of Emerging Variants with Dynamical Social
  Distancing.
\newblock In: Mondaini RP, editor. Trends in Biomathematics: Modeling
  Epidemiological, Neuronal, and Social Dynamics: Selected Works from the
  BIOMAT Consortium Lectures, Rio de Janeiro, Brazil, 2022. Cham: Springer
  Nature Switzerland; 2023. p. 215-32.

\bibitem{Karimizadeh:2023}
Karimizadeh Z, Dowran R, Mokhtari-Azad T, Shafiei-Jandaghi NZ.
\newblock {The reproduction rate of severe acute respiratory syndrome
  coronavirus 2 different variants recently circulated in human: a narrative
  review}.
\newblock European Journal of Medical Research. 2023;28(1):94.
\newblock Available from: \url{https://doi.org/10.1186/s40001-023-01047-0}.

\bibitem{Byambasuren:2020}
Byambasuren O, Cardona M, Bell K, Clark J, McLaws ML, Glasziou P.
\newblock {Estimating the extent of asymptomatic COVID-19 and its potential for
  community transmission: systematic review and meta-analysis}.
\newblock Official Journal of the Association of Medical Microbiology and
  Infectious Disease Canada. 2020;5(4):223-34.

\bibitem{CDNA:2024}
{Australian Centre for Disease Control}.
\newblock {Coronavirus Disease 2019 (COVID-19): CDNA National Guidelines for
  Public Health Units}.
\newblock Canberra, ACT, Australia: {Australian Government Department for
  Health and Aged Care}; 2024. 8.
\newblock Available from:
  \url{https://www.health.gov.au/sites/default/files/2024-06/coronavirus-covid-19-cdna-national-guidelines-for-public-health-units_0.pdf}.

\bibitem{Stein:2023}
Stein C, Nassereldine H, Sorensen RJ, Amlag JO, Bisignano C, Byrne S, et~al.
\newblock {Past SARS-CoV-2 infection protection against re-infection: a
  systematic review and meta-analysis}.
\newblock The Lancet. 2023;401(10379):833-42.

\bibitem{Eales:2025}
Eales O, Teo M, Price DJ, Hao T, Ryan GE, Senior KL, et~al.
\newblock Temporal trends in te st-seeking behaviour during the {COVID-19}
  pandemic.
\newblock Mathematics in Medical and Life Sciences. 2025;2(1):2521858.

\bibitem{Kucharski:2020}
Kucharski AJ, Klepac P, Conlan AJ, Kissler SM, Tang ML, Fry H, et~al.
\newblock Effectiveness of isolation, testing, contact tracing, and physical
  distancing on reducing transmission of SARS-CoV-2 in different settings: a
  mathematical modelling study.
\newblock The Lancet infectious diseases. 2020;20(10):1151-60.

\bibitem{Bedson:2021}
Bedson J, Skrip LA, Pedi D, Abramowitz S, Carter S, Jalloh MF, et~al.
\newblock {A review and agenda for integrated disease models including social
  and behavioural factors}.
\newblock Nature human behaviour. 2021;5(7):834-46.

\bibitem{Borjas:2020}
Borjas GJ.
\newblock {Demographic determinants of testing incidence and COVID-19
  infections in New York City neighborhoods}.
\newblock National Bureau of Economic Research; 2020.

\bibitem{Mena:2021}
Mena GE, Martinez PP, Mahmud AS, Marquet PA, Buckee CO, Santillana M.
\newblock {Socioeconomic status determines COVID-19 incidence and related
  mortality in Santiago, Chile}.
\newblock Science. 2021;372(6545):eabg5298.

\bibitem{Bajaj:2024}
Bajaj S, Chen S, Creswell R, Naidoo R, Tsui JL, Kolade O, et~al.
\newblock {COVID-19 testing and reporting behaviours in England across
  different sociodemographic groups: a population-based study using testing
  data and data from community prevalence surveillance surveys}.
\newblock The Lancet Digital Health. 2024;6(11):e778-90.

\bibitem{Guajardo:2025}
{Uribe Guajardo} MG, Moore C, Giannopoulos V, Liu H, Tickle A, Adily P, et~al.
\newblock {The impact of contextual socioeconomic and demographic
  characteristics of residents on COVID-19 outcomes during public health
  restrictions in Sydney, Australia}.
\newblock Australian and New Zealand Journal of Public Health.
  2025;49(2):100228.
\newblock Available from:
  \url{https://www.sciencedirect.com/science/article/pii/S1326020025000093}.

\bibitem{Embrett:2022}
Embrett M, Sim SM, Caldwell HA, Boulos L, Yu Z, Agarwal G, et~al.
\newblock {Barriers to and strategies to address COVID-19 testing hesitancy: a
  rapid scoping review}.
\newblock BMC public health. 2022;22(1):750.

\bibitem{Zhang:2022}
Zhang JC, Christensen KL, Leuchter RK, Vangala S, Han M, Croymans DM.
\newblock {Examining the role of COVID-19 testing availability on intention to
  isolate: A Randomized hypothetical scenario}.
\newblock Plos one. 2022;17(2):e0262659.

\bibitem{UKHSA:2022}
UKHSA. Weekly statistics for {NHS} {Test} and {Trace} ({England}): 2 to 15
  {June} 2022; 2022.
\newblock Available from:
  \url{https://www.gov.uk/government/publications/weekly-statistics-for-nhs-test-and-trace-england-2-to-15-june-2022}.

\bibitem{Green:2021}
Green MA, García-Fiñana M, Barr B, Burnside G, Cheyne CP, Hughes D, et~al.
\newblock Evaluating social and spatial inequalities of large scale rapid
  lateral flow {SARS}-{CoV}-2 antigen testing in {COVID}-19 management: {An}
  observational study of {Liverpool}, {UK} ({November} 2020 to {January} 2021).
\newblock The Lancet Regional Health - Europe. 2021 Jul;6:100107.
\newblock Available from:
  \url{https://www.sciencedirect.com/science/article/pii/S2666776221000843}.

\bibitem{NitaBharti:2021}
{Nita Bharti}.
\newblock {Linking human behaviors and infectious disease}.
\newblock Proceedings of the National Academy of Sciences of the United States
  of America. 2021 2;118(6).

\bibitem{Bruin:2022}
de~Bruin M, Suk JE, Baggio M, Blomquist SE, Falcon M, Forjaz MJ, et~al.
\newblock {Behavioural insights and the evolving COVID-19 pandemic}.
\newblock Eurosurveillance. 2022;27(18):pii=2100615.
\newblock Available from:
  \url{https://www.eurosurveillance.org/content/10.2807/1560-7917.ES.2022.27.18.2100615}.

\end{thebibliography}
\end{document}